\documentclass[english,10pt,twocolumn,twoside]{IEEEtran}
\usepackage[T1]{fontenc}
\usepackage[latin9]{inputenc}
\usepackage{array}
\usepackage{bm}
\usepackage{multirow}
\usepackage{amsthm}
\usepackage{amsmath}
\usepackage{amssymb}
\usepackage{graphicx}

\makeatletter

\providecommand{\tabularnewline}{\\}

\theoremstyle{plain}
\newtheorem{thm}{\protect\theoremname}
\theoremstyle{plain}
\newtheorem{lem}[thm]{\protect\lemmaname}

\usepackage{colortbl}
 \usepackage{cite}

\makeatother

\usepackage{babel}
\providecommand{\lemmaname}{Lemma}
\providecommand{\theoremname}{Theorem}

\begin{document}

\title{A Unifying Framework for Adaptive Radar Detection in Homogeneous
plus Structured Interference-Part II: Detectors Design}

\author{D.~Ciuonzo,\IEEEmembership{~Member,~IEEE,} A.~De~Maio,~\IEEEmembership{Fellow,~IEEE,}
and D. Orlando, \IEEEmembership{Senior~Member,~IEEE}\thanks{Manuscript received 17th July 2015.\protect \\
D. Ciuonzo and A. De Maio are with University of Naples \textquotedbl{}Federico
II\textquotedbl{}, DIETI, Via Claudio 21, 80125 Naples, Italy. E-mail:
domenico.ciuonzo@ieee.org; ademaio@unina.it. \protect \\
D. Orlando is with Università degli Studi \textquotedblleft Niccolò
Cusano\textquotedblright , Via Don Carlo Gnocchi 3, 00166 Roma, Italy.
E-mail: danilo.orlando@unicusano.it.\protect \\
}}
\maketitle
\begin{abstract}
This paper deals with the problem of adaptive multidimensional/multichannel
signal detection in homogeneous Gaussian disturbance with unknown
covariance matrix and structured (unknown) deterministic interference.
The aforementioned problem extends the well-known Generalized Multivariate
Analysis of Variance (GMANOVA) tackled in the open literature. In
a companion paper, we have obtained the Maximal Invariant Statistic
(MIS) for the problem under consideration, as an enabling tool for
the design of suitable detectors which possess the Constant False-Alarm
Rate (CFAR) property. Herein, we focus on the development of several
theoretically-founded detectors for the problem under consideration.
First, all the considered detectors are shown to be function of the
MIS, thus proving their CFARness property. Secondly, coincidence or
statistical equivalence among some of them in such a general signal
model is proved. Thirdly, strong connections to well-known simpler
scenarios found in adaptive detection literature are established.
Finally, simulation results are provided for a comparison of the proposed
receivers. \end{abstract}

\begin{IEEEkeywords}
Adaptive Radar Detection, CFAR, Invariance Theory, Maximal Invariants,
Double-subspace Model, GMANOVA, Coherent Interference.
\end{IEEEkeywords}

\section{Introduction}

\subsection{Motivation and Related Works}

\IEEEPARstart{T}{he problem} of adaptive detection has been object
of great interest in the last decades. Many works appeared in the
open literature, dealing with the design and performance analysis
of suitable detectors in several specific settings (see for instance
\cite{Gini2001} and references therein). 

As introduced in a companion paper, herein we focus on a signal model
which generalizes that of GMANOVA \cite{Dogandzic2003} by considering
an additional unknown double-subspace structured deterministic interference.
Such model is here denoted as I-GMANOVA. The I-GMANOVA model is very
general and comprises many adaptive detection setups as special instances,
ranging from point-like targets (resp. interference) \cite{Kelly1986}
to extended ones \cite{Conte2001}, from a single-steering assumption
to a vector subspace one \cite{Raghavan1996,A.DeMaio2014}, and the
GMANOVA model itself \cite{Dogandzic2003}, only to mention a few
examples. We recall that attractive modifications of GMANOVA have
also appeared in the recent literature \cite{Xu2006,Xu2006a}, focusing
on the design of computationally-efficient approximate ML estimators
when the unknown signal matrix is constrained to be diagonal \cite{Xu2006}
or block-diagonal \cite{Xu2006a}.

In the case of composite hypothesis testing, the three widely-used
design criteria are the Generalized Likelihood Ratio Test (GLRT),
the Rao test, and the Wald test \cite{Kay1998}. Their use is well-established
in the context of adaptive detection literature \cite{Conte2001,Conte2003,A.DeMaio2004,Bandiera2007,A.DeMaio2007};
more important they are known to share the same asymptotic performance
\cite{Kay1998}. However, in the finite-sample case their performance
differ and their relative assessment depends on the specific hypothesis
testing model being considered. Such statement holds true unless some
specific instances occur, such as in \cite{A.DeMaio2008}, where it
is proved that they are statistically equivalent in the case of point-like
targets and a partially-homogeneous scenario. Other than the aforementioned
detectors, in the context of radar adaptive detection it is also customary
to consider their two-step variations, with the two-step GLRT (2S-GLRT)
being the most common. Those are typically obtained by designing the
detector under the assumption of the a known disturbance covariance
matrix and replacing it with a sample estimate based on the so-called
secondary (or signal-free) data \cite{Kelly1989}. 

Furthermore, a few interesting alternative detectors for composite
hypothesis testing are the so-called Durbin (naive) test \cite{Durbin1970}
and the Terrell (Gradient) test \cite{Terrell2002}. These detectors
have been shown to be asymptotically efficient as the aforementioned
well-known criteria. The same rationale applies to the Lawley-Hotelling
(LH) test \cite{Lawley1938}. Though these detectors are well-known
in the statistics field, the development and application of these
decision rules is \emph{less frequently encountered in radar adaptive
detection literature}, e.g., \cite{Kelly1989,A.DeMaio2010}. The reason
is that an important prerequisite for a wide-spread application of
an adaptive detection algorithm consists in showing its CFARness with
respect to the nuisance parameters; in this respect, the assessment
of such property in radar adaptive detection literature has been somewhat
lacking.

Of course, the use of GMANOVA model in the context adaptive radar
detection is not new and dates back to the the milestone study in
\cite{Kelly1989}, where the development and analysis of the GLRT
was first proposed. A similar work was then presented years later
in \cite{Burgess1996}, where the focus was on the design of a compression
matrix prior to the detection process, aimed at reducing the computational
burden and minimizing the performance loss with respect to the standard
processing. More recently, GLRT, Rao and Wald tests were developed
and compared under the GMANOVA model \cite{Liu2014}, along with some
other heuristic detectors. Unfortunately, albeit the CFARness of the
proposed detectors was proved, no clear connection to the MIS was
established. More importantly, no (structured) deterministic interference
was considered in all the aforementioned works; the proposed I-GMANOVA
model is aimed at filling such gap.

We point out that the closest study to ours (in terms of interference
accounting) is the work in \cite{Bandiera2007}, where range-spread
and vector subspace targets and interference are considered; however
the sole GLRT and 2S-GLRT are derived and analyzed. Similarly, a Rao
test (and its two-step version) is recently obtained in \cite{Liu2015}.
It is worth noticing that the model considered by the aforementioned
works is included in the I-GMANOVA model, and can be readily obtained
by assuming a canonical form for the right-subspace matrix of both
the signal and the interference.

Summarizing, in our opinion the huge (but scattered) literature on
adaptive detection in many case-specific signal models and the presence
of several detectors (developed on theoretically-solid assumptions)
for generic composite hypothesis testing problems lacks a comprehensive
and systematic analysis. First, such analysis may help the generic
designer in readily obtaining a plethora of suitable adaptive detectors
in some relevant scenarios which can be fitted into the considered
I-GMANOVA model. Secondly, the development of detectors closed-form
expressions under I-GMANOVA model may allow to easily claim some general
statistical equivalence results than those already noticed in some
special instances (see e.g., \cite{A.DeMaio2004,A.DeMaio2010,Liu2013}).
Thirdly, the available explicit expression for each detector allows
for a systematic analysis of its (possible) CFARness (under a quite
general signal model). The latter study is greatly simplified when
knowledge of the explicit form of the MIS is available for the considered
problem; in this respect, the derivation of the MIS and its analysis,
object of a companion paper, fulfills this need.

\subsection{Summary of the contributions and Paper Organization}

The main contributions of the second part of this work are thus related
to detectors development and CFAR property analysis and can summarized
as follows:
\begin{itemize}
\item Starting from the canonical form obtained in our companion paper,
for the general model under investigation we derive closed-form expressions
for the ($i$) GLRT, ($ii$) Rao test, ($iii)$ Wald test, ($iv$)
Gradient test ($v$) Durbin test, ($vi$) two-step GLRT (2S-GLRT),
and ($vii$) LH test. As an interesting byproduct of our derivation,
we show that Durbin test is \emph{statistically equivalent} to the
Rao test for the considered (adaptive) detection problem, thus extending
the findings in \cite{A.DeMaio2010}, obtained for the simpler case
of a point-like target without interference. Similarly, we demonstrate
the statistical equivalence between Wald test and 2S-GLRT, thus extending
the works in \cite{A.DeMaio2004} and \cite{Liu2013}, concerning
the special instances of point-like targets (no interference) and
multidimensional signals, respectively;
\item The general expressions of the receivers are exploited to analyze
special cases of interest, such as: ($a$) vector subspace detection
of point-like targets (with possible structured interference) \cite{Kelly1986,A.DeMaio2004,A.DeMaio2014},
($b$) multidimensional signals \cite{Conte2003,Liu2013}, ($c$)
range-spread (viz. extended) targets \cite{Conte2001,Shuai2012,Raghavan2013},
and ($d$) standard GMANOVA (i.e., without structured interference)
\cite{Kelly1989,Liu2014}. In such special instances, possible coincidence
or statistical equivalence is investigated among the considered detectors;
\item Exploiting the matrix pair form of the MIS obtained in part one, we
show that \emph{all} the considered detector can be expressed as a
function of the MIS, thus proving their CFARness with respect to both
the covariance of the disturbance and the deterministic (structured)
interference;
\item Finally, a simulation results section is provided to compare the proposed
detectors in terms of the relevant parameters and underline the common
trends among them.
\end{itemize}
The remainder of the paper is organized as follows: in Sec.~\ref{sec: Problem formulation},
we describe the hypothesis testing problem under investigation; in
Sec.~\ref{sec: Detectors design}, we obtain the general expressions
for the detectors considered in this paper and we express them as
a function of the MIS; in Sec.~\ref{sec: Detectors in special cases},
we particularize the obtained expressions to the aforementioned special
instances of adaptive detection problems; finally, in Sec.~\ref{sec: Simulation results}
we compare the obtained detectors through simulation results and in
Sec.~\ref{sec: Conclusions} we draw some concluding remarks and
indicate future research directions. Proofs and derivations are confined
to an additional document containing supplemental material\footnote{\emph{Notation} - Lower-case (resp. Upper-case) bold letters denote
vectors (resp. matrices), with $a_{n}$ (resp. $A_{n,m}$) representing
the $n$-th (resp. the $(n,m)$-th) element of the vector $\bm{a}$
(resp. matrix $\bm{A}$); $\mathbb{R}^{N}$, $\mathbb{C}^{N}$, and
$\mathbb{H}^{N\times N}$ are the sets of $N$-dimensional vectors
of real numbers, of complex numbers, and of $N\times N$ Hermitian
matrices, respectively; upper-case calligraphic letters and braces
denote finite sets; $\mathbb{E}\{\cdot\}$, $(\cdot)^{T}$, $(\cdot)^{\dagger}$,
$\mathrm{Tr}\left[\cdot\right]$, $\left\Vert \cdot\right\Vert $,
$\Re\{\cdot\}$ and $\Im\{\cdot\}$, denote expectation, transpose,
Hermitian, matrix trace, Euclidean norm, real part, and imaginary
part operators, respectively; $\bm{0}_{N\times M}$ (resp. $\bm{I}_{N}$)
denotes the $N\times M$ null (resp. identity) matrix; $\bm{0}_{N}$
(resp. $\bm{1}_{N}$) denotes the null (resp. ones) column vector
of length $N$; $\mathrm{vec}(\bm{M})$ stacks the first to the last
column of the matrix $\bm{M}$ one under another to form a long vector;
$\det(\bm{A})$ and $||\bm{A}||_{F}$ denote the determinant and Frobenius
norm of matrix $\bm{A}$; $\bm{A}\otimes\bm{B}$ indicates the Kronecker
product between $\bm{A}$ and $\bm{B}$ matrices; $\frac{\partial f(\bm{x})}{\partial\bm{x}}$
denotes the gradient of scalar valued function $f(\bm{x})$ w.r.t.
vector $\bm{x}$ arranged in a column vector, while $\frac{\partial f(\bm{x})}{\partial\bm{x}^{T}}$
its transpose (i.e. a row vector); the symbol ``$\sim$'' means
``distributed as''; $\bm{x}\sim\mathcal{C}\mathcal{N}_{N}(\bm{\mu},\bm{\Sigma})$
denotes a complex (proper) Gaussian-distributed vector $\bm{x}$ with
mean vector $\bm{\mu}\in\mathbb{C}^{N\times1}$ and covariance matrix
$\bm{\Sigma}\in\mathbb{C}^{N\times N}$; $\bm{X}\sim\mathcal{C}\mathcal{N}_{N\times M}(\bm{A},\bm{B},\bm{C})$
denotes a complex (proper) Gaussian-distributed matrix $\bm{X}$ with
mean $\bm{A}\in\mathbb{C}^{N\times M}$ and $\mathrm{Cov}[\mathrm{vec}(\bm{X})]=\bm{B}\otimes\bm{C}$;
$\bm{P}_{A}$ denotes the orthogonal projection of the full-column-rank
matrix $\bm{A}$, that is $\bm{P}_{\bm{A}}\triangleq[\bm{A}(\bm{A}^{\dagger}\bm{A})^{-1}\bm{A}^{\dagger}]$,
while $\bm{P}_{A}^{\perp}$ its complement, that is $\bm{P}_{A}^{\perp}\triangleq(\bm{I}-\bm{P}_{\bm{A}})$.}.

\section{Problem Formulation \label{sec: Problem formulation}}

In a companion paper, we have shown that the considered problem admits
an equivalent (but simpler) formulation by exploiting the so-called
``canonical form'', that is:

\begin{equation}
\begin{cases}
\mathcal{H}_{0}: & \bm{Z}=\bm{A}\,\begin{bmatrix}\bm{B}_{t,0}^{T} & \bm{0}_{M\times r}\end{bmatrix}^{T}\,\bm{C}+\bm{N}\\
\mathcal{H}_{1}: & \bm{Z}=\bm{A}\,\bm{B}_{s}\,\bm{C}+\bm{N}
\end{cases}\label{eq: Transformed data - hypothesis testing problem}
\end{equation}
where we have assumed that a data matrix $\bm{Z}\in\mathbb{C}^{N\times K}$
has been collected. Also, we have adopted the following definitions:
\begin{itemize}
\item $\bm{A}\triangleq\begin{bmatrix}\bm{E}_{t} & \bm{E}_{r}\end{bmatrix}\in\mathbb{C}^{N\times J}$,
where $\bm{E}_{t}\triangleq\begin{bmatrix}\bm{I}_{t} & \bm{0}_{t\times(N-t)}\end{bmatrix}^{T}$
and $\bm{E}_{r}\triangleq\begin{bmatrix}\bm{0}_{r\times t} & \bm{I}_{r} & \bm{0}_{r\times(N-J)}\end{bmatrix}^{T}$
are the (known) left subspaces of the interference and useful signal,
respectively (we have denoted $J\triangleq r+t$);
\item $\bm{B}_{s}\triangleq\begin{bmatrix}\bm{B}_{t,1}^{T} & \bm{B}^{T}\end{bmatrix}^{T}$,
where $\bm{B}_{t,i}\in\mathbb{C}^{t\times M}$ and $\bm{B}\in\mathbb{C}^{r\times M}$
are the (unknown) interference (under $\mathcal{H}_{i}$) and useful
signal matrices, respectively;
\item $\bm{C}\triangleq\begin{bmatrix}\bm{I}_{M} & \bm{0}_{M\times(K-M)}\end{bmatrix}\in\mathbb{C}^{M\times K}$
is the (known) right subspace matrix associated to \emph{both} signal
and interference in canonical form;
\item $\bm{N}$ is a disturbance matrix such that $\bm{N}\sim\mathcal{CN}_{N\times K}(\bm{0}_{N\times K},\bm{I}_{K},\bm{R})$,
where $\bm{R}\in\mathbb{C}^{N\times N}$ is an (unknown) positive
definite covariance matrix \cite{Kelly1989}.
\end{itemize}
We recall that the detection problem in (\ref{eq: Transformed data - hypothesis testing problem})
is tantamount to testing the null hypothesis $\bm{B}=\bm{0}_{r\times M}$
(viz. $||\bm{B}||_{F}=0$, denoted with $\mathcal{H}_{0}$) against
the alternative that $\bm{B}$ is unrestricted (viz. $||\bm{B}||_{F}>0$,
denoted with $\mathcal{H}_{1}$), along with the set of nuisance parameters
$\bm{B}_{t,i}$ and $\bm{R}$. 

In the present manuscript we will consider decision rules which declare
$\mathcal{H}_{1}$ (resp. $\mathcal{H}_{0}$) if $\Phi(\bm{Z})>\eta$
(resp. $\Phi(\bm{Z})<\eta$), where $\Phi(\cdot)\in\mathbb{C}^{N\times K}\rightarrow\mathbb{R}$
indicates the generic form of a statistic processing the received
data $\bm{Z}$ and $\eta$ denotes the threshold to be set in order
to achieve a predetermined probability of false alarm ($P_{fa}$).

As a preliminary step towards the derivation of suitable detectors
for the problem at hand, we also give the following auxiliary definitions:
\begin{itemize}
\item $\bm{b}_{R}\in\mathbb{R}^{rM\times1}$ and $\bm{b}_{I}\in\mathbb{R}^{rM\times1}$
are obtained as $\bm{b}_{R}\triangleq\Re\{\bm{b}\}$ and $\bm{b}_{I}\triangleq\Im\{\bm{b}\}$,
respectively, where we have defined $\bm{b}\triangleq\mathrm{vec}(\bm{B})$;
\item $\bm{\theta}_{r}\triangleq\left[\begin{array}{cc}
\bm{b}_{R}^{T} & \bm{b}_{I}^{T}\end{array}\right]^{T}\in\mathbb{R}^{2rM\times1}$ is the vector collecting the parameters of interest;
\item $\bm{\theta}_{s}\triangleq\begin{bmatrix}\bm{\theta}_{s,a}^{T} & \bm{\theta}_{s,b}^{T}\end{bmatrix}^{T}\in\mathbb{R}^{(2tM+N^{2})\times1}$
is the vector of nuisance parameters containing: ($a$) $\bm{\theta}_{s,a}\triangleq\left[\begin{array}{cc}
\bm{b}_{t,R}^{T} & \bm{b}_{t,I}^{T}\end{array}\right]^{T}\in\mathbb{R}^{2tM\times1}$ where $\bm{b}_{t,R}$ and $\bm{b}_{t,I}$ are the vectors obtained
as $\bm{b}_{t,R}\triangleq\Re\{\bm{b}_{t}\}$ and $\bm{b}_{t,I}\triangleq\Im\{\bm{b}_{t}\}$,
respectively, where $\bm{b}_{t}\triangleq\mathrm{vec}(\bm{B}_{t})$
(i.e. $\bm{B}_{t,i}$ under $\mathcal{H}_{i}$); ($b$) $\bm{\theta}_{s,b}$
contains in a given order\footnote{More specifically, $\bm{\theta}_{s,b}\triangleq\bm{\Xi}(\bm{R})$,
where $\bm{\Xi}(\cdot)$ denotes the one-to-one mapping providing
$\bm{\theta}_{s,b}$ from $\bm{R}$.} the real and imaginary parts of the off-diagonal entries together
with the diagonal elements of $\bm{R}$; 
\item $\bm{\theta}\triangleq\left[\begin{array}{cc}
\bm{\theta}_{r}^{T} & \bm{\theta}_{s}^{T}\end{array}\right]^{T}\in\mathbb{R}^{(2JM+N^{2})\times1}$ is the overall unknown parameter vector;
\item $\widehat{\bm{\theta}}_{0}\triangleq\left[\begin{array}{cc}
\bm{\theta}_{r,0}^{T} & \widehat{\bm{\theta}}_{s,0}^{T}\end{array}\right]^{T}$, with $\widehat{\bm{\theta}}_{s,0}$ denoting the Maximum Likelihood
(ML) estimate of $\bm{\theta}_{s}$ under $\mathcal{H}_{0}$ and $\bm{\theta}_{r,0}=\bm{0}_{2rM}$
(that is, the true value of $\bm{\theta}_{r}$ under $\mathcal{H}_{0}$);
\item $\widehat{\bm{\theta}}_{1}\triangleq\left[\begin{array}{cc}
\widehat{\bm{\theta}}_{r,1}^{T} & \widehat{\bm{\theta}}_{s,1}^{T}\end{array}\right]^{T}$, with $\widehat{\bm{\theta}}_{r,1}$ and $\widehat{\bm{\theta}}_{s,1}$
denoting the ML estimates of $\bm{\theta}_{r}$ and $\bm{\theta}_{s}$,
respectively, under $\mathcal{H}_{1}$.
\end{itemize}
The probability density function (pdf) of $\bm{Z}$, when the hypothesis
$\mathcal{H}_{1}$ is in force, is denoted with $f_{1}(\cdot)$ and
it is given in closed form as:
\begin{gather}
f_{1}(\bm{Z};\bm{B}_{s},\bm{R})=\pi^{-NK}\mathrm{det}(\bm{R})^{-K}\nonumber \\
\times\exp\left(-\mathrm{Tr}\left[\bm{R}^{-1}(\bm{Z}-\bm{A}\bm{B}_{s}\bm{C})(\bm{Z}-\bm{A}\bm{B}_{s}\bm{C})^{\dagger}\right]\right)\,,\label{eq: f_1()}
\end{gather}
while the corresponding pdf under $\mathcal{H}_{0}$, denoted in the
following with $f_{0}(\cdot)$, is similarly obtained when replacing
$\bm{A}\bm{B}_{s}\bm{C}$ with $\bm{E}_{t}\bm{B}_{t,0}\bm{C}$ in
Eq. (\ref{eq: f_1()}). In the following, in order for our analysis
to apply, we will assume that the condition $(K-M)\geq N$ holds.
Such condition is typically satisfied in practical adaptive detection
setups \cite{Kelly1989}.

\subsection{MIS for the considered problem\label{sub: MIS recall}}

In what follows, we recall the MIS for the hypothesis testing under
investigation, obtained in our companion paper. The mentioned statistic
will be exploited in Sec. \ref{sec: Detectors design} to ascertain
the CFARness of each considered detector. Before proceeding further,
let
\begin{equation}
\bm{V}_{c,1}\triangleq\begin{bmatrix}\bm{I}_{M}\\
\bm{0}_{(K-M)\times M}
\end{bmatrix},\quad\,\bm{V}_{c,2}\triangleq\begin{bmatrix}\bm{0}_{M\times(K-M)}\\
\bm{I}_{K-M}
\end{bmatrix},
\end{equation}
and observe that $\bm{P}_{\bm{C}^{\dagger}}=(\bm{V}_{c,1}\bm{V}_{c,1}^{\dagger})$
and $\bm{P}_{\bm{C}^{\dagger}}^{\perp}=(\bm{V}_{c,2}\bm{V}_{c,2}^{\dagger})$.
Given these definitions, we denote (as in Part I): ($i$) $\bm{Z}_{c}\triangleq\,(\bm{Z}\bm{V}_{c,1})\in\mathbb{C}^{N\times M}$,
$(ii$) $\bm{Z}_{c,\perp}\triangleq\,(\bm{Z}\,\bm{V}_{c,2})\in\mathbb{C}^{N\times(K-M)}$
and ($iii$) $\bm{S}_{c}\triangleq\,(\bm{Z}_{c,\perp}\bm{Z}_{c,\perp}^{\dagger})=(\bm{Z}\bm{P}_{\bm{C}^{\dagger}}^{\perp}\bm{Z}^{\dagger})\in\mathbb{C}^{N\times N}$. 

It has been shown in our companion paper that the MIS is given by:
\begin{gather}
\bm{T}(\bm{Z}_{c},\bm{S}_{c})=\begin{cases}
\begin{bmatrix}\bm{T}_{a}\triangleq\left\{ \bm{Z}_{2.3}^{\dagger}\,\bm{S}_{2.3}^{-1}\,\bm{Z}_{2.3}\right\} \\
\bm{T}_{b}\triangleq\left\{ \bm{Z}_{3}^{\dagger}\,\bm{S}_{33}^{-1}\bm{Z}_{3}\right\} 
\end{bmatrix} & J<N\\
\bm{Z}_{2}^{\dagger}\,\bm{S}_{22}^{-1}\bm{Z}_{2} & J=N
\end{cases}\label{eq: MIS_final}
\end{gather}
where $\bm{Z}_{2.3}\triangleq(\bm{Z}_{2}-\bm{S}_{23}\bm{S}_{33}^{-1}\bm{Z}_{3})$
and $\bm{S}_{2.3}\triangleq(\bm{S}_{22}-\bm{S}_{23}\,\bm{S}_{33}^{-1}\bm{S}_{32})$.
Also, we have exploited the following partitioning for matrices $\bm{Z}_{c}$
and $\bm{S}_{c}$:
\begin{equation}
\bm{Z}_{c}=\begin{bmatrix}\bm{Z}_{1}\\
\bm{Z}_{2}\\
\bm{Z}_{3}
\end{bmatrix},\quad\bm{S}_{c}=\begin{bmatrix}\bm{S}_{11} & \bm{S}_{12} & \bm{S}_{13}\\
\bm{S}_{21} & \bm{S}_{22} & \bm{S}_{23}\\
\bm{S}_{31} & \bm{S}_{32} & \bm{S}_{33}
\end{bmatrix}\,,\label{eq: Z_c S_c (block definition)}
\end{equation}
where $\bm{Z}_{1}\in\mathbb{C}^{t\times M}$, $\bm{Z}_{2}\in\mathbb{C}^{r\times M}$,
and $\bm{Z}_{3}\in\mathbb{C}^{(N-J)\times M}$, respectively. Furthermore,
$\bm{S}_{ij}$, $(i,j)\in\{1,2,3\}\times\{1,2,3\}$, is a sub-matrix
whose dimensions can be obtained replacing $1$, $2$ and $3$ with
$t$, $r$ and $(N-J)$, respectively\footnote{Hereinafter, in the case $J=N$, the ``3-components'' are no longer
present in the partitioning.}. Additionally, for notational convenience, we also give the following
definitions that will be used throughout the manuscript:
\begin{equation}
\bm{Z}_{23}\triangleq\begin{bmatrix}\bm{Z}_{2}\\
\bm{Z}_{3}
\end{bmatrix},\quad\bm{S}_{2}\triangleq\begin{bmatrix}\bm{S}_{22} & \bm{S}_{23}\\
\bm{S}_{32} & \bm{S}_{33}
\end{bmatrix}\,.\label{eq: Z2:3 and S2 def}
\end{equation}
Finally we recall that, for the I-GMANOVA model, the induced maximal
invariant equals $\bm{T}_{\mathrm{p}}\triangleq\bm{B}^{\dagger}\,\bm{R}_{2.3}^{-1}\,\bm{B}\in\mathbb{C}^{M\times M}$,
where $\bm{R}_{2.3}$ is analogously defined as $\bm{S}_{2.3}$ when
$\bm{S}_{c}$ is replaced with the true covariance $\bm{R}$.

\section{Detectors design \label{sec: Detectors design}}

In this section we will consider several decision statistics designed
according to well-founded design criteria. Initially, we will concentrate
on the derivation of the well-known GLRT (including its two-step version),
Rao and Wald tests \cite{Kay1998}. 

Then, we will devise the explicit form of recently used detection
statistics, such as the Gradient (Terrell) test \cite{Terrell2002},
the Durbin (naive) test \cite{Durbin1970}, which have been shown
to be asymptotically distributed as the three aforementioned detectors
(under very mild conditions). Finally, for the sake of completeness,
we will obtain the LH test for the problem at hand, following the
lead of \cite{Kelly1989}.

\subsection{GLR \label{sub: GLR general}}

The generic form of the GLR in terms of the complex-valued unknowns
is given by \cite{Kay1998}:
\begin{equation}
\frac{\max_{\{\bm{B}_{s},\bm{R}\}}f_{1}(\bm{Z};\,\bm{B}_{s},\bm{R})}{\max_{\{\bm{B}_{t,0},\bm{R}\}}f_{0}(\bm{Z};\,\bm{B}_{t,0},\bm{R})}\label{eq: GLRT generic_interferene-1}
\end{equation}
First, it can be readily shown that the ML estimate of $\bm{R}$ under
$\mathcal{H}_{1}$ (resp. under $\mathcal{H}_{0}$), parametrized
by $\bm{B}_{s}$ (resp. $\bm{B}_{t,0}$) is:
\begin{gather}
\hat{\bm{R}}_{1}(\bm{B}_{s})\triangleq\,K^{-1}\,(\bm{Z}-\bm{A}\,\bm{B}_{s}\bm{C})(\bm{Z}-\bm{A}\,\bm{B}_{s}\bm{C})^{\dagger}\label{eq: cond ML estimate R_star under H1}\\
\hat{\bm{R}}_{0}(\bm{B}_{t,0})\triangleq\,K^{-1}\,(\bm{Z}-\bm{E}_{t}\,\bm{B}_{t,0}\bm{C})(\bm{Z}-\bm{E}_{t}\,\bm{B}_{t,0}\bm{C})^{\dagger}\label{eq: cond ML estimate R_star under H0}
\end{gather}
After substitution of Eqs. (\ref{eq: cond ML estimate R_star under H1})
and (\ref{eq: cond ML estimate R_star under H0}) in $f_{1}(\cdot)$
and $f_{0}(\cdot)$, respectively, the concentrated likelihoods are
expressed as:
\begin{gather}
f_{1}(\bm{Z};\bm{B}_{s},\hat{\bm{R}}_{1}(\bm{B}_{s}))=\left(K/(\pi e)\right)^{KN}\nonumber \\
\times\det\left[(\bm{Z}-\bm{A}\,\bm{B}_{s}\bm{C})(\bm{Z}-\bm{A}\,\bm{B}_{s}\bm{C})^{\dagger}\right]^{-K}\label{eq: concentrated likelihood H1_interference}\\
f_{0}(\bm{Z};\bm{B}_{t,0},\hat{\bm{R}}_{0}(\bm{B}_{t,0}))=\left(K/(\pi e)\right)^{KN}\nonumber \\
\times\det\left[(\bm{Z}-\bm{E}_{t}\,\bm{B}_{t,0}\bm{C})(\bm{Z}-\bm{E}_{t}\,\bm{B}_{t,0}\bm{C})^{\dagger}\right]{}^{-K}\label{eq: concentrated likelihood H0_interference}
\end{gather}
Then the ML estimates of $\bm{B}_{s}$ and $\bm{B}_{t,0}$ under $\mathcal{H}_{1}$
and $\mathcal{H}_{0}$, respectively, are obtained as the solutions
to the following optimization problems:
\begin{gather}
\widehat{\bm{B}}_{s}\triangleq\arg\min_{\bm{B}_{s}}\det[(\bm{Z}-\bm{A}\,\bm{B}_{s}\bm{C})(\bm{Z}-\bm{A}\,\bm{B}_{s}\bm{C})^{\dagger}]\,;\\
\widehat{\bm{B}}_{t,0}\triangleq\arg\min_{\bm{B}_{t,0}}\det[(\bm{Z}-\bm{E}_{t}\bm{B}_{t,0}\bm{C})(\bm{Z}-\bm{E}_{t}\bm{B}_{t,0}\bm{C})^{\dagger}]\,.
\end{gather}
It has been shown in \cite{Kelly1989} that the optimizers have the
closed form:
\begin{eqnarray}
\widehat{\bm{B}}_{s} & = & (\bm{A}^{\dagger}\,\bm{S}_{c}^{-1}\,\bm{A})^{-1}\bm{A}^{\dagger}\,\bm{S}_{c}^{-1}\,\bm{Z}\,\bm{C}^{\dagger}(\bm{C}\bm{C}^{\dagger})^{-1};\label{eq: ML estimate B_s under H1}\\
\widehat{\bm{B}}_{t,0} & = & (\bm{E}_{t}^{\dagger}\,\bm{S}_{c}^{-1}\,\bm{E}_{t})^{-1}\bm{E}_{t}^{\dagger}\,\bm{S}_{c}^{-1}\bm{Z}\,\bm{C}^{\dagger}(\bm{C}\bm{C}^{\dagger})^{-1}.\label{eq: ML estimate Bt0 under H0}
\end{eqnarray}
Substituting Eqs.~(\ref{eq: ML estimate B_s under H1}) and (\ref{eq: ML estimate Bt0 under H0})
into (\ref{eq: concentrated likelihood H1_interference}) and (\ref{eq: concentrated likelihood H0_interference}),
respectively, provides (after lengthy manipulations):
\begin{gather}
f_{1}(\bm{Z};\,\widehat{\bm{B}}_{s},\hat{\bm{R}}_{1})=\left(K/(\pi e)\right)^{KN}\,\det[\bm{S}_{c}]^{-K}\,\nonumber \\
\times\det\left[\bm{I}_{M}+(\bm{Z}_{{\scriptscriptstyle W1}}\bm{V}_{c,1})^{\dagger}\,\bm{P}_{\bm{A}_{1}}^{\perp}\,(\bm{Z}_{{\scriptscriptstyle W1}}\,\bm{V}_{c,1})\right]^{-K}\label{eq: final likelihood H1_interference}\\
f_{0}(\bm{Z};\,\widehat{\bm{B}}_{t,0},\hat{\bm{R}}_{0})=\left(K/(\pi e)\right)^{KN}\,\det[\bm{S}_{c}]^{-K}\,\nonumber \\
\times\det\left[\bm{I}_{M}+(\bm{Z}_{{\scriptscriptstyle W1}}\bm{V}_{c,1})^{\dagger}\,\bm{P}_{\bm{A}_{0}}^{\perp}\,(\bm{Z}_{{\scriptscriptstyle W1}}\,\bm{V}_{c,1})\right]^{-K}\label{eq: final likelihood H0_interference}
\end{gather}
where we have defined $\bm{A}_{1}\triangleq(\bm{S}_{c}^{-1/2}\bm{A})$,
$\bm{A}_{0}\triangleq(\bm{S}_{c}^{-1/2}\bm{E}_{t})$ and $\bm{Z}_{{\scriptscriptstyle W1}}\triangleq(\bm{S}_{c}^{-1/2}\bm{Z})$,
respectively. Finally, substituting Eqs.~(\ref{eq: final likelihood H1_interference})
and (\ref{eq: final likelihood H0_interference}) into Eq. (\ref{eq: GLRT generic_interferene-1})
and after taking the $k$-th root, the following explicit statistic
is obtained:
\begin{align}
t_{\mathrm{glr}} & \triangleq\frac{\det[\bm{I}_{M}+(\bm{Z}_{{\scriptscriptstyle W1}}\bm{V}_{c,1})^{\dagger}\,\bm{P}_{\bm{A}_{0}}^{\perp}(\bm{Z}_{{\scriptscriptstyle W1}}\bm{V}_{c,1})]}{\det[\bm{I}_{M}+(\bm{Z}_{{\scriptscriptstyle W1}}\bm{V}_{c,1})^{\dagger}\bm{P}_{\bm{A}_{1}}^{\perp}(\bm{Z}_{{\scriptscriptstyle W1}}\bm{V}_{c,1})]}\label{eq: GLRT_final form KF}\\
 & =\frac{\det[\bm{I}_{K}+\bm{Z}_{{\scriptscriptstyle W1}}^{\dagger}\,\bm{P}_{\bm{A}_{0}}^{\perp}\bm{Z}_{{\scriptscriptstyle W1}}\bm{P}_{\bm{C}^{\dagger}}]}{\det[\bm{I}_{K}+\bm{Z}_{{\scriptscriptstyle W1}}^{\dagger}\,\bm{P}_{\bm{A}_{1}}^{\perp}\,\bm{Z}_{{\scriptscriptstyle W1}}\bm{P}_{\bm{C}^{\dagger}}]}\,,
\end{align}
where the last expression follows from Sylvester's determinant theorem
\cite{Bernstein2009}. Furthermore, we observe that Eq. (\ref{eq: GLRT_final form KF})
can be also re-arranged in the following useful equivalent forms (again
obtained via Sylvester's determinant theorem):
\begin{gather}
t_{\mathrm{glr}}=\det[\bm{I}_{M}-\bm{D}_{0}^{-1/2}(\bm{Z}_{{\scriptscriptstyle W1}}\bm{V}_{c,1})^{\dagger}\bm{\mathcal{P}}_{\Delta}(\bm{Z}_{{\scriptscriptstyle W1}}\bm{V}_{c,1})\bm{D}_{0}^{-1/2}]^{-1}\label{eq: GLRT_final form (KF alternative)}\\
=\det[\bm{I}_{M}+\bm{D}_{1}^{-1/2}(\bm{Z}_{{\scriptscriptstyle W1}}\bm{V}_{c,1})^{\dagger}\bm{\mathcal{P}}_{\Delta}(\bm{Z}_{{\scriptscriptstyle W1}}\bm{V}_{c,1})\bm{D}_{1}^{-1/2}]\label{eq: GLRT_final form (Wilks' Lambda statistic)}
\end{gather}
where $\bm{\mathcal{P}}_{\Delta}\triangleq(\bm{P}_{\bm{A}_{1}}-\,\bm{P}_{\bm{A}_{0}})$
and $\bm{D}_{i}\triangleq[\bm{I}_{M}+(\bm{Z}_{{\scriptscriptstyle W1}}\bm{V}_{c,1})^{\dagger}\,\bm{P}_{\bm{A}_{i}}^{\perp}\,(\bm{Z}_{{\scriptscriptstyle W1}}\bm{V}_{c,1})]$,
respectively. Finally, it is worth noticing that Eq. (\ref{eq: GLRT_final form (Wilks' Lambda statistic)})
is in the well-known \emph{Wilks' Lambda statistic} form \cite{Wilks1932}.
Moreover, the latter expression generalizes the GLR in \cite{Kelly1989}
to the interference scenario (i.e., $t\neq0$).

For the sake of completeness, we also report the closed form ML estimates
of $\bm{R}$ obtained under $\mathcal{H}_{0}$ and $\mathcal{H}_{1}$
(after back-substitution of (\ref{eq: ML estimate B_s under H1})
and (\ref{eq: ML estimate Bt0 under H0}) in Eqs. (\ref{eq: cond ML estimate R_star under H1})
and (\ref{eq: cond ML estimate R_star under H0}), respectively):
\begin{align}
\hat{\bm{R}}_{1}(\widehat{\bm{B}}{}_{s})=\; & K^{-1}[\bm{S}_{c}+(\bm{Z}-\bm{S}_{c}^{1/2}\,\bm{P}_{\bm{A}_{1}}\,\bm{Z}_{{\scriptscriptstyle W1}})\bm{P}_{\bm{C}^{\dagger}}\nonumber \\
 & \times(\bm{Z}-\bm{S}_{c}^{1/2}\,\bm{P}_{\bm{A}_{1}}\,\bm{Z}_{{\scriptscriptstyle W1}})^{\dagger}]\,,\label{eq: Closed form MLE covariance H_1}\\
\hat{\bm{R}}_{0}(\widehat{\bm{B}}_{t,0})=\; & K^{-1}[\bm{S}_{c}+(\bm{Z}-\bm{S}_{c}^{1/2}\,\bm{P}_{\bm{A}_{0}}\,\bm{Z}_{{\scriptscriptstyle W1}})\bm{P}_{\bm{C}^{\dagger}}\nonumber \\
 & \times(\bm{Z}-\bm{S}_{c}^{1/2}\,\bm{P}_{\bm{A}_{0}}\,\bm{Z}_{{\scriptscriptstyle W1}})^{\dagger}]\,,\label{eq: Closed form MLE covariance H_0}
\end{align}
and underline that we will use the short-hand notation $\hat{\bm{R}}_{i}$
in what follows. Finally, before proceeding further, we state some
useful properties of ML covariance estimates (later exploited in this
paper) in the form of the following lemma.
\begin{lem}
The ML estimates of $\bm{R}$ under $\mathcal{H}_{1}$ and $\mathcal{H}_{0}$
satisfy the following equalities:\label{lem: ML covariance estimates properties}
\begin{align}
\hat{\bm{R}}_{1}^{-1}\bm{A} & =K\,\bm{S}_{c}^{-1}\bm{A}=K\,\begin{bmatrix}\bm{S}_{c}^{-1}\bm{E}_{t} & \bm{S}_{c}^{-1}\bm{E}_{r}\end{bmatrix}\label{eq: Lem1 R1}\\
\hat{\bm{R}}_{0}^{-1}\bm{E}_{t} & =K\,\bm{S}_{c}^{-1}\bm{E}_{t}\label{eq: Lem1 R0}
\end{align}
\end{lem}
\begin{IEEEproof}
Provided as supplementary material.
\end{IEEEproof}

\subsubsection*{CFARness of GLRT\label{sub: GLRT MIS}}

Using the expression in Eq. (\ref{eq: GLRT_final form KF}), we here
verify that $t_{\mathrm{glr}}$ can be expressed in terms of the MIS
(cf. Eq.~(\ref{eq: MIS_final})), thus proving its CFARness. Indeed,
it can be shown that\footnote{The proof of the aforementioned equalities\emph{ }is non-trivial and
thus provided as supplementary material.}
\begin{align}
(\bm{Z}_{{\scriptscriptstyle W1}}\bm{V}_{c,1})^{\dagger}\bm{P}_{\bm{A}_{0}}^{\perp}(\bm{Z}_{{\scriptscriptstyle W1}}\bm{V}_{c,1}) & =\bm{Z}_{2.3}^{\dagger}\,\bm{S}_{2.3}^{-1}\bm{Z}_{2.3}+\bm{Z}_{3}^{\dagger}\,\bm{S}_{33}^{-1}\bm{Z}_{3},\label{eq: MIS_E1}\\
(\bm{Z}_{{\scriptscriptstyle W1}}\bm{V}_{c,1})^{\dagger}\bm{P}_{\bm{A}_{1}}^{\perp}(\bm{Z}_{{\scriptscriptstyle W1}}\bm{V}_{c,1}) & =\bm{Z}_{3}^{\dagger}\,\bm{S}_{33}^{-1}\,\bm{Z}_{3},\label{eq: MIS_E2}
\end{align}
from which it follows
\begin{equation}
t_{\mathrm{glr}}=\frac{\det[\bm{I}_{M}+\bm{T}_{a}+\bm{T}_{b}]}{\det[\bm{I}_{M}+\bm{T}_{b}]},
\end{equation}
 which demonstrates invariance of the GLR(T) with respect to the nuisance
parameters and thus ensures CFAR property.

\subsection{Rao statistic\label{sub: Rao general}}

The generic form for the Rao statistic is given by \cite{Kay1998}:
\begin{gather}
\left.\frac{\partial\ln f_{1}(\bm{Z};\bm{\theta})}{\partial\bm{\theta}_{r}^{T}}\right|_{\bm{\theta}=\widehat{\bm{\theta}}_{0}}[\bm{I}^{-1}(\widehat{\bm{\theta}}_{0})]_{\bm{\theta}_{r},\bm{\theta}_{r}}\left.\frac{\partial\ln f_{1}(\bm{Z};\bm{\theta})}{\partial\bm{\theta}_{r}}\right|_{\bm{\theta}=\widehat{\bm{\theta}}_{0}}\label{eq: Rao statistic (implicit form)}
\end{gather}
where 
\begin{align}
\bm{I}(\bm{\theta}) & \triangleq\,\mathbb{E}\left\{ \frac{\partial\ln f_{1}(\bm{Z};\bm{\theta})}{\partial\bm{\theta}}\frac{\partial\ln f_{1}(\bm{Z};\bm{\theta})}{\partial\bm{\theta}^{T}}\right\} ,
\end{align}
denotes the Fisher Information Matrix (FIM) and $[\bm{I}^{-1}(\bm{\theta})]_{\mathrm{\bm{\theta}}_{r},\bm{\theta}_{r}}$
indicates the sub-matrix obtained by selecting from the FIM inverse
only the elements corresponding to the vector $\bm{\theta}_{r}$.
It is shown (the proof is provided as supplementary material) that
the aforementioned statistic is given in closed form as:
\begin{equation}
t_{\mathrm{rao}}\triangleq\mathrm{Tr}\left[\bm{Z}_{d,0}^{\dagger}\,\widehat{\bm{R}}_{0}^{-1}\,\bm{E}_{r}\,\widehat{\bm{\Gamma}}_{22}^{\circ}\,\bm{E}_{r}^{\dagger}\,\widehat{\bm{R}}_{0}^{-1}\,\bm{Z}_{d,0}\,\bm{P}_{\bm{C}^{\dagger}}\right]\label{eq: Rao_final form (implicit)}
\end{equation}
where we have partitioned $\widehat{\bm{\Gamma}}^{\circ}\triangleq(\bm{A}^{\dagger}\,\widehat{\bm{R}}_{0}^{-1}\,\bm{A})^{-1}$
as:
\begin{equation}
\widehat{\bm{\Gamma}}^{\circ}=\begin{bmatrix}\widehat{\bm{\Gamma}}_{11}^{\circ} & \widehat{\bm{\Gamma}}_{12}^{\circ}\\
\widehat{\bm{\Gamma}}_{21}^{\circ} & \widehat{\bm{\Gamma}}_{22}^{\circ}
\end{bmatrix}.\label{eq: partitioning Gamma_0 (Rao)}
\end{equation}
and, $\widehat{\bm{\Gamma}}_{ij}^{\circ}$, $(i,j)\in\{1,2\}\times\{1,2\}$,
is a sub-matrix whose dimensions can be obtained replacing $1$ and
$2$ with $t$ and $r$, respectively. Additionally, we have defined:
\begin{gather}
\bm{Z}_{d,0}\triangleq\left(\bm{Z}-\bm{S}_{c}^{1/2}\,\bm{P}_{\bm{A}_{0}}\,\bm{S}_{c}^{-1/2}\,\bm{Z}\,\bm{P}_{\bm{C}^{\dagger}}\right)\,.\label{eq: Zd0 (Rao test)}
\end{gather}
Eq. (\ref{eq: Rao_final form (implicit)}) can be rewritten in a more
familiar (and convenient) way. Indeed, it is proved\footnote{The proof is provided as supplementary material.}
that:
\begin{equation}
\widehat{\bm{R}}_{0}^{-1/2}\bm{Z}_{d,0}\,\bm{P}_{\bm{C}^{\dagger}}=\bm{P}_{\bar{\bm{A}}_{0}}^{\perp}\bm{Z}_{{\scriptscriptstyle W0}}\,\bm{P}_{\bm{C}^{\dagger}},\label{eq: R1}
\end{equation}
where $\bar{\bm{A}}_{0}\triangleq(\widehat{\bm{R}}_{0}^{-1/2}\bm{E}_{t})$
and $\bm{Z}_{{\scriptscriptstyle W0}}\triangleq(\widehat{\bm{R}}_{0}^{-1/2}\bm{Z})$,
respectively, and also 
\begin{equation}
\bm{P}_{\bar{\bm{A}}_{0}}^{\perp}\,\widehat{\bm{R}}_{0}^{-1/2}\,\bm{E}_{r}\,\bm{\Gamma}_{22}^{\circ}\,\bm{E}_{r}^{\dagger}\,\widehat{\bm{R}}_{0}^{-1/2}\,\bm{P}_{\bar{\bm{A}}_{0}}^{\perp}=\left(\bm{P}_{\bar{\bm{A}}_{1}}-\bm{P}_{\bar{\bm{A}}_{0}}\right),\label{eq: R2}
\end{equation}
where $\bar{\bm{A}}_{1}\triangleq\widehat{\bm{R}}_{0}^{-1/2}\bm{A}$
and $\bar{\bm{A}}_{0}\triangleq\widehat{\bm{R}}_{0}^{-1/2}\bm{E}_{t}$,
respectively. Therefore, an alternative form of $t_{\mathrm{rao}}$
is obtained substituting\emph{ }Eq. (\ref{eq: R1})  into Eq.~(\ref{eq: Rao_final form (implicit)})
and exploiting (\ref{eq: R2}), thus leading to the compact expression:
\begin{gather}
t_{\mathrm{rao}}=\mathrm{Tr}\left[\bm{Z}_{{\scriptscriptstyle W0}}^{\dagger}\,(\bm{P}_{\bar{\bm{A}}_{1}}-\bm{P}_{\bar{\bm{A}}_{0}})\,\bm{Z}_{{\scriptscriptstyle W0}}\,\bm{P}_{\bm{C}^{\dagger}}\right]\,.\label{eq: Rao test - final form (alternative)}
\end{gather}

\subsubsection*{CFARness of Rao Test}

We now express the Rao statistic as a function of the MIS, aiming
at showing its CFARness. To this end, we first notice that Eq. (\ref{eq: Rao test - final form (alternative)})
can be rewritten as:
\begin{equation}
t_{\mathrm{rao}}=\mathrm{Tr}\left[(\bm{Z}_{{\scriptscriptstyle W0}}\bm{V}_{c,1})^{\dagger}(\bm{P}_{\bar{\bm{A}}_{0}}^{\perp}-\bm{P}_{\bar{\bm{A}}_{1}}^{\perp})\,(\bm{Z}_{{\scriptscriptstyle W0}}\bm{V}_{c,1})\right]\,.
\end{equation}
Moreover, exploiting the equalities\footnote{Their proof is provided as supplementary material for this manuscript.}
\begin{gather}
(\bm{Z}_{{\scriptscriptstyle W0}}\bm{V}_{c,1})^{\dagger}\,\bm{P}_{\bar{\bm{A}}_{0}}^{\perp}(\bm{Z}_{{\scriptscriptstyle W0}}\bm{V}_{c,1})=K\,\left\{ \bm{Z}_{23}^{\dagger}\,\bm{S}_{2}^{-1}\,\bm{Z}_{23}\right.\nonumber \\
\left.-\bm{Z}_{23}^{\dagger}\,\bm{S}_{2}^{-1}\,\bm{Z}_{23}(\bm{I}_{M}+\bm{Z}_{23}^{\dagger}\,\bm{S}_{2}^{-1}\,\bm{Z}_{23})^{-1}\bm{Z}_{23}^{\dagger}\,\bm{S}_{2}^{-1}\,\bm{Z}_{23}\right\} ,\label{eq: MIS_E3}\\
(\bm{Z}_{{\scriptscriptstyle W0}}\bm{V}_{c,1})^{\dagger}\,\bm{P}_{\bar{\bm{A}}_{1}}^{\perp}\,(\bm{Z}_{{\scriptscriptstyle W0}}\bm{V}_{c,1})=K\,\left\{ \bm{Z}_{3}^{\dagger}\,\bm{S}_{33}^{-1}\,\bm{Z}_{3}\right.\nonumber \\
\left.-\bm{Z}_{3}^{\dagger}\,\bm{S}_{33}^{-1}\,\bm{Z}_{3}(\bm{I}_{M}+\bm{Z}_{3}^{\dagger}\,\bm{S}_{33}^{-1}\,\bm{Z}_{3})^{-1}\bm{Z}_{3}^{\dagger}\,\bm{S}_{33}^{-1}\,\bm{Z}_{3}\right\} ,\label{eq: MIS_E4}
\end{gather}
Eq. (\ref{eq: Rao test - final form (alternative)}) can be rewritten
as:
\begin{align}
t_{\mathrm{rao}}=\; & K\,\mathrm{Tr}[\bm{T}_{a}-(\bm{T}_{a}+\bm{T}_{b})\nonumber \\
 & \times(\bm{I}_{M}+\bm{T}_{a}+\bm{T}_{b})^{-1}(\bm{T}_{a}+\bm{T}_{b})\nonumber \\
 & +\bm{T}_{b}(\bm{I}_{M}+\bm{T}_{b})^{-1}\bm{T}_{b}]\label{eq: Rao_MIS}
\end{align}
which is only function of the MIS, thus proving its CFARness.

\subsection{Wald statistic\label{sub: Wald general}}

The generic form for the Wald statistic is given by \cite{Kay1998}:
\begin{gather}
(\hat{\bm{\theta}}_{r,1}-\bm{\theta}_{r,0})^{T}\{[\bm{I}^{-1}(\widehat{\bm{\theta}}_{1})]_{\bm{\theta}_{r},\bm{\theta}_{r}}\}^{-1}\,(\hat{\bm{\theta}}_{r,1}-\bm{\theta}_{r,0})\,.\label{eq: Wald_generic}
\end{gather}
It is shown (the proof is provided as supplementary material) that
the aforementioned statistic is given in closed form as:
\begin{gather}
t_{\mathrm{wald}}\triangleq\mathrm{Tr}\left[\bm{Z}_{{\scriptscriptstyle W1}}^{\dagger}K\bm{P}_{\bm{A}_{0}}^{\perp}\bm{S}_{c}^{-1/2}\bm{E}_{r}\,\widehat{\bm{\Gamma}}_{22}^{1}\,\bm{E}_{r}^{\dagger}\bm{S}_{c}^{-1/2}\bm{P}_{\bm{A}_{0}}^{\perp}\bm{Z}_{{\scriptscriptstyle W1}}\,\bm{P}_{\bm{C}^{\dagger}}\right]\label{eq: Wald_final form}
\end{gather}
where $\widehat{\bm{\Gamma}}_{ij}^{1}$ indicates the ($i,j)$-th
sub-matrix of $\widehat{\bm{\Gamma}}^{1}\triangleq(\bm{A}^{\dagger}\,\widehat{\bm{R}}_{1}^{-1}\,\bm{A})^{-1}$,
obtained using the same partitioning as in Eq.~(\ref{eq: partitioning Gamma_0 (Rao)})
for matrix $\widehat{\bm{\Gamma}}^{\circ}$. The above expression
can be rewritten in a more compact way, as shown in what follows.
Indeed, the inner matrix in Eq. (\ref{eq: Wald_final form}) is rewritten
as\footnote{The proof is provided as supplementary material.}:
\begin{gather}
K\,\bm{P}_{\bm{A}_{0}}^{\perp}\bm{S}_{c}^{-1/2}\,\bm{E}_{r}\,\widehat{\bm{\Gamma}}_{22}^{1}\,\bm{E}_{r}^{\dagger}\bm{S}_{c}^{-1/2}\bm{P}_{\bm{A}_{0}}^{\perp}=\bm{\mathcal{P}}_{\Delta}\label{eq: W1 - Wald statistic}
\end{gather}
which then gives:
\begin{align}
t_{\mathrm{wald}} & =\mathrm{Tr}\left[\bm{Z}_{{\scriptscriptstyle W1}}^{\dagger}\,\bm{\mathcal{P}}_{\Delta}\,\bm{Z}_{{\scriptscriptstyle W1}}\,\bm{P}_{\bm{C}^{\dagger}}\right]\label{eq: Wald_final form (alternative)}
\end{align}

\subsubsection*{CFARness of Wald Test}

Finally we  prove CFARness of Wald statistic. First, it is apparent
that Eq.~(\ref{eq: Wald_final form (alternative)}) can be rewritten
as:
\begin{gather}
t_{\mathrm{wald}}=\mathrm{Tr}\left[(\bm{Z}_{{\scriptscriptstyle W1}}\bm{V}_{c,1})^{\dagger}(\bm{P}_{\bm{A}_{0}}^{\perp}-\bm{P}_{\bm{A}_{1}}^{\perp})\,\bm{Z}_{{\scriptscriptstyle W1}}\bm{V}_{c,1}\right]\,.\label{eq: Wald _Pre MIS}
\end{gather}
Secondly, exploiting Eqs. (\ref{eq: MIS_E1}) and (\ref{eq: MIS_E2})
(as in the GLR case), Eq. (\ref{eq: Wald _Pre MIS}) is rewritten
as:
\begin{align}
t_{\mathrm{wald}}=\; & \mathrm{Tr}\left[\bm{Z}_{2.3}^{\dagger}\,\bm{S}_{2.3}^{-1}\,\bm{Z}_{2.3}\right]=\mathrm{Tr}[\,\bm{T}_{a}\,]\,.\label{eq: Wald - MIS}
\end{align}
Therefore $t_{\mathrm{wald}}$  depends on the data matrix  uniquely
through the MIS (actually, only through the first component).

\subsection{Gradient statistic\label{sub: Gradient general}}

The Gradient (Terrell) test requires the evaluation of the following
statistic \cite{Terrell2002,Lemonte2012}:

\begin{equation}
\left.\frac{\partial\ln f_{1}(\bm{Z};\bm{\theta})}{\partial\bm{\theta}_{r}^{T}}\right|_{\bm{\theta}=\hat{\bm{\theta}}_{0}}(\hat{\bm{\theta}}_{r,1}-\bm{\theta}_{r,0})\label{eq: Grad stat implicit form}
\end{equation}
The appeal of Eq. (\ref{eq: Grad stat implicit form}) arises from
the fact that it does not require neither to invert the FIM nor to
evaluate a compressed likelihood function under both hypotheses (as
opposed to GLR, Wald, and Rao statistics). As a consequence, this
formal simplicity can make the Gradient statistic easy to compute.
Moreover, under some mild technical conditions, such test is asymptotically
equivalent to the GLR, Rao and Wald statistics \cite{Terrell2002}. 

It is shown (the proof is provided as supplementary material) that
the Gradient statistic is given in closed form as:
\begin{gather}
t_{\mathrm{grad}}\triangleq\Re\left\{ \mathrm{Tr}\left[\bm{Z}_{{\scriptscriptstyle W1}}^{\dagger}K\bm{P}_{\bm{A}_{0}}^{\perp}\bm{S}_{c}^{-1/2}\bm{E}_{r}\,\widehat{\bm{\Gamma}}_{22}^{1}\,\bm{E}_{r}^{\dagger}\widehat{\bm{R}}_{0}^{-1}\,\bm{Z}_{d,0}\,\bm{P}_{\bm{C}^{\dagger}}\right]\right\} \label{eq: GT (final form)}
\end{gather}
where $\bm{Z}_{d,0}$ is given in Eq.~(\ref{eq: Zd0 (Rao test)}).
The expression in Eq. (\ref{eq: GT (final form)}) can be cast in
a more compact form, as shown below. First, we notice that the following
equality holds\footnote{The proof is deferred to supplementary material.}:
\begin{equation}
\widehat{\bm{R}}_{0}^{-1}\,\bm{Z}_{d,0}\,\bm{P}_{\bm{C}^{\dagger}}=\bm{S}_{c}^{-1/2}\bm{P}_{\bm{A}_{0}}^{\perp}(\bm{S}_{c}^{1/2}\,\widehat{\bm{R}}_{0}^{-1})\,\bm{Z}\,\bm{P}_{\bm{C}^{\dagger}}\label{eq: G.1}
\end{equation}
which, after substitution in Eq. (\ref{eq: GT (final form)}) and
exploitation of Eq.~(\ref{eq: W1 - Wald statistic}), gives the final
form
\begin{gather}
t_{\mathrm{grad}}=\Re\{\mathrm{Tr}[\bm{Z}_{{\scriptscriptstyle W1}}^{\dagger}\,\bm{\bm{\mathcal{P}}}_{\Delta}\,(\bm{S}_{c}^{1/2}\,\widehat{\bm{R}}_{0}^{-1/2})\,\bm{Z}_{{\scriptscriptstyle W0}}\,\bm{P}_{\bm{C}^{\dagger}}]\},\label{eq: GT (final form) (alternative)}
\end{gather}
where identical steps as for Wald test have been exploited.

\subsubsection*{CFARness of Gradient Test\label{sub: CFARness Gradient statistic}}

First, Eq. (\ref{eq: GT (final form) (alternative)}) can be readily
rewritten as:
\begin{gather}
t_{\mathrm{grad}}=\Re\left\{ \mathrm{Tr}\left[(\bm{Z}_{{\scriptscriptstyle W1}}\bm{V}_{c,1})^{\dagger}\,(\bm{P}_{\bm{A}_{0}}^{\perp}-\bm{P}_{\bm{A}_{1}}^{\perp})\right.\right.\nonumber \\
\left.\left.\quad\times(\bm{S}_{c}^{1/2}\,\widehat{\bm{R}}_{0}^{-1/2})\,\bm{Z}_{{\scriptscriptstyle W0}}\bm{V}_{c,1}\right]\right\} \,.
\end{gather}
Moreover, exploiting the following equalities\footnote{The proof is provided as supplementary material.}
\begin{gather}
\left((\bm{Z}_{{\scriptscriptstyle W1}}\bm{V}_{c,1})^{\dagger}\,\bm{P}_{\bm{A}_{0}}^{\perp}\,\bm{S}_{c}^{1/2}\,\widehat{\bm{R}}_{0}^{-1/2}\bm{Z}_{{\scriptscriptstyle W0}}\bm{V}_{c,1}\right)\nonumber \\
=K\,(\bm{T}_{a}+\bm{T}_{b})[\bm{I}_{M}-(\bm{I}_{M}+\bm{T}_{a}+\bm{T}_{b})^{-1}(\bm{T}_{a}+\bm{T}_{b})]\label{eq: MIS_E5}\\
\left((\bm{Z}_{{\scriptscriptstyle W1}}\bm{V}_{c,1})^{\dagger}\,\bm{P}_{\bm{A}_{1}}^{\perp}\,\bm{S}_{c}^{1/2}\,\widehat{\bm{R}}_{0}^{-1/2}\bm{Z}_{{\scriptscriptstyle W0}}\bm{V}_{c,1}\right)\nonumber \\
=K\,\bm{T}_{b}[\bm{I}_{M}-(\bm{I}_{M}+\bm{T}_{a}+\bm{T}_{b})^{-1}(\bm{T}_{a}+\bm{T}_{b})]\label{eq: MIS_E6}
\end{gather}
 it thus follows that:
\begin{align}
t_{\mathrm{grad}}= & \;\Re\left\{ \mathrm{Tr}\left[K\,\bm{T}_{a}\left(\bm{I}_{M}-(\bm{I}_{M}+\bm{T}_{a}+\bm{T}_{b})^{-1}\right.\right.\right.\nonumber \\
 & \left.\left.\left.\quad\qquad\times(\bm{T}_{a}+\bm{T}_{b})\right)\right]\right\} \label{eq: Gradient test MIS}
\end{align}
which shows that also the Gradient test satisfies the CFAR property.

\subsection{Durbin statistic\label{sub: Durbin general}}

The Durbin test (also referred to as ``Naive test'') consists in
the evaluation of the following decision statistic \cite{Durbin1970}:
\begin{gather}
(\hat{\bm{\theta}}_{r,01}-\bm{\theta}_{r,0})^{T}\left\{ \left[\bm{I}\left(\widehat{\bm{\theta}}_{0}\right)\right]_{\bm{\theta}_{r},\bm{\theta}_{r}}\left[\bm{I}^{-1}\left(\widehat{\bm{\theta}}_{0}\right)\right]_{\bm{\theta}_{r},\bm{\theta}_{r}}\times\right.\nonumber \\
\left.\left[\bm{I}\left(\widehat{\bm{\theta}}_{0}\right)\right]_{\bm{\theta}_{r},\bm{\theta}_{r}}\right\} \,(\hat{\bm{\theta}}_{0,1}-\bm{\theta}_{r,0})\,,\label{eq: Durbin test (generic form)}
\end{gather}
where the estimate $\hat{\bm{\theta}}_{r,01}$ is defined as:
\begin{equation}
\hat{\bm{\theta}}_{r,01}\triangleq\arg\max_{\bm{\theta}_{r}}\,f_{1}(\bm{Z};\bm{\theta}_{r},\hat{\bm{\theta}}_{s,0})\,.\label{eq: Durbin_ML estimate-1-1}
\end{equation}
In general, the Durbin statistic is asymptotically equivalent to GLR,
Rao and Wald statistics, as shown in \cite{Durbin1970}. However,
for the considered problem, a stronger result holds with respect to
the Rao statistic, as stated by the following theorem.
\begin{thm}
The Durbin statistic for the hypothesis testing model considered in
Eq. (\ref{eq: Transformed data - hypothesis testing problem}) is
statistically equivalent to the Rao statistic. Therefore, the test
is also CFAR.\label{thm: Durbin-Rao equivalence}\end{thm}
\begin{IEEEproof}
Provided as supplementary material.
\end{IEEEproof}
It is worth noticing that the present result generalizes the statistical
equivalence observed between Rao and Durbin statistics for the simpler
scenario of point-like targets and single-steering assumption in \cite{A.DeMaio2010}.
On the other hand, Thm. \ref{thm: Durbin-Rao equivalence} proves
that such result holds for the (very general) hypothesis testing problem
considered in this work.

\subsection{Two-step GLR (2S-GLR)\label{sub: 2S-GLR general}}

It is also worth considering a two-step GLR (2S-GLR), which first
consists in evaluating the GLR statistic under the assumption that
 $\bm{R}$ is known and then plugging-in a reasonable estimate of
$\bm{R}$. The GLR statistic for known $\bm{R}$ can be expressed
in implicit form as \cite{Kelly1989}:
\begin{equation}
\frac{\max_{\bm{B}_{s}}f_{1}(\bm{Z};\bm{B}_{s},\bm{R})}{\max_{\bm{B}_{t,0}}f_{0}(\bm{Z};\bm{B}_{t,0},\bm{R})}\,.\label{eq: 2S-GLRT implicit- interference}
\end{equation}
The ML estimates of $\bm{B}_{s}$ and $\bm{B}_{t,0}$ are more easily
obtained from optimizing the logarithm of $f_{1}(\cdot)$ and $f_{0}(\cdot)$,
respectively, that is: 
\begin{align}
 & -K\,\ln(\pi^{N}\det[\bm{R}])\nonumber \\
 & -\mathrm{Tr}[\bm{R}^{-1}(\bm{Z}-\bm{A}\bm{B}_{s}\bm{C})(\bm{Z}-\bm{A}\bm{B}_{s}\bm{C})^{\dagger}]\label{eq: 2S-GLRT objective numerator}\\
 & -K\,\ln(\pi^{N}\det[\bm{R}])\nonumber \\
 & -\mathrm{Tr}[\bm{R}^{-1}(\bm{Z}-\bm{E}_{t}\,\bm{B}_{t,0}\,\bm{C})(\bm{Z}-\bm{E}_{t}\,\bm{B}_{t,0}\,\bm{C})^{\dagger}]\label{eq: 2S-GLRT objective denominator}
\end{align}
Maximization of Eqs. (\ref{eq: 2S-GLRT objective numerator}) and
(\ref{eq: 2S-GLRT objective denominator}) with respect to $\bm{B}_{s}$
and $\bm{B}_{t,0}$, respectively, can be obtained following the same
steps employed in \cite{Kelly1989} and thus it is omitted for brevity.
Therefore, after optimization, the following statistic is obtained
(as the logarithm of Eq. (\ref{eq: 2S-GLRT implicit- interference})):
\begin{equation}
\mathrm{Tr}[\bm{Z}^{\dagger}\,\bm{R}^{-1/2}\,(\bm{P}_{\breve{\bm{A}}_{1}}-\bm{P}_{\breve{\bm{A}}_{0}})\,\bm{R}^{-1/2}\,\bm{Z}\,\bm{P}_{\bm{C}^{\dagger}}]\label{eq: 2S-GLRT-interference (R known)}
\end{equation}
where we have defined $\breve{\bm{A}}_{1}\triangleq(\bm{R}^{-1/2}\bm{A})$
and $\breve{\bm{A}}_{0}\triangleq(\bm{R}^{-1/2}\bm{E}_{t})$, respectively.
We recall that the expression in Eq. (\ref{eq: 2S-GLRT-interference (R known)})
depends on $\bm{R}$. We now turn our attention on finding an estimate
for the covariance $\bm{R}$. Clearly, in order to obtain a meaningful
estimate to be be plugged in both the numerator and denominator of
Eq. (\ref{eq: 2S-GLRT implicit- interference}), such estimate should
be based only on signal-free data (also commonly denoted as ``secondary
data''). 

It is not difficult to show that the covariance estimate  based only
on secondary data is given by\footnote{It should be noted that the same result would be obtained by considering
$\widehat{\bm{R}}_{1}^{-1}$ (i.e., the ML estimate under $\mathcal{H}_{1}$)
as the signal-free covariance estimate. Indeed, since Eq. (\ref{eq: 2S-GLRT-interference (R known)})
depends only on $\bm{R}$ through the quantities $(\bm{R}^{-1}\bm{A})$
and ($\bm{R}^{-1}\bm{E}_{t})$ an thus Lem. \ref{lem: ML covariance estimates properties}
could be exploited to obtain the same final statistic.}:
\begin{equation}
\widehat{\bm{R}}_{sd}=(K-M)^{-1}\,\bm{S}_{c}\,.\label{eq: secondary data_cov matrix}
\end{equation}
 Thus, substitution of Eq. (\ref{eq: secondary data_cov matrix})
into Eq. (\ref{eq: 2S-GLRT-interference (R known)}) leads to the
final form of 2S-GLR:
\begin{gather}
\mathrm{Tr}[\bm{Z}^{\dagger}\,\sqrt{K-M}\bm{S}_{c}^{-1/2}\,\bm{\mathcal{P}}_{\Delta}\,\bm{S}_{c}^{-1/2}\,\sqrt{K-M}\,\bm{Z}\,\bm{P}_{\bm{C}^{\dagger}}]\propto\nonumber \\
t_{\mathrm{2s-glr}}\triangleq\mathrm{Tr}\left[\bm{Z}^{\dagger}\,\bm{S}_{c}^{-1/2}\,\bm{\mathcal{P}}_{\Delta}\,\bm{S}_{c}^{-1/2}\,\bm{Z}\,\bm{P}_{\bm{C}^{\dagger}}\right]\label{eq: 2S-GLRT final form - interference-1}
\end{gather}
From direct comparison of Eqs. (\ref{eq: Wald_final form (alternative)})
and (\ref{eq: 2S-GLRT final form - interference-1}), a general equivalence
result is obtained, stated in the form of the following lemma.
\begin{lem}
The 2S-GLR statistic is statistically equivalent to the Wald statistic.
Therefore, the test is also CFAR.
\end{lem}
The aforementioned lemma extends the statistical equivalence observed
between 2S-GLR and Wald statistics in the simpler cases of point-like
targets \cite{A.DeMaio2004}, range-spread targets \cite{Shuai2012}
and multidimensional signals \cite{Liu2013}.

\subsection{Lawley-Hotelling (LH) statistic\label{sub: LH general}}

Finally, for the sake of a complete comparison, we also consider (and
generalize) the simpler statistic proposed in \cite[pag. 37]{Kelly1989}
as a reasonable approximation to GLR. Indeed, Wilks' Lambda form of
GLR in Eq.~(\ref{eq: GLRT_final form (Wilks' Lambda statistic)})
can be rewritten as:
\begin{align}
t_{\mathrm{glr}}=\, & \det[\bm{I}_{M}+\bm{D}_{1}^{-1/2}\,\{\widehat{\bm{B}}_{s}^{\dagger}(\bm{A}^{\dagger}\bm{S}_{c}^{-1}\bm{A})\,\widehat{\bm{B}}_{s}\nonumber \\
 & -\widehat{\bm{B}}_{t,0}^{\dagger}(\bm{E}_{t}^{\dagger}\bm{S}_{c}^{-1}\bm{E}_{t})\,\widehat{\bm{B}}_{t,0}\}\,\bm{D}_{1}^{-1/2}]\label{eq: Wilks statistic rewritten LH}
\end{align}
where we exploited $(\bm{C}\bm{C}^{\dagger})^{-1}=(\bm{C}\bm{C}^{\dagger})^{-1/2}$
and closed-form estimates for $\widehat{\bm{B}}_{s}$ and $\widehat{\bm{B}}_{t,0}^{\dagger}$
(cf. Eqs. (\ref{eq: ML estimate B_s under H1}) and (\ref{eq: ML estimate Bt0 under H0}),
respectively). As the number of samples $K$ grows large, we can invoke
approximation $\bm{S}_{c}\approx(K-M)\bm{R}$, that is, the sample
covariance based on secondary data will accurately approximate the
true covariance matrix. Accordingly, we can safely approximate
\begin{align}
\bm{D}_{1} & \approx\bm{I}_{M}\,,\nonumber \\
\widehat{\bm{B}}_{s} & \approx(\bm{A}^{\dagger}\bm{R}^{-1}\bm{A})^{-1}\,\bm{A}^{\dagger}\bm{R}^{-1}\bm{C}^{\dagger}(\bm{C}\bm{C}^{\dagger})^{-1}\,,\nonumber \\
\widehat{\bm{B}}_{t,0} & \approx(\bm{E}_{t}^{\dagger}\bm{R}^{-1}\bm{E}_{t})^{-1}\,\bm{E}^{\dagger}\bm{R}^{-1}\bm{C}^{\dagger}(\bm{C}\bm{C}^{\dagger})^{-1}\,,
\end{align}
since $(\bm{A}^{\dagger}\bm{S}_{c}^{-1}\bm{A})\approx(K-M)^{-1}(\bm{A}^{\dagger}\bm{R}^{-1}\bm{A})$
and $(\bm{E}_{t}^{\dagger}\bm{S}_{c}^{-1}\bm{E}_{t})\approx(K-M)^{-1}(\bm{E}_{t}^{\dagger}\bm{R}^{-1}\bm{E}_{t})$,
respectively. Therefore, based on these approximations, it is apparent
that the second contribution within the determinant in Eq.~(\ref{eq: Wilks statistic rewritten LH})
will be a vanishing term as the number of observations increases.
Hence, GLR statistic will be given by the determinant of an \emph{identity
matrix plus a small perturbing term}.

Additionally, we remark that when $\bm{\Upsilon}\in\mathbb{H}^{M\times M}$
is a small perturbing matrix, $\det[\bm{I}_{M}+\bm{\Upsilon}]$ can
be (accurately) approximated at first order as $\prod_{i=1}^{M}(1+\upsilon_{i})\approx1+\sum_{i=1}^{M}\upsilon_{i}=1+\mathrm{Tr}[\bm{\Upsilon}]$,
where $\upsilon_{i}$ denotes the $i$-th eigenvalue of $\bm{\Upsilon}$.
Based on those reasons, we formulate the LH statistic as:
\begin{align}
t_{\mathrm{lh}}\triangleq & \,\mathrm{Tr}\left[\bm{D}_{1}^{-1/2}\,(\bm{Z}_{{\scriptscriptstyle W1}}\bm{V}_{c,1})^{\dagger}\,\mathcal{\bm{P}}_{\Delta}\,(\bm{Z}_{{\scriptscriptstyle W1}}\bm{V}_{c,1})\,\bm{D}_{1}^{-1/2}\right]\label{eq: Lawley-Hotelling final form}\\
= & \,\mathrm{Tr}\left[(\bm{Z}_{{\scriptscriptstyle W1}}\bm{V}_{c,1})^{\dagger}\,\mathcal{\bm{P}}_{\Delta}\,(\bm{Z}_{{\scriptscriptstyle W1}}\bm{V}_{c,1})\,\bm{D}_{1}^{-1}\right]\nonumber 
\end{align}

\subsubsection*{CFARness of LH statistic}

The CFARness is proved by using Eqs. (\ref{eq: MIS_E1}) and (\ref{eq: MIS_E2})
from Sec. \ref{sub: GLRT MIS} within Eq. (\ref{eq: Lawley-Hotelling final form}),
thus obtaining:
\begin{equation}
t_{\mathrm{lh}}=\mathrm{Tr}\left[\bm{T}_{a}(\bm{I}_{M}+\bm{T}_{b})^{-1}\right]\,.\label{eq: LH_MIS form}
\end{equation}
Finally, in Tab. \ref{tab: Detectors comparison} it is shown a recap
table, summarizing all the considered detectors and their respective
expressions in terms of the MIS in Eq. (\ref{eq: MIS_final}).

\begin{center}
\begin{table*}
\centering{}\medskip{}
\protect\caption{Detectors comparison and their functional dependence of the MIS (viz.
CFARness). Auxiliary definitions: $\bm{T}_{a+b}\triangleq(\bm{T}_{a}+\bm{T}_{b})$
and $\bm{D}_{i}\triangleq[\bm{I}_{M}+(\bm{Z}_{{\scriptscriptstyle W1}}\bm{V}_{c,1})^{\dagger}\,\bm{P}_{\bm{A}_{i}}^{\perp}\,(\bm{Z}_{{\scriptscriptstyle W1}}\bm{V}_{c,1})]$.
\label{tab: Detectors comparison}}
\begin{tabular}{c||c||c}
\hline 
\noalign{\vskip\doublerulesep}
Detector & Standard Expression & MIS function\tabularnewline[\doublerulesep]
\hline 
\hline 
\noalign{\vskip0.1cm}
GLR & $\det[\bm{D}_{0}]/\det[\bm{D}_{1}]$ & $\det[\bm{I}_{M}+\bm{T}_{a+b}]/\det[\bm{I}_{M}+\bm{T}_{b}]$\tabularnewline[0.1cm]
\hline 
\noalign{\vskip0.1cm}
\multirow{1}{*}{Rao/Durbin} & $\mathrm{Tr}[\bm{Z}_{{\scriptscriptstyle W0}}^{\dagger}\,(\bm{P}_{\bar{\bm{A}}_{1}}-\bm{P}_{\bar{\bm{A}}_{0}})\,\bm{Z}_{{\scriptscriptstyle W0}}\,\bm{P}_{\bm{C}^{\dagger}}]$ & \multirow{1}{*}{$K\mathrm{\,Tr}[\bm{T}_{a}-\bm{T}_{a+b}(\bm{I}_{M}+\bm{T}_{a+b})^{-1}\bm{T}_{a+b}+\bm{T}_{b}(\bm{I}_{M}+\bm{T}_{b})^{-1}\bm{T}_{b}]$}\tabularnewline[0.1cm]
\hline 
\noalign{\vskip0.1cm}
Wald/2S-GLR & $\mathrm{Tr}[\bm{Z}_{{\scriptscriptstyle W1}}^{\dagger}(\bm{P}_{\bm{A}_{1}}-\bm{P}_{\bm{A}_{0}})\,\bm{Z}_{{\scriptscriptstyle W1}}\,\bm{P}_{\bm{C}^{\dagger}}]$ & $\mathrm{Tr}[\bm{T}_{a}]$\tabularnewline[0.1cm]
\hline 
\noalign{\vskip0.1cm}
Gradient & $\Re\left\{ \mathrm{Tr}[\bm{Z}_{{\scriptscriptstyle W1}}^{\dagger}\,(\bm{P}_{\bm{A}_{1}}-\bm{P}_{\bm{A}_{0}})\,(\bm{S}_{c}^{1/2}\,\widehat{\bm{R}}_{0}^{-1/2})\,\bm{Z}_{{\scriptscriptstyle W0}}\,\bm{P}_{\bm{C}^{\dagger}}]\right\} $ & $\Re\left\{ \mathrm{Tr}\left[K\,\bm{T}_{a}\left(\bm{I}_{M}-(\bm{I}_{M}+\bm{T}_{a+b})^{-1}\bm{T}_{a+b}\right)\right]\right\} $\tabularnewline[0.1cm]
\hline 
\noalign{\vskip0.1cm}
LH & $\mathrm{Tr}[(\bm{Z}_{{\scriptscriptstyle W1}}\bm{V}_{c,1})^{\dagger}\,(\bm{P}_{\bm{A}_{1}}-\bm{P}_{\bm{A}_{0}})\,(\bm{Z}_{{\scriptscriptstyle W1}}\bm{V}_{c,1})\,\bm{D}_{1}^{-1}]$ & $\mathrm{Tr}\left[\bm{T}_{a}(\bm{I}_{M}+\bm{T}_{b})^{-1}\right]\,.$\tabularnewline[0.1cm]
\hline 
\end{tabular}
\end{table*}

\par\end{center}

\section{Detectors in Special Cases \label{sec: Detectors in special cases}}

\subsection{Adaptive (Vector Subspace) Detection of a Point-like Target\label{sub: Vector subspace detection}}

In the present case we start from general formulation in Eq.~(\ref{eq: Transformed data - hypothesis testing problem})
and assume that: ($i$) $t=0$ (i.e., there is no interference, thus
$J=r$ and $\bm{A}=\begin{bmatrix}\bm{I}_{r} & \bm{0}_{r\times(N-r)}\end{bmatrix}^{T}\in\mathbb{C}^{N\times r}$);
($ii$) $M=1$, i.e., the matrix $\bm{B}$ collapses to a vector $\bm{b}\in\mathbb{C}^{J\times1}$
and $(iii)$ $\bm{c}\triangleq\begin{bmatrix}1 & 0 & \cdots & 0\end{bmatrix}\in\mathbb{C}^{1\times K}$
(i.e., a row vector). Such case has been extensively dealt in adaptive
detection literature \cite{Kelly1989,Bose1995,Scharf1994,Raghavan1996,Kelly1986}.
The hypothesis testing in canonical form is then: 
\begin{equation}
\begin{cases}
\mathcal{H}_{0}: & \bm{Z}=\bm{N}\\
\mathcal{H}_{1}: & \bm{Z}=\bm{A}\,\bm{b}\,\bm{c}+\bm{N}
\end{cases}
\end{equation}
Clearly, since in this case $M=1$ holds, $(K-1)$ vector components
are assumed signal-free, that is, $\bm{Z}$ admits the partitioning
$\bm{Z}=\begin{bmatrix}\bm{z}_{p} & \bm{Z}_{s}\end{bmatrix}=\begin{bmatrix}\bm{z}_{c} & \bm{Z}_{c,\perp}\end{bmatrix}$,
where $\bm{z}_{p}$ denotes the signal vector related to the cell
under test and the columns of $\bm{Z}_{s}$ represent the secondary
(training) data. Also, $\bm{P}_{\bm{A}_{0}}=\bm{0}_{N\times N}$ (resp.
$\bm{P}_{\bm{A}_{0}}^{\perp}=\bm{I}_{N}$) holds, because of the absence
of the structured interference. In the latter case, it can be shown
that the simplified projector form holds:
\begin{equation}
\bm{P}_{\bm{C}^{\dagger}}=\begin{bmatrix}1 & \bm{0}_{K-1}^{T}\\
\bm{0}_{K-1} & \bm{0}_{(K-1)\times(K-1)}
\end{bmatrix}\label{eq: C projection  - vector subspace detection}
\end{equation}
Given the results in Eq. (\ref{eq: C projection  - vector subspace detection}),
it can be shown that $\bm{S}_{c}=\bm{Z}_{s}\bm{Z}_{s}^{\dagger}$
and $\hat{\bm{R}}_{0}=\frac{1}{K}\bm{S}_{0}$, where $\bm{S}_{0}\triangleq(\bm{z}_{p}\,\bm{z}_{p}^{\dagger}+\bm{Z}_{s}\bm{Z}_{s}^{\dagger})$
hold, respectively. In some cases we will also use the Sherman-Woodbury
formula \cite{Bernstein2009} applied to $\bm{S}_{0}^{-1}$, that
is:
\begin{equation}
\bm{S}_{0}^{-1}=\bm{S}_{c}^{-1}-\frac{\bm{S}_{c}^{-1}\bm{z}_{p}\,\bm{z}_{p}^{\dagger}\bm{S}_{c}^{-1}}{1+\bm{z}_{p}^{\dagger}\,\bm{S}_{c}^{-1}\bm{z}_{p}}\,.\label{eq: Sherman-Woodbury}
\end{equation}

\subsubsection*{GLR}

In the specific case of $M=1$, the following form of the GLR is obtained
from Eq. (\ref{eq: GLRT_final form (KF alternative)}):
\begin{equation}
t_{\mathrm{glr}}=\frac{1}{1-\eta},\quad\eta\triangleq\frac{\bm{z}_{p}^{\dagger}\,\bm{S}_{c}^{-1/2}\bm{P}_{\bm{A}_{1}}\bm{S}_{c}^{-1/2}\bm{z}_{p}}{1+\bm{z}_{p}^{\dagger}\,\bm{S}_{c}^{-1}\bm{z}_{p}},\label{eq: GLRT vector subspace detection}
\end{equation}
since we have exploited $\bm{D}_{0}\rightarrow d_{0}=(1+\bm{z}_{p,1}^{\dagger}\bm{z}_{p,1})$
and $(\bm{Z}_{{\scriptscriptstyle W1}}\bm{V}_{c,1})^{\dagger}\,\bm{\mathcal{P}}_{\Delta}\,(\bm{Z}_{{\scriptscriptstyle W1}}\bm{V}_{c,1})\rightarrow(\bm{z}_{p,1}^{\dagger}\bm{P}_{\bm{A}_{1}}\bm{z}_{p,1})$,
where $\bm{z}_{p,1}\triangleq(\bm{S}_{c}^{-1/2}\bm{z}_{p})$. Clearly,
$t_{\mathrm{glr}}$ is an increasing function of $\eta$, the latter
thus being an equivalent form of the statistic and coinciding with
the so-called multi-rank signal model GLR described in \cite{Kelly1986,Kelly1989}.

\subsubsection*{Rao/Durbin statistic}

For the present scenario, Eq. (\ref{eq: Rao test - final form (alternative)})
specializes into:
\begin{gather}
t_{\mathrm{rao}}=\mathrm{Tr}[\bm{Z}_{{\scriptscriptstyle W0}}^{\dagger}\bm{P}_{\bar{\bm{A}}_{1}}\bm{Z}_{{\scriptscriptstyle W0}}\bm{P}_{\bm{C}^{\dagger}}]=\mathrm{Tr}[\bm{z}_{p}^{\dagger}(\hat{\bm{R}}_{0}^{-1/2}\,\bm{P}_{\bar{\bm{A}}_{1}}\,\hat{\bm{R}}_{0}^{-1/2})\bm{z}_{p}]\nonumber \\
\propto\bm{z}_{p}^{\dagger}\,\bm{S}_{0}^{-1}\bm{A}(\bm{A}^{\dagger}\bm{S}_{0}^{-1}\bm{A})^{-1}\bm{A}^{\dagger}\bm{S}_{0}^{-1}\bm{z}_{p}\triangleq\eta_{\mathrm{rao}}\label{eq: intermediate Rao_vector subspace detection}
\end{gather}
Eq. (\ref{eq: intermediate Rao_vector subspace detection}) can be
further simplified by exploiting the Woodbury identity in (\ref{eq: Sherman-Woodbury})
(and similar steps as in \cite{A.DeMaio2007}), thus obtaining the
following simplified form of the Rao statistic:
\begin{equation}
\eta_{\mathrm{rao}}=\frac{1}{1+\bm{z}_{p,1}^{\dagger}\bm{z}_{p,1}}\left[\frac{\bm{z}_{p,1}^{\dagger}\bm{P}_{\bm{A}_{1}}\bm{z}_{p,1}}{1+\bm{z}_{p,1}\bm{P}_{\bm{A}_{1}}^{\perp}\bm{z}_{p,1}}\right]\,.\label{eq: Rao vector subspace final}
\end{equation}
Finally, for $r=1$ (i.e., a single-steering case) ($\bm{A}\rightarrow\bm{a}\in\mathbb{C}^{N\times1}$),
Eq. (\ref{eq: Rao vector subspace final}) reduces to:
\begin{gather}
\eta_{\mathrm{rao}}=\frac{\left|\bm{z}_{p}^{\dagger}\bm{S}_{c}^{-1}\bm{a}\right|^{2}/\left(\bm{a}^{\dagger}\bm{S}_{c}^{-1}\bm{a}\right)}{[1+\bm{z}_{p}^{\dagger}\bm{S}_{c}^{-1}\bm{z}_{p}]\left[1+\bm{z}_{p}^{\dagger}\bm{S}_{c}^{-1}\bm{z}_{p}-\frac{\left|\bm{z}_{p}^{\dagger}\bm{S}_{c}^{-1}\bm{a}\right|^{2}}{\left(\bm{a}^{\dagger}\bm{S}_{c}^{-1}\bm{a}\right)}\right]}\,,
\end{gather}
which coincides with the well-known Rao statistic for the single-steering
case developed in \cite{A.DeMaio2007}.

\subsubsection*{Wald/2S-GLR statistic}

Starting from Eq. (\ref{eq: Wald_final form (alternative)}), we particularize
the Wald statistic as follows:
\begin{gather}
t_{\mathrm{wald}}=\mathrm{Tr}[\bm{Z}_{{\scriptscriptstyle W1}}^{\dagger}\bm{P}_{\bm{A}_{1}}\bm{Z}_{{\scriptscriptstyle W1}}\bm{P}_{\bm{C}^{\dagger}}]=\mathrm{Tr}[\bm{z}_{p,1}^{\dagger}\,\bm{P}_{\bm{A}_{1}}\,\bm{z}_{p,1}]\label{eq: Wald_vector subspace detection}\\
=\bm{z}_{p}^{\dagger}\,\bm{S}_{c}^{-1}\bm{A}(\bm{A}^{\dagger}\bm{S}_{c}^{-1}\bm{A})^{-1}\bm{A}^{\dagger}\bm{S}_{c}^{-1}\bm{z}_{p}\nonumber 
\end{gather}
In the special case $r=1$ (i.e., a single-steering case), Eq. (\ref{eq: Wald_vector subspace detection})
becomes:
\begin{equation}
t_{\mathrm{wald}}=\frac{\left|\bm{z}_{p}^{\dagger}\bm{S}_{c}^{-1}\bm{a}\right|^{2}}{\bm{a}^{\dagger}\bm{S}_{c}^{-1}\bm{a}},
\end{equation}
which is recognized as the well-known \emph{Adaptive Matched Filter}
(AMF) \cite{A.DeMaio2004,Fuhrmann1992}.

\subsubsection*{Gradient statistic}

In this case the gradient statistic in Eq.~(\ref{eq: GT (final form) (alternative)})
specializes into:
\begin{align}
t_{\mathrm{grad}} & =\Re\left\{ \mathrm{Tr}\left[\bm{Z}_{{\scriptscriptstyle W1}}^{\dagger}\bm{P}_{\bm{A}_{1}}(\bm{S}_{c}^{1/2}\widehat{\bm{R}}_{0}^{-1/2})\bm{Z}_{{\scriptscriptstyle W0}}\bm{P}_{\bm{C}^{\dagger}}\right]\right\} \\
 & =\Re\left\{ \bm{z}_{p,1}^{\dagger}\,\bm{P}_{\bm{A}_{1}}(\bm{S}_{c}^{1/2}\widehat{\bm{R}}_{0}^{-1/2})\,\bm{z}_{p,0}\right\} \nonumber \\
 & =K\,\Re\left\{ \bm{z}_{p}^{\dagger}\,\bm{S}_{c}^{-1}\bm{A}(\bm{A}^{\dagger}\bm{S}_{c}^{-1}\bm{A})^{-1}\bm{A}^{\dagger}\bm{S}_{0}^{-1}\bm{z}_{p}\right\} \nonumber 
\end{align}
where $\bm{z}_{p,0}\triangleq(\widehat{\bm{R}}_{0}^{-1/2}\bm{z}_{p})$.
It is interesting to note that, exploiting Eq. (\ref{eq: Sherman-Woodbury}),
the gradient statistic is rewritten as: 
\begin{align}
t_{\mathrm{grad}} & =K\,\Re\left\{ \frac{\left(\bm{A}^{\dagger}\bm{S}_{c}^{-1}\bm{z}_{p}\right)^{\dagger}(\bm{A}^{\dagger}\bm{S}_{c}^{-1}\bm{A})^{-1}\left(\bm{A}^{\dagger}\bm{S}_{c}^{-1}\bm{z}_{p}\right)}{1+\bm{z}_{p}^{\dagger}\,\bm{S}_{c}^{-1}\bm{z}_{p}}\right\} \nonumber \\
 & =K\,\frac{\left(\bm{A}^{\dagger}\bm{S}_{c}^{-1}\bm{z}_{p}\right)^{\dagger}(\bm{A}^{\dagger}\bm{S}_{c}^{-1}\bm{A})^{-1}\left(\bm{A}^{\dagger}\bm{S}_{c}^{-1}\bm{z}_{p}\right)}{1+\bm{z}_{p}^{\dagger}\,\bm{S}_{c}^{-1}\bm{z}_{p}}\label{eq: GT (vector subspace)}
\end{align}
where in last line we have omitted $\Re\{\cdot\}$ since Eq. (\ref{eq: GT (vector subspace)})
is formed by Hermitian quadratic forms (at both numerator and denominator);
thus it is \emph{always real-valued}. Therefore Eq.~(\ref{eq: GT (vector subspace)})
is \emph{statistically equivalent} to Kelly's GLR in Eq.~(\ref{eq: GLRT vector subspace detection}).

\subsubsection*{LH statistic}

We recall that for point-like targets, the condition $M=1$ holds.
Therefore the LH test is\emph{ statistically equivalent} to the GLRT
since the operators $\mathrm{Tr[\cdot]}$ and $\det[\cdot]$ are non-influential
when applied to a scalar value. This follows since the expressions
in Eqs.~(\ref{eq: GLRT_final form (Wilks' Lambda statistic)}) and
(\ref{eq: Lawley-Hotelling final form}) are thus related by a monotone
transformation.

\subsection{Adaptive Vector Subspace Detection with Structured Interference}

In the present case we start from general formulation in Eq.~(\ref{eq: Transformed data - hypothesis testing problem})
and assume that: ($i$) $M=1$, i.e., the matrices $\bm{B}$ and $\bm{B}_{t,i}$
collapse to the vectors $\bm{b}\in\mathbb{C}^{r\times1}$ and $\bm{b}_{t,i}\in\mathbb{C}^{t\times1}$,
respectively; $(ii)$ $\bm{c}\triangleq\begin{bmatrix}1 & 0 & \cdots & 0\end{bmatrix}\in\mathbb{C}^{1\times K}$
(i.e., a row vector). Such case has been dealt in \cite{A.DeMaio2014}.
Given the aforementioned assumptions, the problem in canonical form
is given as: 
\begin{equation}
\begin{cases}
\mathcal{H}_{0}: & \bm{Z}=\bm{A}\,\begin{bmatrix}\bm{b}_{t,0}^{T} & \bm{0}_{r}^{T}\end{bmatrix}^{T}\,\bm{c}+\bm{N}\\
\mathcal{H}_{1}: & \bm{Z}=\bm{A}\,\begin{bmatrix}\bm{b}_{t,1}^{T} & \bm{b}^{T}\end{bmatrix}^{T}\,\bm{c}+\bm{N}
\end{cases}
\end{equation}
Clearly, since in this case $M=1$ holds, $(K-1)$ vector components
are assumed signal-free, that is, $\bm{Z}$ admits the partitioning
$\bm{Z}=\begin{bmatrix}\bm{z}_{p} & \bm{Z}_{s}\end{bmatrix}=\begin{bmatrix}\bm{z}_{c} & \bm{Z}_{c,\perp}\end{bmatrix}$,
where $\bm{z}_{p}$ denotes the signal vector related to the cell
of interest and the columns of $\bm{Z}_{s}$ represent the secondary
(or training) data. In the latter case, it can be shown that the same
simplified projector form in Eq. (\ref{eq: C projection  - vector subspace detection})
holds. Given the results in Eq.~(\ref{eq: C projection  - vector subspace detection})
, it can be shown that $\bm{S}_{c}=\bm{Z}_{s}\bm{Z}_{s}^{\dagger}$
and $\hat{\bm{R}}_{0}=\frac{1}{K}\bm{S}_{0}$, where $\bm{S}_{0}\triangleq(\bm{S}_{c}+\bm{S}_{c}^{1/2}\bm{P}_{\bm{A}_{0}}^{\perp}\bm{S}_{c}^{-1/2}\bm{z}_{p}\bm{z}_{p}^{\dagger}\bm{S}_{c}^{-1/2}\bm{P}_{\bm{A}_{0}}^{\perp}\bm{S}_{c}^{1/2})$
hold, respectively. In some cases we will also use the Sherman-Woodbury
formula \cite{Bernstein2009} applied to $\bm{S}_{0}^{-1}$ and consider
the product $\bm{S}_{0}^{-1}\bm{z}_{p}$, which provides:
\begin{eqnarray}
\bm{S}_{0}^{-1}\bm{z}_{p} & = & \bm{S}_{c}^{-1}\bm{z}_{p}-(\bm{S}_{c}^{-1/2}\bm{P}_{\bm{A}_{0}}^{\perp}\bm{S}_{c}^{-1/2})\bm{z}_{p}\label{eq: Sherman-Woodbury-vector subspace interf}\\
 &  & \times\frac{\bm{z}_{p}^{\dagger}\bm{S}_{c}^{-1/2}\bm{P}_{\bm{A}_{0}}^{\perp}\bm{S}_{c}^{-1/2}\bm{z}_{p}}{1+\bm{z}_{p}^{\dagger}\,\bm{S}_{c}^{-1/2}\bm{P}_{\bm{A}_{0}}^{\perp}\bm{S}_{c}^{-1/2}\bm{z}_{p}}\nonumber 
\end{eqnarray}

\subsubsection*{GLR}

In the specific case of $M=1$, the following form of the GLRT is
obtained from Eq. (\ref{eq: GLRT_final form (KF alternative)}):
\begin{equation}
t_{\mathrm{glr}}=\frac{1}{1-\eta},\quad\eta\triangleq\frac{\bm{z}_{p,1}^{\dagger}\,(\bm{P}_{\bm{A}_{1}}-\bm{P}_{\bm{A}_{0}})\,\bm{z}_{p,1}}{1+\bm{z}_{p,1}^{\dagger}\,\bm{P}_{\bm{A}_{0}}^{\perp}\,\bm{z}_{p,1}},\label{eq: GLRT _vector subspace interference}
\end{equation}
since we have exploited $\bm{D}_{0}\rightarrow d_{0}=(1+\bm{z}_{p}^{\dagger}\bm{S}_{c}^{-1/2}\,\bm{P}_{\bm{A}_{0}}^{\perp}\,\bm{S}_{c}^{-1/2}\bm{z}_{p})$
and $(\bm{Z}_{{\scriptscriptstyle W1}}\bm{V}_{c,1})^{\dagger}\,\bm{\mathcal{P}}_{\Delta}\,(\bm{Z}_{{\scriptscriptstyle W1}}\bm{V}_{c,1})\rightarrow\bm{z}_{p,1}^{\dagger}(\bm{P}_{\bm{A}_{1}}-\bm{P}_{\bm{A}_{0}})\,\bm{z}_{p,1}$,
where $\bm{z}_{p,1}\triangleq(\bm{S}_{c}^{-1/2}\bm{z}_{p})$. Clearly,
Eq. (\ref{eq: GLRT _vector subspace interference}) is an increasing
function of $\eta$, which can be thus seen as an equivalent form
of the GLR.

\subsubsection*{Rao/Durbin statistic}

For the present scenario, Eq. (\ref{eq: Rao test - final form (alternative)})
specializes into:
\begin{gather}
t_{\mathrm{rao}}=\bm{z}_{p,0}^{\dagger}\,(\bm{P}_{\bar{\bm{A}}_{1}}-\bm{P}_{\bar{\bm{A}}_{0}})\,\bm{z}_{p,0}\label{eq: intermediate Rao_vector subspace interference}
\end{gather}
where $\bm{z}_{p,0}\triangleq(\widehat{\bm{R}}_{0}^{-1/2}\bm{z}_{p})$.

\subsubsection*{Wald/2S-GLR statistic}

Starting from Eq. (\ref{eq: Wald_final form (alternative)}), we particularize
the Wald statistic as follows:
\begin{gather}
t_{\mathrm{wald}}=\bm{z}_{p,1}^{\dagger}\,(\bm{P}_{\bm{A}_{1}}-\bm{P}_{\bm{A}_{0}})\,\bm{z}_{p,1}\label{eq: Wald Test _vector subspace interference}
\end{gather}

\subsubsection*{Gradient statistic}

In this case the gradient statistic in Eq.~(\ref{eq: GT (final form) (alternative)})
specializes into:
\begin{gather}
t_{\mathrm{grad}}=\Re\left\{ \bm{z}_{p,1}^{\dagger}\,(\bm{P}_{\bm{A}_{1}}-\bm{P}_{\bm{A}_{0}})\,(\bm{S}_{c}^{1/2}\widehat{\bm{R}}_{0}^{-1/2})\,\bm{z}_{p,0}\right\} \label{eq: GT _vector subspace interference}
\end{gather}
We now rewrite Eq. (\ref{eq: GT _vector subspace interference}) as:
\begin{equation}
t_{\mathrm{grad}}=K\,\Re\left\{ \bm{z}_{p}^{\dagger}\,\bm{S}_{c}^{-1/2}\,(\bm{P}_{\bm{A}_{1}}-\bm{P}_{\bm{A}_{0}})\,(\bm{S}_{c}^{1/2}\bm{S}_{0}^{-1})\,\bm{z}_{p}\right\} \label{eq: GT_vectro subspace interf rewritten}
\end{equation}
Exploiting Eq. (\ref{eq: Sherman-Woodbury-vector subspace interf})
and observing that $(\bm{P}_{\bm{A}_{1}}-\bm{P}_{\bm{A}_{0}})\bm{P}_{\bm{A}_{0}}^{\perp}=\bm{P}_{\bm{A}_{1}}-\bm{P}_{\bm{A}_{0}}$
holds, Eq. (\ref{eq: GT_vectro subspace interf rewritten}) is expressed
as:
\begin{eqnarray}
t_{\mathrm{grad}} & = & K\,\Re\left\{ \frac{\bm{z}_{p}^{\dagger}\,\bm{S}_{c}^{-1/2}\,(\bm{P}_{\bm{A}_{1}}-\bm{P}_{\bm{A}_{0}})\,\bm{S}_{c}^{-1/2}\,\bm{z}_{p}}{1+\bm{z}_{p}^{\dagger}\,\bm{S}_{c}^{-1/2}\,\bm{P}_{\bm{A}_{0}}^{\perp}\,\bm{S}_{c}^{-1/2}\,\bm{z}_{p}}\right\} \nonumber \\
 & = & K\,\frac{\bm{z}_{p}^{\dagger}\,\bm{S}_{c}^{-1/2}\,(\bm{P}_{\bm{A}_{1}}-\bm{P}_{\bm{A}_{0}})\,\bm{S}_{c}^{-1/2}\,\bm{z}_{p}}{1+\bm{z}_{p}^{\dagger}\,\bm{S}_{c}^{-1/2}\,\bm{P}_{\bm{A}_{0}}^{\perp}\,\bm{S}_{c}^{-1/2}\,\bm{z}_{p}},\label{eq: GT_vector subspace inter_final}
\end{eqnarray}
where in last line we have omitted $\Re\{\cdot\}$ since Eq. (\ref{eq: GT_vector subspace inter_final})
is formed by Hermitian quadratic forms (at both numerator and denominator);
thus it is \emph{always real-valued}. Therefore Eq.~(\ref{eq: GT_vector subspace inter_final})
is \emph{statistically equivalent} to GLR in Eq.~(\ref{eq: GLRT _vector subspace interference}).

\subsubsection*{LH statistic}

As in the case of no-interference in Sec.~\ref{sub: Vector subspace detection},
the condition $M=1$ holds. Therefore the LH statistic is\emph{ statistically
equivalent} to the GLR.

\subsection{Multidimensional Signals}

In the present case we start from formulation in Eq.~(\ref{eq: Transformed data - hypothesis testing problem})
and assume that: ($i$) $t=0$ (i.e. there is no interference, meaning
$J=r$), ($ii$) $\bm{A}=\bm{E}_{r}=\bm{I}_{N}$ (thus $J=r=N)$ and
($iii$) $\bm{C}\triangleq\begin{bmatrix}\bm{I}_{M} & \bm{0}_{M\times(K-M)}\end{bmatrix}$.
Such case has been dealt in \cite{Liu2013,Conte2003}. Thus, the hypothesis
testing in canonical form is given by: 
\begin{equation}
\begin{cases}
\mathcal{H}_{0}: & \bm{Z}=\bm{N}\\
\mathcal{H}_{1}: & \bm{Z}=\bm{B}\,\bm{C}+\bm{N}
\end{cases}\label{eq: hypothesis testing _multidim signals}
\end{equation}
Clearly, since in this case $J=N$ holds, $(K-M)$ vector components
are assumed signal-free, that is, $\bm{Z}$ admits the partitioning
$\bm{Z}=\begin{bmatrix}\bm{Z}_{M} & \bm{Z}_{s}\end{bmatrix}=\begin{bmatrix}\bm{Z}_{c} & \bm{Z}_{c,\perp}\end{bmatrix}$,
where $\bm{Z}_{M}$ denotes the signal matrix collecting the cells
containing the useful signals and the columns of $\bm{Z}_{s}$ are
the training data. In the latter case, it can be shown that the simplified
projector form holds:
\begin{align}
\bm{P}_{\bm{C}^{\dagger}} & =\,\begin{bmatrix}\bm{I}_{M} & \bm{0}_{M\times(K-M)}\\
\bm{0}_{(K-M)\times M} & \bm{0}_{(K-M)\times(K-M)}
\end{bmatrix}\label{eq: C projection case (C)}
\end{align}
Given the results in Eq. (\ref{eq: C projection  - vector subspace detection}),
it can be shown that $\bm{S}_{c}=\bm{Z}_{s}\bm{Z}_{s}^{\dagger}$
and $\hat{\bm{R}}_{0}=\frac{1}{K}\bm{S}_{0}$, where $\bm{S}_{0}\triangleq(\bm{Z}_{M}\bm{Z}_{M}^{\dagger}+\bm{Z}_{s}\bm{Z}_{s}^{\dagger})$
holds, respectively. Also, it is not difficult to show that $\bm{P}_{\bm{A}_{1}}=\bm{P}_{\bar{\bm{A}}_{1}}=\bm{I}_{N}$
and $\bm{P}_{\bm{A}_{1}}^{\perp}=\bm{P}_{\bar{\bm{A}}_{1}}^{\perp}=\bm{0}_{N\times N}$,
respectively.

\subsubsection*{GLR}

In order to specialize GLR expression we start from Eq. (\ref{eq: GLRT_final form KF}).
Indeed, it can be easily shown that:
\begin{align}
t_{\mathrm{glr}} & =\frac{\det[\bm{I}_{M}+(\bm{Z}_{{\scriptscriptstyle W1}}\bm{V}_{c,1})^{\dagger}(\bm{Z}_{{\scriptscriptstyle W1}}\bm{V}_{c,1})]}{\det[\bm{I}_{M}+(\bm{Z}_{{\scriptscriptstyle W1}}\bm{V}_{c,1})^{\dagger}\bm{P}_{\bm{A}_{1}}^{\perp}(\bm{Z}_{{\scriptscriptstyle W1}}\bm{V}_{c,1})]}\nonumber \\
 & =\det[\bm{I}_{M}+\bm{Z}_{M}^{\dagger}\bm{S}_{c}^{-1}\bm{Z}_{M}]\nonumber \\
 & =\det[\bm{I}_{M}+\bm{S}_{c}^{-1/2}\bm{Z}_{M}\,\bm{Z}_{M}^{\dagger}\,\bm{S}_{c}^{-1/2}]\nonumber \\
 & =\det[\bm{S}_{c}+\bm{Z}_{M}\bm{Z}_{M}^{\dagger}]\,/\,\det[\bm{S}_{c}]
\end{align}
where we have exploited $\bm{P}_{\bm{A}_{1}}^{\perp}=\bm{0}_{N\times N}$
and Sylvester's determinant theorem in third and fourth lines, respectively.
It is apparent that the latter expressions coincide with those in
\cite[Eqs. (18) and (20)]{Conte2003}, respectively.

\subsubsection*{Rao/Durbin statistic}

For the present setup Eq. (\ref{eq: Rao test - final form (alternative)})
specializes into:
\begin{align}
t_{\mathrm{rao}}=\mathrm{Tr}[\bm{Z}_{{\scriptscriptstyle W0}}^{\dagger}\bm{P}_{\bar{\bm{A}}_{1}}\bm{Z}_{{\scriptscriptstyle W0}}\bm{P}_{\bm{C}^{\dagger}}] & =\mathrm{Tr}[\bm{Z}{}^{\dagger}\widehat{\bm{R}}_{0}^{-1}\bm{Z}\,\bm{P}_{\bm{C}^{\dagger}}]\nonumber \\
=K\,\mathrm{Tr}[(\bm{Z}\,\bm{P}_{\bm{C}^{\dagger}})^{\dagger}\bm{S}_{0}^{-1}(\bm{Z}\,\bm{P}_{\bm{C}^{\dagger}})] & =K\,\mathrm{Tr}[\bm{Z}_{M}^{\dagger}\bm{S}_{0}^{-1}\bm{Z}_{M}]\label{eq: Rao test multidim signals}
\end{align}
which coincides with the specific result obtained in \cite{Liu2013},
which was originally derived as a modified two-step GLRT procedure
in \cite{Conte2003}.

\subsubsection*{Wald/2S-GLR statistic}

Starting from Eq. (\ref{eq: Wald_final form (alternative)}), we particularize
the Wald statistic as follows:
\begin{align}
t_{\mathrm{wald}}=\mathrm{Tr}[\bm{Z}_{{\scriptscriptstyle W1}}^{\dagger}\bm{P}_{\bm{A}_{1}}\bm{Z}_{{\scriptscriptstyle W1}}\bm{P}_{\bm{C}^{\dagger}}] & =\mathrm{Tr}[\bm{Z}{}^{\dagger}\bm{S}_{c}^{-1}\bm{Z}\,\bm{P}_{\bm{C}^{\dagger}}]\nonumber \\
=\mathrm{Tr}[(\bm{Z}\,\bm{P}_{\bm{C}^{\dagger}})^{\dagger}\bm{S}_{c}^{-1}(\bm{Z}\,\bm{P}_{\bm{C}^{\dagger}})] & =\mathrm{Tr}[\bm{Z}_{M}^{\dagger}\bm{S}_{c}^{-1}\bm{Z}_{M}]\label{eq: Wald test (multidim signals)}
\end{align}
which coincides with the specific result obtained in \cite{Liu2013}.

\subsubsection*{Gradient statistic}

In this case the gradient statistic in Eq.~(\ref{eq: GT (final form) (alternative)})
specializes into:
\begin{gather}
t_{\mathrm{grad}}=\Re\left\{ \mathrm{Tr}\left[\bm{Z}_{{\scriptscriptstyle W1}}^{\dagger}\bm{P}_{\bm{A}_{1}}(\bm{S}_{c}^{1/2}\widehat{\bm{R}}_{0}^{-1/2})\bm{Z}_{{\scriptscriptstyle W0}}\bm{P}_{\bm{C}^{\dagger}}\right]\right\} \nonumber \\
=K\,\Re\left\{ \mathrm{Tr}\left[\bm{Z}^{\dagger}\bm{S}_{0}^{-1}\,\bm{Z}\,\bm{P}_{\bm{C}^{\dagger}}\right]\right\} =K\,\mathrm{Tr}\left[\bm{Z}_{M}^{\dagger}\,\bm{S}_{0}^{-1}\,\bm{Z}_{M}\right]\label{eq: GT _multidim signals}
\end{gather}
It is interesting to observe that in this specific scenario, \emph{Gradient
statistic coincides with Rao statistic} in Eq. (\ref{eq: Rao test multidim signals}).

\subsubsection*{LH statistic}

In this specific instance, LH statistic in Eq.~(\ref{eq: Lawley-Hotelling final form})
specializes into:
\begin{gather}
t_{\mathrm{lh}}=\mathrm{Tr}\left[(\bm{Z}_{{\scriptscriptstyle W1}}\bm{V}_{c,1})^{\dagger}\bm{P}_{\bm{A}_{1}}(\bm{Z}_{{\scriptscriptstyle W1}}\bm{V}_{c,1})\times\right.\label{eq: LH _multidim signals}\\
\left.\left(\bm{I}_{M}+(\bm{Z}_{{\scriptscriptstyle W1}}\bm{V}_{c,1})^{\dagger}\bm{P}_{\bm{A}_{1}}^{\perp}(\bm{Z}_{{\scriptscriptstyle W1}}\bm{V}_{c,1})\right)^{-1}\right]=\mathrm{Tr}\left[\bm{Z}_{M}^{\dagger}\bm{S}_{c}^{-1}\bm{Z}_{M}\right]\nonumber 
\end{gather}
since $\bm{P}_{\bm{A}_{1}}^{\perp}=\bm{0}_{N\times N}$ (resp. $\bm{P}_{\bm{A}_{1}}=\bm{I}_{N}$)
for multidimensional signal setup. From inspection of the last line,
it is apparent that \emph{LH statistic coincides with Wald/2S-GLR
statistic} in Eq.~(\ref{eq: Wald test (multidim signals)}) for this
specific scenario.

\subsection{Range-spread Targets}

In the present case we start from general formulation in Eq.~(\ref{eq: Transformed data - hypothesis testing problem})
and assume that: ($i$) $t=0$ (i.e., there is no interference, thus
$J=r$); ($ii$) $r=1$, thus the matrices $\bm{A}$ and $\bm{B}$
collapse to $\bm{a}\triangleq\begin{bmatrix}1 & 0 & \cdots & 0\end{bmatrix}^{T}\in\mathbb{C}^{N\times1}$
and $\bm{b}\in\mathbb{C}^{1\times M}$ (i.e., a row vector), respectively;
($iii$) $\bm{C}\triangleq\begin{bmatrix}\bm{I}_{M} & \bm{0}_{M\times K-M}\end{bmatrix}$.
Such case has been dealt in \cite{Shuai2012,Conte2001,Raghavan2013}.
Therefore, the hypothesis testing in canonical form is given by:
\begin{equation}
\begin{cases}
\mathcal{H}_{0}: & \bm{Z}=\bm{N}\\
\mathcal{H}_{1}: & \bm{Z}=\bm{a}\,\bm{b}\,\bm{C}+\bm{N}
\end{cases}\,.
\end{equation}
Additionally, $(K-M)$ vector components are assumed signal-free,
that is, $\bm{Z}$ admits the partitioning $\bm{Z}=\begin{bmatrix}\bm{Z}_{e} & \bm{Z}_{s}\end{bmatrix}$
where $\bm{Z}_{e}\in\mathbb{C}^{N\times M}$ comprises the cells containing
the extended target and $\bm{Z}_{s}\in\mathbb{C}^{N\times(K-M)}$
collects the secondary (training) data. In the latter case, the following
simplified projector form holds:
\begin{eqnarray}
\bm{P}_{\bm{C}^{\dagger}} & = & \begin{bmatrix}\bm{I}_{M} & \bm{0}_{M\times(K-M)}\\
\bm{0}_{(K-M)\times M} & \bm{0}_{(K-M)\times M}
\end{bmatrix}\,.\label{eq: C projection case (B)}
\end{eqnarray}
Based on the structure of Eq. (\ref{eq: C projection case (B)}),
it follows that $\bm{S}_{c}=\bm{Z}_{s}\bm{Z}_{s}^{\dagger}$ and $\widehat{\bm{R}}_{0}=\frac{1}{K}\bm{S}_{0}$,
where $\bm{S}_{0}\triangleq(\bm{Z}_{e}\bm{Z}_{e}+\bm{Z}_{s}\bm{Z}_{s}^{\dagger})$.
Moreover, it can be shown that $\bm{P}_{\bm{a}_{1}}$ and $\bm{P}_{\bar{\bm{a}}_{1}}$
(where we have analogously defined $\bm{a}_{1}\triangleq(\bm{S}_{c}^{-1/2}\bm{a})$
and $\bar{\bm{a}}_{1}\triangleq(\widehat{\bm{R}}_{0}^{-1/2}\bm{a})$)
assumes the following simplified expression:
\begin{equation}
\bm{P}_{\bm{a}_{1}}=\frac{\bm{S}_{c}^{-1/2}\bm{a}\bm{a}^{\dagger}\bm{S}_{c}^{-1/2}}{\bm{a}^{\dagger}\bm{S}_{c}^{-1}\bm{a}};\qquad\bm{P}_{\bar{\bm{a}}_{1}}=\frac{\bm{S}_{0}^{-1/2}\bm{a}\bm{a}^{\dagger}\bm{S}_{0}^{-1/2}}{\bm{a}^{\dagger}\bm{S}_{0}^{-1}\bm{a}}\,.\label{eq: A projection (case B)}
\end{equation}
In some cases, we will use the Woodbury identity applied to $\bm{S}_{0}^{-1}$,
that is:
\begin{equation}
\bm{S}_{0}^{-1}=\bm{S}_{c}^{-1}-\bm{S}_{c}^{-1}\bm{Z}_{e}\,(\bm{I}_{M}+\bm{Z}_{e}^{\dagger}\bm{S}_{c}^{-1}\bm{Z}_{e})^{-1}\,\bm{Z}_{e}^{\dagger}\,\bm{S}_{c}^{-1}\,.\label{eq: Woodbury identity}
\end{equation}

\subsubsection*{GLR}

Aiming at particularizing the expression of the GLR for the present
case, we follow the same derivation as in \cite{Kelly1989} to show
that Eq. (\ref{eq: GLRT_final form (KF alternative)}) can be specialized
exploiting the equalities
\begin{align}
(\bm{Z}_{{\scriptscriptstyle W1}}\bm{V}_{c,1})^{\dagger}\bm{P}_{\bm{a}_{1}}(\bm{Z}_{{\scriptscriptstyle W1}}\bm{V}_{c,1}) & =\frac{(\bm{Z}_{e}^{\dagger}\bm{S}_{c}^{-1}\bm{a})(\bm{Z}_{e}^{\dagger}\bm{S}_{c}^{-1}\bm{a})^{\dagger}}{(\bm{a}^{\dagger}\bm{S}_{c}^{-1}\bm{a})}\,,\label{eq: projection GLR (range spread)}\\
\bm{D}_{0} & =\bm{I}_{M}+\bm{Z}_{e}^{\dagger}\bm{S}_{c}^{-1}\bm{Z}_{e}\,,\label{eq: D_0 (range spread)}
\end{align}
thus obtaining $t_{\mathrm{glr}}=[1/(1-\eta^{'})]$, where:
\begin{equation}
\eta^{'}\triangleq\frac{(\bm{a}^{\dagger}\bm{S}_{c}^{-1}\bm{Z}_{e})\,[\bm{I}_{M}+\bm{Z}_{e}^{\dagger}\bm{S}_{c}^{-1}\bm{Z}_{e}]^{-1}\,(\bm{Z}_{e}^{\dagger}\bm{S}_{c}^{-1}\bm{a})}{(\bm{a}^{\dagger}\bm{S}_{c}^{-1}\bm{a})}\,.\label{eq: GLRT range-spread KF}
\end{equation}
The result in Eq. (\ref{eq: GLRT range-spread KF}) is obtained after
substitution of Eqs. (\ref{eq: projection GLR (range spread)}) and
(\ref{eq: D_0 (range spread)}) into Eq. (\ref{eq: GLRT_final form (KF alternative)})
and exploiting Sylvester's determinant theorem. Such GLR form\footnote{It is worth pointing out that an alternative (equivalent) form of
GLR was obtained in \cite{Conte2001,Wang1991} for the range-spread
case. The aforementioned expression can be simply obtained starting
from general formula in Eq. (\ref{eq: GLRT_final form KF}), straightforward
application of Sylvester's determinant theorem and exploitation of
the simplified assumptions of range-spread scenario.} (as $t_{\mathrm{glr}}$ is a monotone function of $\eta^{'}$) corresponds
to that found in \cite{Kelly1989}.

\subsubsection*{Rao/Durbin statistic}

In the present case Eq. (\ref{eq: Rao test - final form (alternative)})
specializes into:
\begin{gather}
t_{\mathrm{rao}}=\mathrm{Tr}[\bm{Z}_{{\scriptscriptstyle W0}}^{\dagger}\bm{P}_{\bar{\bm{a}}_{1}}\bm{Z}_{{\scriptscriptstyle W0}}\bm{P}_{\bm{C}^{\dagger}}]\nonumber \\
=\mathrm{Tr}[(\bm{Z}\bm{P}_{\bm{C}^{\dagger}})^{\dagger}\widehat{\bm{R}}_{0}^{-1/2}\bm{P}_{\bar{\bm{a}}_{1}}\widehat{\bm{R}}_{0}^{-1/2}(\bm{Z}\bm{P}_{\bm{C}^{\dagger}})]\nonumber \\
=\mathrm{Tr}[\bm{Z}_{e}^{\dagger}\,\widehat{\bm{R}}_{0}^{-1/2}\bm{P}_{\bar{\bm{a}}_{1}}\widehat{\bm{R}}_{0}^{-1/2}\bm{Z}_{e}]\nonumber \\
=\frac{K\,\mathrm{Tr}[\bm{Z}_{e}^{\dagger}\,\bm{S}_{0}^{-1}\bm{a}\bm{a}^{\dagger}\bm{S}_{0}^{-1}\bm{Z}_{e}]}{(\bm{a}^{\dagger}\bm{S}_{0}^{-1}\bm{a})}=K\frac{\left\Vert \bm{Z}_{e}^{\dagger}\bm{S}_{0}^{-1}\bm{a}\right\Vert ^{2}}{(\bm{a}^{\dagger}\bm{S}_{0}^{-1}\bm{a})}\label{eq: Rao test_range spread}
\end{gather}
Eq. (\ref{eq: Rao test_range spread}) is recognized as the result
found in \cite{Shuai2012}.

\subsubsection*{Wald/2S-GLR statistic}

Starting from Eq. (\ref{eq: Wald_final form (alternative)}), we particularize
the test as follows:
\begin{gather}
t_{\mathrm{wald}}=\mathrm{Tr}[\bm{Z}_{{\scriptscriptstyle W1}}^{\dagger}\bm{P}_{\bm{a}_{1}}\bm{Z}_{{\scriptscriptstyle W1}}\bm{P}_{\bm{C}^{\dagger}}]\nonumber \\
=\mathrm{Tr}[(\bm{Z}\bm{P}_{\bm{C}^{\dagger}})^{\dagger}\bm{S}_{c}^{-1/2}\bm{P}_{\bm{a}_{1}}\bm{S}_{c}^{-1/2}(\bm{Z}\bm{P}_{\bm{C}^{\dagger}})]\nonumber \\
=\mathrm{Tr}[\bm{Z}_{e}^{\dagger}\,\bm{S}_{c}^{-1/2}\bm{P}_{\bm{a}_{1}}\bm{S}_{c}^{-1/2}\,\bm{Z}_{e}]\nonumber \\
=\frac{\mathrm{Tr}[\bm{Z}_{e}^{\dagger}\,\bm{S}_{c}^{-1}\bm{a}\bm{a}^{\dagger}\bm{S}_{c}^{-1}\bm{Z}_{e}]}{(\bm{a}^{\dagger}\bm{S}_{c}^{-1}\bm{a})}=\frac{\left\Vert \bm{Z}_{e}^{\dagger}\bm{S}_{c}^{-1}\bm{a}\right\Vert ^{2}}{(\bm{a}^{\dagger}\bm{S}_{c}^{-1}\bm{a})}\label{eq: Wald test range-spread}
\end{gather}
which agrees with the result in \cite{Shuai2012} and can be shown
to coincide with the generalized AMF proposed in \cite{Conte2001},
thus extending the theoretical findings in \cite{A.DeMaio2004}.

\subsubsection*{Gradient statistic}

In this case Eq. (\ref{eq: GT (final form) (alternative)}) reduces
to:
\begin{gather}
t_{\mathrm{grad}}=\Re\left\{ \mathrm{Tr}\left[\bm{Z}_{{\scriptscriptstyle W1}}^{\dagger}\bm{P}_{\bm{a}_{1}}(\bm{S}_{c}^{1/2}\widehat{\bm{R}}_{0}^{-1/2})\bm{Z}_{{\scriptscriptstyle W0}}\bm{P}_{\bm{C}^{\dagger}}\right]\right\} \nonumber \\
=K\frac{\Re\left\{ \mathrm{Tr}\left[\bm{Z}_{e}^{\dagger}\,\bm{S}_{c}^{-1}\,\bm{a}\bm{a}^{\dagger}\bm{S}_{0}^{-1}\bm{Z}_{e}\right]\right\} }{(\bm{a}^{\dagger}\bm{S}_{c}^{-1}\bm{a})}\nonumber \\
=K\frac{\Re\left\{ \left[\bm{Z}_{e}^{\dagger}\bm{S}_{c}^{-1}\bm{a}\right]^{\dagger}\left[\bm{Z}_{e}^{\dagger}\bm{S}_{0}^{-1}\bm{a}\right]\right\} }{(\bm{a}^{\dagger}\bm{S}_{c}^{-1}\bm{a})}\label{eq: GT range-spread implicit}
\end{gather}
where we have used $\Re\{\mathrm{Tr}[\bm{m}\bm{n}^{\dagger}]\}=\Re[\bm{m}^{\dagger}\bm{n}]$,
with $\bm{m}$ and $\bm{n}$ being two column vectors of proper size.
Furthermore, by exploiting Eq. (\ref{eq: Woodbury identity}), the
following equality holds
\begin{equation}
\left(\bm{Z}_{e}^{\dagger}\,\bm{S}_{0}^{-1}\bm{a}\right)=[\bm{I}_{M}+\bm{Z}_{e}^{\dagger}\,\bm{S}_{c}^{-1}\bm{Z}_{e}]^{-1}(\bm{Z}_{e}^{\dagger}\,\bm{S}_{c}^{-1}\bm{a})
\end{equation}
which, substituted into Eq. (\ref{eq: GT range-spread implicit}),
gives:
\begin{gather}
t_{\mathrm{grad}}=K\frac{\left[\bm{Z}_{e}^{\dagger}\bm{S}_{c}^{-1}\bm{a}\right]^{\dagger}[\bm{I}_{M}+\bm{Z}_{e}^{\dagger}\bm{S}_{c}^{-1}\bm{Z}_{e}]^{-1}\left[\bm{Z}_{e}^{\dagger}\bm{S}_{c}^{-1}\bm{a}\right]}{(\bm{a}^{\dagger}\bm{S}_{c}^{-1}\bm{a})}\label{eq: grad stat _like GLR (range spread)}
\end{gather}
where we have omitted $\Re\{\cdot\}$ since Eq. (\ref{eq: grad stat _like GLR (range spread)})
is an Hermitian quadratic form (i.e., it is always real-valued). Therefore,
the \emph{Gradient statistic is statistically equivalent to the GLR
in Eq.~(\ref{eq: GLRT range-spread KF})}.

\subsubsection*{LH statistic}

In this case the general LH statistic form in Eq. (\ref{eq: Lawley-Hotelling final form})
specializes into:
\begin{align}
t_{\mathrm{lh}} & =\mathrm{Tr}\left[(\bm{Z}_{{\scriptscriptstyle W1}}\bm{V}_{c,1})^{\dagger}\bm{P}_{\bm{a}_{1}}(\bm{Z}_{{\scriptscriptstyle W1}}\bm{V}_{c,1})\bm{D}_{1}^{-1}\right]\nonumber \\
 & =\mathrm{Tr}\left[\frac{\bm{Z}_{e}^{\dagger}\,\bm{S}_{c}^{-1}\bm{a}\bm{a}^{\dagger}\bm{S}_{c}^{-1}\bm{Z}_{e}}{(\bm{a}^{\dagger}\bm{S}_{c}^{-1}\bm{a})}\bm{D}_{1}^{-1}\right]\label{eq: LH (range spread)}
\end{align}
where $\bm{D}_{1}=\bm{I}_{M}+(\bm{Z}_{{\scriptscriptstyle W1}}\bm{V}_{c,1})^{\dagger}\bm{P}_{\bm{a}_{1}}^{\perp}(\bm{Z}_{{\scriptscriptstyle W1}}\bm{V}_{c,1})$
in this specific case. Matrix $\bm{D}_{1}$ can be further rewritten
as:
\begin{gather}
\bm{D}_{1}=(\bm{I}_{M}+\bm{Z}_{e}^{\dagger}\,\bm{S}_{c}^{-1}\,\bm{Z}_{e})-\frac{\bm{Z}_{e}^{\dagger}\,\bm{S}_{c}^{-1}\bm{a}\bm{a}^{\dagger}\bm{S}_{c}^{-1}\bm{Z}_{e}}{(\bm{a}^{\dagger}\bm{S}_{c}^{-1}\bm{a})}\,
\end{gather}
Applying the Woodbury identity on $\bm{D}_{1}^{-1}$, we obtain:
\begin{align}
\bm{D}_{1}^{-1}= & \left\{ \bm{D}_{0}^{-1}+\right.\nonumber \\
 & \left.\frac{\bm{D}_{0}^{-1}\left(\bm{Z}_{e}^{\dagger}\bm{S}_{c}^{-1}\bm{a}\right)(\bm{Z}_{e}^{\dagger}\bm{S}_{c}^{-1}\bm{a})^{\dagger}\bm{D}_{0}^{-1}}{\left(\bm{a}^{\dagger}\bm{S}_{c}^{-1}\bm{a}\right)\left[1-\frac{(\bm{Z}_{e}^{\dagger}\bm{S}_{c}^{-1}\bm{a})^{\dagger}\bm{D}_{0}^{-1}(\bm{Z}_{e}^{\dagger}\bm{S}_{c}^{-1}\bm{a})}{\bm{a}^{\dagger}\bm{S}_{c}^{-1}\bm{a}}\right]}\right\} 
\end{align}
where we exploited the definition of $\bm{D}_{0}$ in Eq. (\ref{eq: D_0 (range spread)}).
Thus, after substitution into Eq. (\ref{eq: LH (range spread)}),
we obtain
\begin{equation}
t_{\mathrm{lh}}=\eta^{'}+\frac{\left(\eta^{'}\right)^{2}}{1-\eta^{'}}=\frac{\eta'}{1-\eta^{'}}\propto\eta^{'}
\end{equation}
with $\eta'$ given by Eq. (\ref{eq: GLRT range-spread KF}). Thus
LH statistic is \emph{statistically equivalent to the GLR} for range-spread
targets.

\subsection{Standard GMANOVA}

In the present case no interference is present ($t=0$, thus $J=r$).
This reduces to the standard adaptive detection problem via the GMANOVA
model considered in \cite{Kelly1989,Burgess1996,Liu2014} and whose
canonical form is:
\begin{equation}
\begin{cases}
\mathcal{H}_{0}: & \bm{Z}=\bm{N}\\
\mathcal{H}_{1}: & \bm{Z}=\bm{A}\bm{B}\bm{C}+\bm{N}
\end{cases}\label{eq: Hypothesis testing formulation - GMANOVA no interf}
\end{equation}
Clearly, under the above assumptions, it holds $\bm{P}_{\bm{A}_{0}}=\bm{P}_{\bar{\bm{A}}_{0}}=\bm{0}_{N\times N}$.
Therefore, the ML covariance estimate under $\mathcal{H}_{0}$ simplifies
into $\widehat{\bm{R}}_{0}=K^{-1}\bm{S}_{0}$, where $\bm{S}_{0}\triangleq\bm{Z}\bm{Z}^{\dagger}$(cf.
Eq. (\ref{eq: Closed form MLE covariance H_0})).

\subsubsection*{GLR}

Direct specialization of Eq. (\ref{eq: GLRT_final form KF}) gives
the explicit statistic:
\begin{align}
t_{\mathrm{glr}} & =\frac{\det[\bm{I}_{M}+(\bm{Z}_{{\scriptscriptstyle W1}}\bm{V}_{c,1})^{\dagger}(\bm{Z}_{{\scriptscriptstyle W1}}\bm{V}_{c,1})]}{\det[\bm{I}_{M}+(\bm{Z}_{{\scriptscriptstyle W1}}\bm{V}_{c,1})^{\dagger}\,\bm{P}_{\bm{A}_{1}}^{\perp}(\bm{Z}_{{\scriptscriptstyle W1}}\bm{V}_{c,1})]}\label{eq: GLRT_GMANOVA no interf}
\end{align}
which coincides with the classical expression\footnote{We point out that Eq. (\ref{eq: GLRT_GMANOVA no interf}) can be also
re-arranged in a similar form as Eq. (\ref{eq: GLRT_final form (Wilks' Lambda statistic)})
(i.e., a Wilks' Lambda statistic form). Such expression, being equal
to $t_{\mathrm{glr}}=\det[\bm{I}_{M}+\bm{D}_{1}^{-1/2}\,(\bm{Z}_{{\scriptscriptstyle W1}}\bm{V}_{c,1})^{\dagger}\,\bm{P}_{\bm{A}_{1}}(\bm{Z}_{{\scriptscriptstyle W1}}\bm{V}_{c,1})\,\bm{D}_{1}^{-1/2}]$,
represents the alternative GLR form obtained in \cite{Kelly1989}.} of GLR obtained in \cite{Kelly1989}.

\subsubsection*{Rao/Durbin statistic}

Direct particularization of Eq. (\ref{eq: Rao test - final form (alternative)})
gives:
\begin{gather}
t_{\mathrm{rao}}=\mathrm{Tr}[\bm{Z}_{{\scriptscriptstyle W0}}^{\dagger}\bm{P}_{\bar{\bm{A}}_{1}}\bm{Z}_{{\scriptscriptstyle W0}}\bm{P}_{\bm{C}^{\dagger}}]\label{eq: Rao test  - GMANOVA no interf}
\end{gather}
which provides the result obtained in \cite{Liu2014}.

\subsubsection*{Wald/2S-GLR statistic}

Direct specialization of Eq. (\ref{eq: Wald_final form (alternative)})
gives leads to:
\begin{gather}
t_{\mathrm{wald}}=\mathrm{Tr}[\bm{Z}_{{\scriptscriptstyle W1}}^{\dagger}\bm{P}_{\bm{A}_{1}}\bm{Z}_{{\scriptscriptstyle W1}}\bm{P}_{\bm{C}^{\dagger}}]\label{eq: Wald test _GMANOVA no interf}
\end{gather}
which is the same result obtained in \cite{Liu2014}.

\subsubsection*{Gradient statistic}

Direct application of Eq. (\ref{eq: GT (final form) (alternative)})
provides:
\begin{gather}
t_{\mathrm{grad}}=\Re\left\{ \mathrm{Tr}\left[\bm{Z}_{{\scriptscriptstyle W1}}^{\dagger}\bm{P}_{\bm{A}_{1}}(\bm{S}_{c}^{1/2}\widehat{\bm{R}}_{0}^{-1/2})\bm{Z}_{{\scriptscriptstyle W0}}\bm{P}_{\bm{C}^{\dagger}}\right]\right\} \label{eq: GT  - GMANOVA no interf}
\end{gather}

\subsubsection*{LH statistic}

In this case the LH statistic specializes into:
\begin{gather}
t_{\mathrm{lh}}=\mathrm{Tr}\left[(\bm{Z}_{{\scriptscriptstyle W1}}\bm{V}_{c,1})^{\dagger}\bm{P}_{\bm{A}_{1}}(\bm{Z}_{{\scriptscriptstyle W1}}\bm{V}_{c,1})\bm{D}_{1}^{-1}\right]\,.\label{eq: Lawley-Hotelling- GMANOVA no interf}
\end{gather}
where $\bm{D}_{1}$ is defined as in Sec. \ref{sub: GLR general}.
Eq. (\ref{eq: Lawley-Hotelling- GMANOVA no interf}) clearly coincides
with the statistic obtained in \cite[pag. 37]{Kelly1989}.

\section{Simulation results \label{sec: Simulation results}}

\begin{figure*}
\begin{centering}
\includegraphics[width=2\columnwidth]{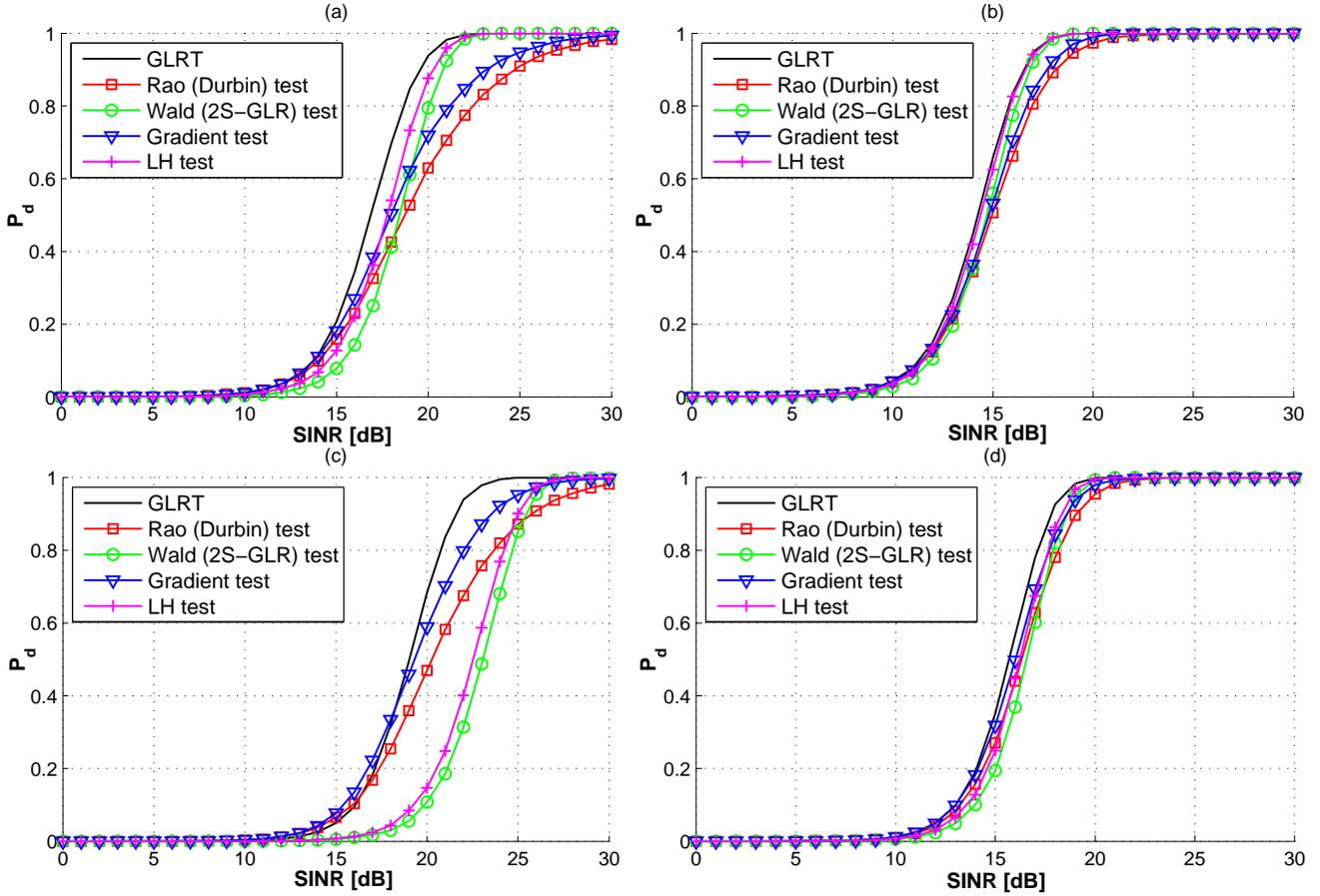}
\par\end{centering}

\protect\caption{$P_{d}$ vs. $\rho$ for all the considered detectors; common parameters:
$M=3$ and $N=8$. Case ($a$) (top-left) $r=2$, $t=4$ and $K=12$;
Case ($b$) (top-right) $r=2$, $t=4$ and $K=19$; Case ($c$) (bottom-left)
$r=4$, $t=2$ and $K=12$; Case ($d$) (bottom-right) $r=4$, $t=2$
and $K=19$. \label{fig: Pd vs SNR M =00003D 3 N =00003D 8 (overall comparison)}}
\end{figure*}

In Fig. \ref{fig: Pd vs SNR M =00003D 3 N =00003D 8 (overall comparison)},
we report $P_{d}$ vs. the SINR for all the considered detectors,
defined as $\rho\triangleq\mathrm{Tr}[\bm{B}^{\dagger}\bm{R}_{2.3}^{-1}\bm{B}]$,
that is, the trace of the induced maximal invariant (cf. Sec. \ref{sub: MIS recall}).
We underline that such term is also proportional to the non-centrality
parameter $\lambda\triangleq(\bm{\theta}_{r,1}-\bm{\theta}_{r,0})^{T}\{[\bm{I}^{-1}(\bm{\theta}_{0})]_{\bm{\theta}_{r},\bm{\theta}_{r}}\}^{-1}(\bm{\theta}_{r,1}-\bm{\theta}_{r,0})$,
representing the synthetic parameter on which the asymptotic performances
of all the considered test depend \cite{Kay1998}. The curves have
been obtained via standard Monte Carlo counting techniques. More specifically,
the thresholds necessary to ensure a preassigned value of $P_{fa}$
have been evaluated exploiting $100/P_{fa}$ independent trials, while
the $P_{d}$ values are estimated over $5\cdot10^{3}$ independent
trials. As to the disturbance, it is modeled as an exponentially-correlated
Gaussian vector with covariance matrix (in canonical form) $\bm{R}=\sigma_{n}^{2}\bm{I}_{N}+\sigma_{c}^{2}\bm{M}_{c}$,
where $\sigma_{n}^{2}>0$ is the thermal noise power, $\sigma_{c}^{2}>0$
is the clutter power, and the ($i,j$)-th element of $\bm{R}_{c}$
is given by $0.95^{|i-j|}$. The clutter-to-noise ratio $\sigma_{c}^{2}/\sigma_{n}^{2}$
is set here to $30\,\mathrm{dB}$, with $\sigma_{n}^{2}=1$. We point
out that the specific value of the deterministic interference $\bm{B}_{t}$
does not need to be specified at each trial considered (for both $P_{fa}$
and $P_{d}$ evaluation); the reason is that the performance of each
detector depends on the unknown parameters solely through the induced
maximal invariant, which is \emph{independent} on $\bm{B}_{t}$ (cf.
Sec. \ref{sub: MIS recall}). Finally, all the numerical examples
assume $P_{fa}=10^{-4}$. 

In order to average the performance of $P_{d}$ with respect to $\bm{B}$,
for each independent trial we generate  the signal matrix as $\bm{B}=\alpha_{B}\bm{B}_{g}$,
where $\bm{B}_{g}\sim\mathcal{CN}(\bm{0}_{r\times M},\bm{I}_{M},\bm{I}_{r})$
and $\alpha_{B}\in\mathbb{R}$. The latter coefficient is  a scaling
factor used to achieve the desired SINR value, that is, $\alpha_{b}\triangleq\sqrt{\rho\,/\,\mathrm{Tr}[\bm{B}_{g}^{\dagger}\,\bm{R}_{2.3}^{-1}\,\bm{B}_{g}]}$.

For our simulations\footnote{Of course, due to the high number of setup parameters involved in
the detection problem (i.e., $N,K,M,r,t$), we do not claim the following
conclusions to be general for any type of setup. Nonetheless, we illustrate
a generic setup in order to show some common trends observed among
the detectors. A general numerical comparison is omitted due to the
lack of space and since performance comparison in some specific setups
(such as those considered in Sec. \ref{sec: Detectors in special cases})
can be found in the related literature. Nonetheless, the supplementary
material attached to this paper contains some additional numerical
results aimed at confirming the statistical equivalence results obtained
for the considered special scenarios.} we assume $M=3$ (i.e., an extended target), $N=8$, and two different
scenarios of signal and interference lying in a vector subspace, that
is ($i$) $r=2$ and $t=4$ (sub-plots ($a$) and ($b$)) and ($ii$)
$r=4$ and $t=2$ (sub-plots ($c$) and ($d$)).  Additionally, for
each of these setups, the cases corresponding to $K=12$ and $K=19$
columns for $\bm{Z}$ have been considered, representing two extreme
case-studies. Indeed, the first case clearly corresponds to a so-called
\emph{sample-starved} scenario (i.e. the number of signal-free data
required to achieve a consistent (invertible) estimate of $\bm{R}$
is just satisfied, that is, $(K-M)=9$) while the second case to a
setup where an adequate number of samples needed to obtain an accurate
estimate for $\bm{R}$ is provided (i.e., in this case $(K-M)=2N=16$,
with a consequent loss of $3\,\mathrm{dB}$ in estimating $\bm{R}$
with the sample covariance approach, with respect to the known covariance
case, as dictated from \cite{Reed1974}).

The following observations can be made from inspection of the results.
First, as $K$ grows large, all the considered detectors converge
to the same performance, corresponding to the non-adaptive case. Differently,
in the sample-starved case (viz. the difference $K-M$ is close to
$N$) a significant difference in detection performance can be observed
among them. First of all, the GLR has the best performance in the
medium-high SNR range. Differently, the Rao and Gradient tests perform
significantly better than Wald and LH tests for a moderate number
of $K$ (i.e., $K=12$) in the case $r>t$ (cf. sub-plot ($c$), corresponding
to $r=4$ and $t=2$). On the other hand, for the same case $K=12$,
but $r=2$ and $t=4$, Wald and LH tests outperform Rao and Gradient
tests  when $\rho$ is  higher than $\approx18\,\mathrm{dB}$. This
is easily explained since both Wald (viz. 2S-GLR) and LH test both
rely on an accurate estimate of true covariance $\bm{R}$ based on
the sole signal-free data (cf. Secs. \ref{sub: Wald general}-\ref{sub: 2S-GLR general}
and \ref{sub: LH general}, respectively). Differently, both Gradient
and Rao tests employ a covariance estimate under the hypothesis $\mathcal{H}_{0}$
(that is, $\widehat{\bm{R}}_{0}$). The latter covariance  estimate
also  relies  on the use of  the additional contributions of $\bm{Z}$
corrupted by the signal $\bm{B}$. Although using them to evaluate
$\widehat{\bm{R}}_{0}$ may be detrimental when the number of signal-free
samples is adequate or the SINR is high (cf. sub-plots ($a$) and
($b$)), when the SINR is low (i.e., the energy spread among the different
columns is not so high) and the number of signal-free samples is not
sufficient to guarantee a reliable estimate of $\bm{R}$, the degradation
of using signal-corrupted terms is overcome by the (beneficial) availability
of additional samples for covariance estimation.

\section{Conclusions\label{sec: Conclusions}}

In this second part of this work, we have derived several detectors
for adaptive detection in a GMANOVA signal model with structured interference
(viz. I-GMANOVA). We derived the GLR, Rao, Wald, 2S-GLR, Durbin, Gradient,
and LH statistics. All the aforementioned statistics have been shown
to be CFAR with respect to the nuisance parameters, by proving that
all can be written in terms of the MIS (obtained in the first part
of this work). For the considered general model, we also established
statistical equivalence between: ($i$) Wald and 2S-GLR statistics
and ($ii$) Durbin and Rao statistics. 

Furthermore, the following statistical-equivalence results have been
proved in the following special setups:
\begin{itemize}
\item For point-like targets (with possible point-like interference), we
have shown that Gradient and LH tests are statistically equivalent
to Kelly's GLRT;
\item For multidimensional signals, we have shown that: ($a$) Rao test
is statistically equivalent to Gradient test and ($b$) Wald test
(2S-GLRT) is statistically equivalent to LH test;
\item For range-spread targets and rank-one subspace ($r=1$), we have shown
that Gradient and LH tests are statistically equivalent to the GLRT.
\end{itemize}
Finally, simulation results were provided to compare the performance
of the aforementioned detectors.

\section{Supplementary Material Organization}

The following additional sections  contain supplemental material for
part II of this work. More specifically, Sec. \ref{sec: Lemma 1}
contains the proof of Lem. \ref{lem: ML covariance estimates properties}
in the paper, while Secs. \ref{sec: Appendix _ Rao statistic derivation},
\ref{sec: Appendix_ Wald test} and \ref{sec: Appendix Gradient statistic}
provide the derivation of Rao, Wald, and Gradient (Terrell) tests,
respectively. Furthermore, Sec. \ref{sec: Appendix_ Durbin statistic}
provides the statistical equivalence between Rao and Durbin tests
(Thm. \ref{thm: Durbin-Rao equivalence}). Additionally, Sec. \ref{sec: Appendix _ Useful equalities}
provides a series of useful equalities for showing the CFARness of
all the considered detectors. Finally, Sec. \ref{sec: Simresults}
provides some numerical results aimedat confirming the special equivalence
results obtained in the manuscript.\setcounter{enumiv}{0}

\section{Proof of Lemma 1 \label{sec: Lemma 1}}

We only provide the proof for Eq. (\ref{eq: Lem1 R1}), as the equality
for $\widehat{\bm{R}}_{0}^{-1}\bm{E}_{t}$ in Eq. (\ref{eq: Lem1 R0})
can be obtained following similar steps. We first rewrite Eq.~(\ref{eq: Closed form MLE covariance H_1})
as:
\begin{equation}
\widehat{\bm{R}}_{1}=K^{-1}\,[\bm{S}_{c}+\bm{S}_{c}^{1/2}\,(\bm{P}_{\bm{A}_{1}}^{\perp}\bm{Z}_{{\scriptscriptstyle W1}})\bm{P}_{\bm{C}^{\dagger}}(\bm{P}_{\bm{A}_{1}}^{\perp}\bm{Z}_{{\scriptscriptstyle W1}})^{\dagger}\bm{S}_{c}^{1/2}]
\end{equation}
Taking the inverse and exploiting Woodbury identity \cite{Bernstein2009}
gives:
\begin{gather}
\widehat{\bm{R}}_{1}^{-1}=K\,[\bm{S}_{c}^{-1}-\bm{S}_{c}^{-1/2}(\bm{P}_{\bm{A}_{1}}^{\perp}\bm{Z}_{{\scriptscriptstyle W1}})\bm{V}_{c,1}\nonumber \\
\times\{\bm{I}_{M}+(\bm{Z}_{{\scriptscriptstyle W1}}\bm{V}_{c,1})^{\dagger}\bm{P}_{\bm{A}_{1}}^{\perp}(\bm{Z}_{{\scriptscriptstyle W1}}\bm{V}_{c,1})\}^{-1}\nonumber \\
\times\bm{V}_{c,1}^{\dagger}\bm{Z}_{{\scriptscriptstyle W1}}^{\dagger}\bm{P}_{\bm{A}_{1}}^{\perp}\bm{S}_{c}^{-1/2}]\label{eq: inverse covariance woodbury}
\end{gather}
It is apparent that the second term in Eq. (\ref{eq: inverse covariance woodbury})
is null when post-multiplied by $\bm{A}$, since $(\bm{P}_{\bm{A}_{1}}^{\perp}\bm{S}_{c}^{-1/2}\bm{A})=\bm{P}_{\bm{A}_{1}}^{\perp}\bm{A}_{1}=\bm{0}_{N\times J}$,
thus leading to the claimed result.

\section{Derivation of Rao statistic\label{sec: Appendix _ Rao statistic derivation}}

In this appendix we report the derivation for Rao statistic in Eq.
(\ref{eq: Rao_final form (implicit)}) . Before proceeding, we define
the auxiliary notation $\bm{b}_{s,R}\triangleq\begin{bmatrix}\bm{b}_{R}^{T} & \bm{b}_{t,R}^{T}\end{bmatrix}^{T}$
and $\bm{b}_{s,I}\triangleq\begin{bmatrix}\bm{b}_{I}^{T} & \bm{b}_{t,I}^{T}\end{bmatrix}^{T}$.
First, it can be shown that:
\begin{eqnarray}
\frac{\partial\ln f_{1}(\bm{Z};\bm{B}_{s},\bm{R})}{\partial\bm{b}_{s,R}} & = & \begin{bmatrix}2\,\Re\{\bm{g}_{A}\}\\
2\,\Re\{\bm{g}_{B}\}
\end{bmatrix}\in\mathbb{R}^{JM\times1};\label{eq: grad_rao_bsr}\\
\frac{\partial\ln f_{1}(\bm{Z};\bm{B}_{s},\bm{R})}{\partial\bm{b}_{s,I}} & = & \begin{bmatrix}2\,\Im\{\bm{g}_{A}\}\\
2\,\Im\{\bm{g}_{B}\}
\end{bmatrix}\in\mathbb{R}^{JM\times1};\label{eq: grad_rao_bsi}
\end{eqnarray}
where: 
\begin{align}
\bm{g}_{A} & \triangleq\mathrm{vec}(\bm{E}_{r}^{\dagger}\,\bm{R}^{-1}\,\bm{Z}_{d}\,\bm{C}^{\dagger})\in\mathbb{C}^{rM\times1};\label{eq: g_a Rao}\\
\bm{g}_{B} & \triangleq\mathrm{vec}(\bm{E}_{t}^{\dagger}\,\bm{R}^{-1}\,\bm{Z}_{d}\,\bm{C}^{\dagger})\in\mathbb{C}^{tM\times1}.\label{eq: g_b Rao}
\end{align}
In Eqs. (\ref{eq: g_a Rao}) and (\ref{eq: g_b Rao}), we have adopted
the simplified notation $\bm{Z}_{d}\triangleq(\bm{Z}-\bm{A}\,\bm{B}_{s}\bm{C})$.
The results in Eqs. (\ref{eq: grad_rao_bsr}) and (\ref{eq: grad_rao_bsi})
are obtained by exploiting the following steps:
\begin{enumerate}
\item Evaluate the complex derivatives $\frac{\partial\ln f_{1}(\bm{Z};\bm{B}_{s},\bm{R})}{\partial\bm{b}}$
and $\frac{\partial\ln f_{1}(\bm{Z};\bm{B}_{s},\bm{R})}{\partial\bm{b}_{t}}$
(as well as $\frac{\partial\ln f_{1}(\bm{Z};\bm{B}_{s},\bm{R})}{\partial\bm{b}^{*}}$
and $\frac{\partial\ln f_{1}(\bm{Z};\bm{B}_{s},\bm{R})}{\partial\bm{b}_{t}^{*}}$)
by standard complex differentiation rules \cite{Hjorungnes2011}; 
\item Exploit that for any $f(\bm{x})\,:\,\mathbb{C}^{p\times1}\rightarrow\mathbb{R}$,
it holds $\frac{\partial f(\bm{x})}{\partial\Re\{\bm{x}\}}=2\,\Re\{\frac{\partial f(\bm{x})}{\partial\bm{x}^{*}}\}$
and $\frac{\partial f(\bm{x})}{\partial\Im\{\bm{x}\}}=2\Im\{\frac{\partial f(\bm{x})}{\partial\bm{x}^{*}}\}$
(see e.g., \cite{Kay1993});
\item Obtain $\frac{\partial\ln f_{1}(\bm{Z};\bm{B}_{s},\bm{R})}{\partial\bm{b}_{s,R}}$
and $\frac{\partial\ln f_{1}(\bm{Z};\bm{B}_{s},\bm{R})}{\partial\bm{b}_{s,I}}$
as composition of the gradients  obtained at step $2$). 
\end{enumerate}
By identical steps, it can be also proved that:
\begin{equation}
\frac{\partial\ln f_{1}(\bm{Z};\bm{B}_{s},\bm{R})}{\partial\bm{\theta}_{r}}=\frac{\partial\ln f_{1}(\bm{Z};\bm{\theta})}{\partial\bm{\theta}_{r}}=\begin{bmatrix}2\,\Re\{\bm{g}_{A}\}\\
2\,\Im\{\bm{g}_{A}\}
\end{bmatrix}\,,\label{eq: Der-Log Rao}
\end{equation}
which gives the explicit expression for the gradient of the log-likelihood
required for evaluation of Rao statistic (cf. Eq.~(\ref{eq: Rao statistic (implicit form)})).
On the other hand, the block ($\bm{\theta}_{r}$,$\bm{\theta}_{r})$
of the inverse of the FIM is evaluated as follows. First, we notice
that \cite{Kay1993}:
\begin{eqnarray}
\left[\bm{I}^{-1}\left(\bm{\theta}\right)\right]_{\bm{\theta}_{r},\bm{\theta}_{r}} & = & \left[\bm{I}_{a}^{-1}\left(\bm{\theta}\right)\right]_{\bm{\theta}_{r},\bm{\theta}_{r}}
\end{eqnarray}
where $\left[\bm{I}_{a}(\bm{\theta})\right]$ is here used to denote
the block of the FIM comprising only the contributions related to
$(\bm{\theta}_{r},\bm{\theta}_{s,a})$. The aforementioned property
follows from the cross-terms ($\bm{\theta}_{r}$,$\bm{\theta}_{s,b})$
and ($\bm{\theta}_{s,a},\,\bm{\theta}_{s,b})$ being null in the (overall)
FIM $\bm{I}(\bm{\theta})$. Additionally, the following equality holds
(recalling that $\begin{bmatrix}\bm{\theta}_{r}^{T} & \bm{\theta}_{s,a}^{T}\end{bmatrix}=\begin{bmatrix}\bm{b}_{R}^{T} & \bm{b}_{I}^{T} & \bm{b}_{t,R}^{T} & \bm{b}_{t,I}^{T}\end{bmatrix}$
and exploiting Eqs. (\ref{eq: grad_rao_bsr}) and (\ref{eq: grad_rao_bsi})):
\begin{equation}
\frac{\partial\ln f_{1}(\bm{Z};\bm{B}_{s},\bm{R})}{\partial\begin{bmatrix}\bm{\theta}_{r}\\
\bm{\theta}_{s,a}
\end{bmatrix}}=\bm{P}\,\begin{bmatrix}\frac{\partial\ln f_{1}(\bm{Z};\bm{B}_{s},\bm{R})}{\partial\bm{b}_{s,R}}\\
\frac{\partial\ln f_{1}(\bm{Z};\bm{B}_{s},\bm{R})}{\partial\bm{b}_{s,I}}
\end{bmatrix}\,;\label{eq: Rao derivative}
\end{equation}
where $\bm{P}\in\mathbb{R}^{2JM\times2JM}$ is a suitable permutation
matrix\footnote{Recall that every permutation matrix is a special orthogonal matrix,
that is, $\bm{P}^{-1}=\bm{P}^{T}$.}, defined as
\begin{equation}
\bm{P}\triangleq\begin{bmatrix}\bm{I}_{rM} & \bm{0}_{rM\times tM} & \bm{0}_{rM\times rM} & \bm{0}_{rM\times tM}\\
\bm{0}_{rM\times rM} & \bm{0}_{rM\times tM} & \bm{I}_{rM} & \bm{0}_{rM\times tM}\\
\bm{0}_{tM\times rM} & \bm{I}_{tM} & \bm{0}_{tM\times rM} & \bm{0}_{tM\times tM}\\
\bm{0}_{tM\times rM} & \bm{0}_{tM\times tM} & \bm{0}_{tM\times rM} & \bm{I}_{tM}
\end{bmatrix}\,.
\end{equation}
Before proceeding, we define the matrix $\bm{\Omega}\triangleq(\bm{A}^{\dagger}\bm{R}^{-1}\bm{A})$
and the partitioning:
\begin{equation}
\bm{\Omega}=\begin{bmatrix}\bm{\Omega}_{11} & \bm{\Omega}_{12}\\
\bm{\Omega}_{21} & \bm{\Omega}_{22}
\end{bmatrix}=\begin{bmatrix}\bm{E}_{t}^{\dagger}\bm{R}^{-1}\bm{E}_{t} & \bm{E}_{t}^{\dagger}\bm{R}^{-1}\bm{E}_{r}\\
\bm{E}_{r}^{\dagger}\bm{R}^{-1}\bm{E}_{t} & \bm{E}_{r}^{\dagger}\bm{R}^{-1}\bm{E}_{r}
\end{bmatrix},\label{eq: Omega_partitioning}
\end{equation}
where, $\bm{\Omega}_{ij}$, $(i,j)\in\{1,2\}\times\{1,2\}$, is a
sub-matrix whose dimensions can be obtained replacing $1$ and $2$
with $t$ and $r$, respectively. Then, the sub-FIM $\bm{I}_{a}\left(\bm{\theta}\right)$
is obtained starting from Eq. (\ref{eq: Rao derivative}) as $\bm{I}_{a}\left(\bm{\theta}\right)=(\bm{P}\,\bm{\Psi}\,\bm{P}^{T}\,)$,
where $\bm{\Psi}$ has the following special structure:
\begin{align}
\bm{\Psi} & \triangleq\begin{bmatrix}2\,\Re\{\bm{K}\} & -2\,\Im\{\bm{K}\}\\
2\,\Im\{\bm{K}\} & 2\,\Re\{\bm{K}\}
\end{bmatrix}\,,\label{eq: Psi_matrix_Rao}
\end{align}
and the matrix $\bm{K}\in\mathbb{C}^{JM\times JM}$ is defined as:
\begin{gather}
\bm{K}\triangleq\begin{bmatrix}(\bm{C}\bm{C}^{\dagger})^{T}\otimes\bm{\Omega}_{22} & (\bm{C}\bm{C}^{\dagger})^{T}\otimes\bm{\Omega}_{21}\\
(\bm{C}\bm{C}^{\dagger})^{T}\otimes\bm{\Omega}_{12} & (\bm{C}\bm{C}^{\dagger})^{T}\otimes\bm{\Omega}_{11}
\end{bmatrix}\,.
\end{gather}
Finally, the inverse $\bm{I}_{a}^{-1}\left(\bm{\theta}\right)$ is
obtained as $\bm{I}_{a}^{-1}\left(\bm{\theta}\right)=(\bm{P}\,\bm{\Psi}^{-1}\,\bm{P}^{T})$
(as $\bm{P}$ is orthogonal). In the latter case, the inverse matrix
$\bm{\Psi}^{-1}$ has the same structure as $\bm{\Psi}$ (cf. Eq.~(\ref{eq: Psi_matrix_Rao}))
except for $\bm{K}$ and the factor $2$ replaced by $\bm{K}^{-1}$
and $\frac{1}{2}$, respectively\footnote{Such result is obtained by exploiting the equality $\bm{\Psi}^{-1}\bm{\Psi}=\bm{I}$
and the real-imaginary parts decompositions of $\bm{\Psi}=\bm{\Psi}_{R}+j\bm{\Psi}_{I}$
and $\bm{\Psi}^{-1}=\bar{\bm{\Psi}}_{R}+j\bar{\bm{\Psi}}_{I}$, from
which it follows the set of equations ($i$) $(\bm{\Psi}_{R}\bar{\bm{\Psi}}_{R}-\bm{\Psi}_{I}\bar{\bm{\Psi}}_{I})=\bm{I}$
and ($ii$) $(\bm{\Psi}_{I}\bar{\bm{\Psi}}_{R}+\bm{\Psi}_{R}\bar{\bm{\Psi}}_{I})=\bm{0}$. }. By exploiting the following block structure of $\bm{K}^{-1}$
\begin{equation}
\bm{K}^{-1}=\begin{bmatrix}\bm{K}^{11} & \bm{K}^{12}\\
\bm{K}^{21} & \bm{K}^{22}
\end{bmatrix}\,,
\end{equation}
with $\bm{K}^{11}\in\mathbb{C}^{rM\times rM}$, $\bm{K}^{12}\in\mathbb{C}^{rM\times tM}$,
$\bm{K}^{21}\in\mathbb{C}^{tM\times rM}$ and $\bm{K}^{22}\in\mathbb{C}^{tM\times tM}$,
respectively, and the structure of $\bm{P}$, it can be shown that:
\begin{equation}
\left[\bm{I}^{-1}\left(\bm{\theta}\right)\right]_{\bm{\theta}_{r},\bm{\theta}_{r}}=\begin{bmatrix}\frac{1}{2}\Re\{\bm{K}^{11}\} & -\frac{1}{2}\Im\{\bm{K}^{11}\}\\
\frac{1}{2}\Im\{\bm{K}^{11}\} & \frac{1}{2}\Re\{\bm{K}^{11}\}
\end{bmatrix}\,.\label{eq: FIM_inverse sub-block Rao}
\end{equation}
Similarly, it is not difficult to show that $\bm{K}^{11}$ is given
in closed-form as:
\begin{align}
\bm{K}^{11}= & \,\left\{ (\bm{C}\bm{C}^{\dagger})^{T}\otimes\bm{\Omega}_{22}-(\bm{C}\bm{C}^{\dagger})^{T}\otimes\bm{\Omega}_{21}\right.\nonumber \\
 & \left.\times\left[(\bm{C}\bm{C}^{\dagger})^{T}\otimes\bm{\Omega}_{11}\right]^{-1}(\bm{C}\bm{C}^{\dagger})^{T}\otimes\bm{\Omega}_{12}\right\} ^{-1}\\
= & \,(\bm{C}\bm{C}^{\dagger})^{-T}\otimes\bm{\Gamma}_{22}\label{eq: Rao FIM block}
\end{align}
where $\bm{\Gamma}_{ij}$ is a sub-matrix obtained from $\bm{\Gamma}\triangleq\bm{\Omega}^{-1}$
exploiting identical partitioning (in terms of size) as done in Eq.~(\ref{eq: Omega_partitioning})
for $\bm{\Omega}$. The compact expression in Eq. (\ref{eq: Rao FIM block})
is obtained from the use of mixed product and associative properties
of Kronecker operator\footnote{The mixed product property states that $(\bm{V}_{1}\otimes\bm{V}_{2})(\bm{V}_{3}\otimes\bm{V}_{4})=(\bm{V}_{1}\bm{V}_{3})\otimes(\bm{V}_{2}\bm{V}_{4})$,
where $\bm{V}_{i}$ are generic matrices of compatible sizes. Differently,
the associative property states that $\bm{V}_{1}\otimes(\bm{V}_{2}+\bm{V}_{3})=\bm{V}_{1}\otimes\bm{V}_{2}+\bm{V}_{1}\otimes\bm{V}_{3}$. }. Then, combining Eqs.~(\ref{eq: Der-Log Rao}), (\ref{eq: FIM_inverse sub-block Rao})
and (\ref{eq: Rao FIM block}) leads to:
\begin{gather}
\left\{ \frac{\partial\ln f_{1}(\bm{Z};\bm{\theta})}{\partial\bm{\theta}_{r}^{T}}\left[\bm{I}^{-1}\left(\bm{\theta}\right)\right]_{\bm{\theta}_{r},\bm{\theta}_{r}}\frac{\partial\ln f_{1}(\bm{Z};\bm{\theta})}{\partial\bm{\theta}_{r}}\right\} \propto\nonumber \\
\mathrm{vec}(\bm{E}_{r}^{\dagger}\,\bm{R}^{-1}\bm{Z}_{d}\,\bm{C}^{\dagger})^{\dagger}\left[(\bm{C}\bm{C}^{\dagger})^{-T}\otimes\bm{\Gamma}_{22}\right]\nonumber \\
\times\mathrm{vec}(\bm{E}_{r}^{\dagger}\,\bm{R}^{-1}\bm{Z}_{d}\,\bm{C}^{\dagger})\label{eq: Rao hermitian symmetric eq}\\
=\mathrm{Tr}\left[\bm{Z}_{d}^{\dagger}\,\bm{R}^{-1}\bm{E}_{r}\,\bm{\Gamma}_{22}\,\bm{E}_{r}^{\dagger}\bm{R}^{-1}\bm{Z}_{d}\bm{P}_{\bm{C}^{\dagger}}\right]\label{eq: Rao test (pre-substitution)}
\end{gather}
where we have exploited the well-known equivalence between a real-valued
Hermitian quadratic form and its real symmetric quadratic counterpart
in Eq. (\ref{eq: Rao hermitian symmetric eq}) and some standard properties
of $\mathrm{vec}(\cdot)$ operator\footnote{More specifically, we have exploited $\mathrm{vec}(\bm{V}_{1}\bm{V}_{2}\bm{V}_{3})=(\bm{V}_{2}^{T}\otimes\bm{V}_{1})\,\mathrm{vec}(\bm{V}_{3})$
and $\mathrm{vec(}\bm{V}_{1})^{\dagger}\,\mathrm{vec}(\bm{V}_{2})=\mathrm{Tr}[\bm{V}_{1}^{\dagger}\bm{V}_{2}]$,
with $\bm{V}_{i}$ being generic matrices.} in obtaining Eq. (\ref{eq: Rao test (pre-substitution)}). Finally,
the substitution $\bm{\theta}=\widehat{\bm{\theta}}_{0}$ provides:
\begin{equation}
\mathrm{Tr}\left[\bm{Z}_{d,0}^{\dagger}\,\widehat{\bm{R}}_{0}^{-1}\,\bm{E}_{r}\,\widehat{\bm{\Gamma}}_{22}^{\circ}\,\bm{E}_{r}^{\dagger}\,\widehat{\bm{R}}_{0}^{-1}\,\bm{Z}_{d,0}\,\bm{P}_{\bm{C}^{\dagger}}\right]\,,\label{eq: Rao_final form (implicit)- appendix}
\end{equation}
where $\widehat{\bm{R}}_{0}$ and $\bm{Z}_{d,0}$ are defined in Eqs.
(\ref{eq: Closed form MLE covariance H_0}) and (\ref{eq: Zd0 (Rao test)}),
respectively. Similarly, $\widehat{\bm{\Gamma}}_{ij}^{\circ}$ denotes
a sub-matrix obtained from $\widehat{\bm{\Gamma}}_{ij}^{\circ}=(\bm{A}^{\dagger}\widehat{\bm{R}}_{0}^{-1}\bm{A})^{-1}$
by exploiting identical partitioning (in terms of size) as done in
Eq.~(\ref{eq: Omega_partitioning}). This provides the explicit expression
for the Rao statistic.

\subsection*{Proof of Eq. (\ref{eq: R1})}

The mentioned result is proved as:
\begin{gather}
\widehat{\bm{R}}_{0}^{-1/2}\bm{Z}_{d,0}\,\bm{P}_{\bm{C}^{\dagger}}\nonumber \\
=\widehat{\bm{R}}_{0}^{-1/2}\left(\bm{Z}-\bm{S}_{c}^{1/2}\,\bm{P}_{\bm{A}_{0}}\,\bm{S}_{c}^{-1/2}\,\bm{Z}\,\bm{P}_{\bm{C}^{\dagger}}\right)\,\bm{P}_{\bm{C}^{\dagger}}\\
=\widehat{\bm{R}}_{0}^{-1/2}\left(\bm{I}_{N}-\bm{S}_{c}^{1/2}\,\bm{P}_{\bm{A}_{0}}\,\bm{S}_{c}^{-1/2}\right)\bm{Z}\,\bm{P}_{\bm{C}^{\dagger}}
\end{gather}
where we have exploited Eq. (\ref{eq: Zd0 (Rao test)}) and $\bm{P}_{\bm{C}^{\dagger}}\bm{P}_{\bm{C}^{\dagger}}=\bm{P}_{\bm{C}^{\dagger}}$,
respectively. The above expression can be further rewritten by exploiting
Eq. (\ref{eq: Lem1 R0}) of Lem. \ref{lem: ML covariance estimates properties}
as:
\begin{gather}
\widehat{\bm{R}}_{0}^{-1/2}\bm{Z}_{d,0}\,\bm{P}_{\bm{C}^{\dagger}}\nonumber \\
=(\widehat{\bm{R}}_{0}^{-1/2}-\widehat{\bm{R}}_{0}^{-1/2}\bm{E}_{t}(\bm{E}_{t}^{\dagger}\bm{S}_{c}^{-1}\bm{E}_{t})^{-1}\bm{E}_{t}^{\dagger}\,\bm{S}_{c}^{-1})\bm{Z}\,\bm{P}_{\bm{C}^{\dagger}}\\
=(\widehat{\bm{R}}_{0}^{-1/2}-\bm{P}_{\bar{\bm{A}}_{0}}\,\widehat{\bm{R}}_{0}^{-1/2})\bm{Z}\,\bm{P}_{\bm{C}^{\dagger}}=\bm{P}_{\bar{\bm{A}}_{0}}^{\perp}\widehat{\bm{R}}_{0}^{-1/2}\bm{Z}\,\bm{P}_{\bm{C}^{\dagger}}
\end{gather}
where we have denoted $\bar{\bm{A}}_{0}\triangleq\widehat{\bm{R}}_{0}^{-1/2}\,\bm{E}_{t}$,
which finally provides Eq. (\ref{eq: R1}) (after having defined $\bm{Z}_{{\scriptscriptstyle W0}}\triangleq\widehat{\bm{R}}_{0}^{-1/2}\bm{Z}$).

\subsection*{Proof of Eq. (\ref{eq: R2})}

First, it can be shown that: 
\begin{equation}
\widehat{\bm{\Gamma}}_{22}^{\circ}\,\bm{E}_{r}^{\dagger}\,\widehat{\bm{R}}_{0}^{-1/2}\,\bm{P}_{\bar{\bm{A}}_{0}}^{\perp}=(\widehat{\bm{\Gamma}}_{21}^{\circ}\,\bm{E}_{t}^{\dagger}+\widehat{\bm{\Gamma}}_{22}^{\circ}\,\bm{E}_{r}^{\dagger})\,\widehat{\bm{R}}_{0}^{-1/2},\label{eq: Rao appendix R.2 eq1}
\end{equation}
Therefore, in view of Eq. (\ref{eq: Rao appendix R.2 eq1}), it holds:
\begin{gather}
\bm{P}_{\bar{\bm{A}}_{0}}^{\perp}\,\widehat{\bm{R}}_{0}^{-1/2}\,\bm{E}_{r}\,\widehat{\bm{\Gamma}}_{22}^{\circ}\,\bm{E}_{r}^{\dagger}\,\widehat{\bm{R}}_{0}^{-1/2}\,\bm{P}_{\bar{\bm{A}}_{0}}^{\perp}\\
=\left[(\widehat{\bm{R}}_{0}^{-1/2}\bm{E}_{t})(\widehat{\bm{\Gamma}}_{21}^{\circ})^{\dagger}+(\widehat{\bm{R}}_{0}^{-1/2}\bm{E}_{r})\,(\widehat{\bm{\Gamma}}_{22}^{\circ})^{\dagger}\right]\,(\widehat{\bm{\Gamma}}_{22}^{\circ})^{-1}\nonumber \\
\,\times\left[\widehat{\bm{\Gamma}}_{21}^{\circ}(\bm{E}_{t}^{\dagger}\widehat{\bm{R}}_{0}^{-1/2})+\widehat{\bm{\Gamma}}_{22}^{\circ}(\bm{E}_{r}^{\dagger}\widehat{\bm{R}}_{0}^{-1/2})\right]\\
=\widehat{\bm{R}}_{0}^{-1/2}\bm{A}(\bm{A}^{\dagger}\widehat{\bm{R}}_{0}^{-1}\bm{A})^{-1}\bm{A}^{\dagger}\widehat{\bm{R}}_{0}^{-1/2}\nonumber \\
-(\widehat{\bm{R}}_{0}^{-1/2}\bm{E}_{t})(\bm{E}_{t}^{\dagger}\widehat{\bm{R}}_{0}^{-1}\bm{E}_{t})^{-1}(\bm{E}_{t}^{\dagger}\widehat{\bm{R}}_{0}^{-1/2})\\
=\left(\bm{P}_{\bar{\bm{A}}_{1}}-\bm{P}_{\bar{\bm{A}}_{0}}\right)\label{eq: final R2}
\end{gather}
where we have exploited the equality $(\bm{E}_{t}^{\dagger}\widehat{\bm{R}}_{0}^{-1}\bm{E}_{t})^{-1}=\widehat{\bm{\Gamma}}_{11}^{\circ}-\widehat{\bm{\Gamma}}_{12}^{\circ}(\widehat{\bm{\Gamma}}_{22}^{\circ})^{-1}\widehat{\bm{\Gamma}}_{21}^{\circ}$
(which can be deduced from Eq. (\ref{eq: Omega_partitioning}) after
substitution $\bm{R}=\widehat{\bm{R}}_{0}$ and from $\widehat{\bm{\Gamma}}^{\circ}$
definition). Finally, in Eq.~(\ref{eq: final R2}) we have further
defined $\bar{\bm{A}}_{1}\triangleq(\widehat{\bm{R}}_{0}^{-1/2}\bm{A})$.

\section{Derivation of Wald statistic \label{sec: Appendix_ Wald test}}

In this section we report the derivation for Wald statistic in Eq.
(\ref{eq: Wald_final form}). In order to prove the aforementioned
result, we build upon the explicit expression of $\left[\bm{I}^{-1}\left(\bm{\theta}\right)\right]_{\bm{\theta}_{r},\bm{\theta}_{r}}$
obtained in Eq. (\ref{eq: FIM_inverse sub-block Rao}). Such result
allows to readily evaluate $\{[\bm{I}^{-1}(\widehat{\bm{\theta}}_{1})]_{\bm{\theta}_{r},\bm{\theta}_{r}}\}^{-1}$
in Eq. (\ref{eq: Wald_generic}) by ($i$) substitution $\bm{R}=\widehat{\bm{R}}_{1}$
and ($ii$) matrix inversion\footnote{We again use the property that the inverse of a block-symmetric matrix
in the form of Eq. (\ref{eq: FIM_inverse sub-block Rao}) gives rise
to a similar structure for its inverse, as exploited for the derivation
of Rao statistic.} as:
\begin{gather}
\{[\bm{I}^{-1}(\widehat{\bm{\theta}}_{1})]_{\bm{\theta}_{r},\bm{\theta}_{r}}\}^{-1}=\begin{bmatrix}2\Re\{(\bm{K}_{1}^{11})^{-1}\} & -2\Im\{(\bm{K}_{1}^{11})^{-1}\}\\
2\Im\{(\bm{K}_{1}^{11})^{-1}\} & 2\Re\{(\bm{K}_{1}^{11})^{-1}\}
\end{bmatrix}\,,
\end{gather}
where $(\bm{K}_{1}^{11})^{-1}\triangleq(\bm{C}\bm{C}^{\dagger})^{T}\otimes(\widehat{\bm{\Gamma}}_{22}^{1})^{-1}$,
and $\widehat{\bm{\Gamma}}_{ij}^{1}$ is a sub-matrix obtained from
$\widehat{\bm{\Gamma}}^{1}\triangleq(\bm{A}^{\dagger}\hat{\bm{R}}_{1}^{-1}\bm{A})^{-1}$
exploiting identical partitioning (in terms of size) as done in Eq.~(\ref{eq: Omega_partitioning})
for $\bm{\Omega}$. 

We have now to evaluate $\bm{\theta}_{r,0}$ and $\hat{\bm{\theta}}_{r,1}$,
respectively. First, we recall that $\bm{\theta}_{r,0}=\bm{0}_{2rM}$,
while $\hat{\bm{\theta}}_{r,1}=\begin{bmatrix}\Re\{\mathrm{vec}(\widehat{\bm{B}})\}^{T} & \Im\{\mathrm{vec}(\widehat{\bm{B}})\}^{T}\end{bmatrix}^{T}$,
with $\widehat{\bm{B}}$ representing the ML estimate of the complex-valued
signal matrix under $\mathcal{H}_{1}$. This estimate can be shown
to be equal to:
\begin{equation}
\widehat{\bm{B}}=K\,\widehat{\bm{\Gamma}}_{22}^{1}\,\bm{E}_{r}^{\dagger}\,\bm{S}_{c}^{-1/2}\bm{P}_{\bm{A}_{0}}^{\perp}\bm{S}_{c}^{-1/2}\,\bm{Z}\,\bm{C}^{\dagger}\,(\bm{C}\bm{C}^{\dagger})^{-1}\,.\label{eq: B_hat_appendix}
\end{equation}
The above result is obtained starting from Eq. (\ref{eq: ML estimate B_s under H1})
and observing that $\widehat{\bm{B}}_{s}=\begin{bmatrix}\widehat{\bm{B}}_{t,1}^{T} & \widehat{\bm{B}}^{T}\end{bmatrix}^{T}$.
Therefore, collecting the above results, Wald statistic is obtained
as 
\begin{gather}
(\hat{\bm{\theta}}_{r,1}-\bm{\theta}_{r,0})^{T}\{[\bm{I}^{-1}(\widehat{\bm{\theta}}_{1})]_{\bm{\theta}_{r},\bm{\theta}_{r}}\}^{-1}\,(\hat{\bm{\theta}}_{r,1}-\bm{\theta}_{r,0})\nonumber \\
=2\,\mathrm{vec}(\widehat{\bm{B}}){}^{\dagger}\,(\bm{K}_{1}^{11})^{-1}\,\mathrm{vec}(\widehat{\bm{B}})\label{eq: hermitian quadratic Wald appendix}\\
\propto\mathrm{vec}\left(\widehat{\bm{\Gamma}}_{22}^{1}\,\bm{E}_{r}^{\dagger}\,\bm{S}_{c}^{-1/2}\bm{P}_{\bm{A}_{0}}^{\perp}\bm{S}_{c}^{-1/2}\,\bm{Z}\,\bm{C}^{\dagger}(\bm{C}\bm{C}^{\dagger})^{-1}\right)^{\dagger}\nonumber \\
\times K\,\left\{ (\bm{C}\bm{C}^{\dagger})^{T}\otimes(\widehat{\bm{\Gamma}}_{22}^{1})^{-1}\right\} \nonumber \\
\times\mathrm{vec}\left(\widehat{\bm{\Gamma}}_{22}^{1}\,\bm{E}_{r}^{\dagger}\,\bm{S}_{c}^{-1/2}\bm{P}_{\bm{A}_{0}}^{\perp}\bm{S}_{c}^{-1/2}\,\bm{Z}\,\bm{C}^{\dagger}(\bm{C}\bm{C}^{\dagger})^{-1}\right)\\
=\mathrm{Tr}\left[\bm{Z}_{{\scriptscriptstyle W1}}^{\dagger}(K\,\bm{P}_{\bm{A}_{0}}^{\perp}\bm{S}_{c}^{-1/2}\bm{E}_{r}\,\widehat{\bm{\Gamma}}_{22}^{1}\,\bm{E}_{r}^{\dagger}\bm{S}_{c}^{-1/2}\bm{P}_{\bm{A}_{0}}^{\perp})\bm{Z}_{{\scriptscriptstyle W1}}\bm{P}_{\bm{C}^{\dagger}}\right]\label{eq: Wald_final form-appendix}
\end{gather}
where we have again exploited (as in the derivation of Rao statistic)
the well-known equivalence between a real-valued Hermitian quadratic
form and its real symmetric quadratic counterpart in Eq.~(\ref{eq: hermitian quadratic Wald appendix})
and some standard properties of $\mathrm{vec}(\cdot)$ operator in
obtaining Eq. (\ref{eq: Wald_final form-appendix}). This provides
the closed form expression for Wald statistic.

\subsection*{Proof of Eq. (\ref{eq: W1 - Wald statistic})}

We first notice that the following equality holds: 
\begin{equation}
\widehat{\bm{\Gamma}}_{22}^{1}\,\bm{E}_{r}^{\dagger}\,\bm{S}_{c}^{-1/2}\,\bm{P}_{\bm{A}_{0}}^{\perp}=(\widehat{\bm{\Gamma}}_{21}^{1}\,\bm{E}_{t}^{\dagger}+\widehat{\bm{\Gamma}}_{22}^{1}\,\bm{E}_{r}^{\dagger})\,\bm{S}_{c}^{-1/2}\,.
\end{equation}
The above result is an almost evident consequence of the application
of matrix inversion formula for a $2\times2$ block matrix. Therefore,
in view of the above equality, we observe that:
\begin{gather}
K\,\bm{P}_{\bm{A}_{0}}^{\perp}\bm{S}_{c}^{-1/2}\,\bm{E}_{r}\,\widehat{\bm{\Gamma}}_{22}^{1}\,\bm{E}_{r}^{\dagger}\bm{S}_{c}^{-1/2}\bm{P}_{\bm{A}_{0}}^{\perp}\nonumber \\
=K\left[(\bm{S}_{c}^{-1/2}\bm{E}_{t})(\widehat{\bm{\Gamma}}_{21}^{1})^{\dagger}+(\bm{S}_{c}^{-1/2}\bm{E}_{r})\,(\widehat{\bm{\Gamma}}_{22}^{1})^{\dagger}\right]\,(\widehat{\bm{\Gamma}}_{22}^{1})^{-1}\,\nonumber \\
\times\left[\widehat{\bm{\Gamma}}_{21}^{1}(\bm{E}_{t}^{\dagger}\bm{S}_{c}^{-1/2})+\widehat{\bm{\Gamma}}_{22}^{1}(\bm{E}_{r}^{\dagger}\bm{S}_{c}^{-1/2})\right]\\
=(\bm{P}_{\bm{A}_{1}}-\bm{P}_{\bm{A}_{0}})=\mathcal{\bm{P}}_{\Delta}
\end{gather}
which thus provides Eq. (\ref{eq: W1 - Wald statistic}).

\section{Derivation of Gradient statistic \label{sec: Appendix Gradient statistic}}

The derivation of Gradient statistic is readily obtained from intermediate
results obtained in derivation of Rao and Wald statistics, in Secs.
\ref{sec: Appendix _ Rao statistic derivation}  and \ref{sec: Appendix_ Wald test},
respectively. Indeed, exploiting Eqs. (\ref{eq: Der-Log Rao}) and
(\ref{eq: B_hat_appendix})  provides:
\begin{gather}
\left.\frac{\partial\ln f_{1}(\bm{Z};\bm{\theta})}{\partial\bm{\theta}_{r}^{T}}\right|_{\bm{\theta}=\hat{\bm{\theta}}_{0}}(\hat{\bm{\theta}}_{r,1}-\bm{\theta}_{r,0})=2\,\Re\{\mathrm{vec}(\widehat{\bm{B}})^{\dagger}\bm{g}_{A}^{\circ}\}\label{eq: real-equivalent Gradient statistic Appendix}\\
\propto K\Re\left\{ \mathrm{Tr}\left[\left(\widehat{\bm{\Gamma}}_{22}^{1}\,\bm{E}_{r}^{\dagger}\,\bm{S}_{c}^{-1/2}\,\bm{P}_{\bm{A}_{0}}^{\perp}\,\bm{S}_{c}^{-1/2}\bm{Z}\,\bm{C}^{\dagger}(\bm{C}\bm{C}^{\dagger})^{-1}\right)^{\dagger}\right.\right.\nonumber \\
\left.\left.\times\bm{E}_{r}^{\dagger}\widehat{\bm{R}}_{0}^{-1}\,\bm{Z}_{d,0}\,\bm{C}^{\dagger}\right]\right\} \label{eq: vec_Gradient test}\\
=\Re\left\{ \mathrm{Tr}\left[\bm{Z}_{{\scriptscriptstyle W1}}^{\dagger}K\bm{P}_{\bm{A}_{0}}^{\perp}\bm{S}_{c}^{-1/2}\bm{E}_{r}\,\widehat{\bm{\Gamma}}_{22}^{1}\,\bm{E}_{r}^{\dagger}\widehat{\bm{R}}_{0}^{-1}\,\bm{Z}_{d,0}\,\bm{P}_{\bm{C}^{\dagger}}\right]\right\} \label{eq: GT (final form)-appendix}
\end{gather}
In Eq. (\ref{eq: real-equivalent Gradient statistic Appendix}) we
have exploited the equivalence between the real part of an inner product
in the complex domain and its real-valued equivalent counterpart\footnote{More specifically, given two complex vectors with real/imaginary parts
decomposition $\bm{v}_{1}=\bm{v}_{1,R}+j\bm{v}_{1,I}$ an $\bm{v}_{2}=\bm{v}_{2,R}+j\bm{v}_{2,I}$,
it holds $\Re\{\bm{v}_{1}^{\dagger}\bm{v}_{2}\}=\Re\{\bm{v}_{2}^{\dagger}\bm{v}_{1}\}=\bm{v}_{1,E}^{T}\bm{v}_{2,E}$,
where $\bm{v}_{i,E}\triangleq\begin{bmatrix}\bm{v}_{i,R}^{T} & \bm{v}_{i,I}^{T}\end{bmatrix}^{T}$.}, along with the definition $\bm{g}_{A}^{\circ}\triangleq\mathrm{vec}(\bm{E}_{r}^{\dagger}\,\widehat{\bm{R}}_{0}^{-1}\,\bm{Z}_{d,0}\,\bm{C}^{\dagger})$.
Furthermore, in obtaining Eq.~(\ref{eq: vec_Gradient test}), we
have exploited standard properties of $\mathrm{vec}(\cdot)$ operator.
Finally, we recall that $\bm{Z}_{d,0}$ is given in Eq.~(\ref{eq: Zd0 (Rao test)}).
This concludes the derivation of Gradient statistic.

\subsection*{Proof of Eq. (\ref{eq: G.1})}

We start by observing that
\begin{equation}
\widehat{\bm{R}}_{0}^{-1}\,\bm{Z}_{d,0}\,\bm{P}_{\bm{C}^{\dagger}}=\widehat{\bm{R}}_{0}^{-1/2}\bm{P}_{\bar{\bm{A}}_{0}}^{\perp}\widehat{\bm{R}}_{0}^{-1/2}\bm{Z}\,\bm{P}_{\bm{C}^{\dagger}},\label{eq: G.1 - first step}
\end{equation}
which readily follows from application of Eq. (\ref{eq: R1}). Then,
we rewrite the matrix $\widehat{\bm{R}}_{0}^{-1/2}\bm{P}_{\bar{\bm{A}}_{0}}^{\perp}\widehat{\bm{R}}_{0}^{-1/2}$
as:
\begin{gather}
\widehat{\bm{R}}_{0}^{-1/2}\bm{P}_{\bar{\bm{A}}_{0}}^{\perp}\widehat{\bm{R}}_{0}^{-1/2}\nonumber \\
=\bm{S}_{c}^{-1/2}\,\left[\bm{S}_{c}^{1/2}\widehat{\bm{R}}_{0}^{-1}-\bm{S}_{c}^{1/2}\widehat{\bm{R}}_{0}^{-1/2}\bm{P}_{\bar{\bm{A}}_{0}}\widehat{\bm{R}}_{0}^{-1/2}\right]\\
=\bm{S}_{c}^{-1/2}\,\left[\bm{I}_{N}-\bm{S}_{c}^{1/2}(\widehat{\bm{R}}_{0}^{-1}\bm{E}_{t})\right.\nonumber \\
\left.\times(\bm{E}_{t}^{\dagger}\widehat{\bm{R}}_{0}^{-1}\bm{E}_{t})^{-1}\bm{E}_{t}^{\dagger}\bm{S}_{c}^{-1/2}\right]\bm{S}_{c}^{1/2}\widehat{\bm{R}}_{0}^{-1}\\
=\bm{S}_{c}^{-1/2}\,\bm{P}_{\bm{A}_{0}}^{\perp}\,\bm{S}_{c}^{1/2}\,\widehat{\bm{R}}_{0}^{-1}\label{eq: G.1 step 2}
\end{gather}
where in Eq. (\ref{eq: G.1 step 2}) we have exploited Lem. \ref{lem: ML covariance estimates properties},
Eq. (\ref{eq: Lem1 R0}). Finally, from straightforward combination
of the results in Eqs. (\ref{eq: G.1 - first step}) and (\ref{eq: G.1 step 2}),
the final result follows.

\section{Equivalence between Rao and Durbin statistics (Theorem \ref{thm: Durbin-Rao equivalence})\label{sec: Appendix_ Durbin statistic}}

In this section we prove statistical equivalence between Rao and Durbin
statistics by explicitly deriving the closed form expression of Durbin
statistic (implicitly expressed in Eq. (\ref{eq: Durbin test (generic form)})).
To this end, analogously as for the Rao and Wald statistics, it can
be shown that:
\begin{eqnarray}
\left[\bm{I}^{-1}\left(\widehat{\bm{\theta}}_{0}\right)\right]_{\bm{\theta}_{r},\bm{\theta}_{r}} & = & \begin{bmatrix}\frac{1}{2}\,\Re\{\bm{T}_{0}\} & -\frac{1}{2}\,\Im\{\bm{T}_{0}\}\\
\frac{1}{2}\,\Im\{\bm{T}_{0}\} & \frac{1}{2}\,\Re\{\bm{T}_{0}\}
\end{bmatrix}\label{eq: Durbin FIM-interference-1}\\
\left[\bm{I}\left(\widehat{\bm{\theta}}_{0}\right)\right]_{\bm{\theta}_{r},\bm{\theta}_{r}} & = & \begin{bmatrix}2\,\Re\{\bar{\bm{T}}_{0}\} & -2\,\Im\{\bar{\bm{T}}_{0}\}\\
2\,\Im\{\bar{\bm{T}}_{0}\} & 2\,\Re\{\bar{\bm{T}}_{0}\}
\end{bmatrix}\label{eq: Durbin FIM interference 2-1}
\end{eqnarray}
where: 
\begin{align}
\bm{T}_{0} & \triangleq(\bm{C}\bm{C}^{\dagger})^{-T}\otimes\widehat{\bm{\Gamma}}_{22}^{\circ}\,;\\
\bar{\bm{T}}_{0} & \triangleq(\bm{C}\bm{C}^{\dagger})^{T}\otimes(\bm{E}_{r}^{\dagger}\,\hat{\bm{R}}_{0}^{-1}\,\bm{E}_{r})\,.
\end{align}
The result in Eq. (\ref{eq: Durbin FIM-interference-1}) is obtained
starting from the explicit expression of $\left[\bm{I}^{-1}\left(\bm{\theta}\right)\right]_{\bm{\theta}_{r},\bm{\theta}_{r}}$
in Eq. (\ref{eq: FIM_inverse sub-block Rao}) and plugging back $\bm{R}=\widehat{\bm{R}}_{0}$.
Differently, the estimate $\hat{\bm{\theta}}_{r,01}$ can be obtained
following the steps described next. Without loss of generality, we
consider maximization of $\ln(\cdot)$ of the objective in Eq.~(\ref{eq: Durbin_ML estimate-1-1})
and evaluate the following estimate:
\begin{gather}
\widehat{\bm{B}}_{0}=\label{eq: alternative_Durbin_max_interference-1}\\
\arg\min_{\bm{B}}\mathrm{Tr}\left[\left(\bm{Z}-\bm{A}\begin{bmatrix}\widehat{\bm{B}}_{t,0}\\
\bm{B}
\end{bmatrix}\bm{C}\right)^{\dagger}\hat{\bm{R}}_{0}^{-1}\,\left(\bm{Z}-\bm{A}\begin{bmatrix}\widehat{\bm{B}}_{t,0}\\
\bm{B}
\end{bmatrix}\bm{C}\right)\right]\nonumber 
\end{gather}
Once we have obtained $\widehat{\bm{B}}_{0}$, $\hat{\bm{\theta}}_{r,01}$
is evaluated through the simple operation $\hat{\bm{\theta}}_{r,01}=\begin{bmatrix}\Re\{\mathrm{vec}(\widehat{\bm{B}}_{0})\}^{T} & \Im\{\mathrm{vec}(\widehat{\bm{B}}_{0})\}^{T}\end{bmatrix}^{T}$.
Thus, it is not difficult to show that the solution to the optimization
problem in Eq. (\ref{eq: alternative_Durbin_max_interference-1})
is given in closed-form as: 
\begin{equation}
\widehat{\bm{B}}_{0}=\left(\bm{E}_{r}^{\dagger}\widehat{\bm{R}}_{0}^{-1}\,\bm{E}_{r}\right)^{-1}\bm{E}_{r}^{\dagger}\,\widehat{\bm{R}}_{0}^{-1}\,\bm{Z}_{d,0}\,\bm{C}^{\dagger}(\bm{C}\bm{C}^{\dagger})^{-1}\label{eq: Durbin closed form estimate-interference-1}
\end{equation}
Substituting Eqs. (\ref{eq: Durbin FIM-interference-1}), (\ref{eq: Durbin FIM interference 2-1})
and (\ref{eq: Durbin closed form estimate-interference-1}) into Eq.
(\ref{eq: Durbin test (generic form)}), provides\footnote{We have exploited the fact that $\bar{\bm{T}}_{0}\,\bm{T}_{0}\,\bar{\bm{T}}_{0}$
is Hermitian and that the product of block-symmetric real counterparts
of Hermitian matrices can be expressed as an equivalent block-symmetric
real counterpart with the component matrix being given by the product
of the aforementioned matrices. Finally, we have used the equivalence
between an Hermitian quadratic form and its real block-symmetric counterpart.}:
\begin{gather}
2\,\mathrm{vec}(\widehat{\bm{B}}_{0})^{\dagger}\,(\bar{\bm{T}}_{0}\,\bm{T}_{0}\,\bar{\bm{T}}_{0})\,\mathrm{vec}(\widehat{\bm{B}}_{0})\nonumber \\
\propto\mathrm{vec}\left[\left(\bm{E}_{r}^{\dagger}\widehat{\bm{R}}_{0}^{-1}\,\bm{E}_{r}\right)^{-1}\bm{E}_{r}^{\dagger}\,\widehat{\bm{R}}_{0}^{-1}\,\bm{Z}_{d,0}\bm{C}^{\dagger}(\bm{C}\bm{C}^{\dagger})^{-1}\right]^{\dagger}\nonumber \\
\times\left\{ (\bm{C}\bm{C}^{\dagger})^{T}\otimes\left[\left(\bm{E}_{r}^{\dagger}\widehat{\bm{R}}_{0}^{-1}\,\bm{E}_{r}\right)\widehat{\bm{\Gamma}}_{22}^{\circ}\left(\bm{E}_{r}^{\dagger}\widehat{\bm{R}}_{0}^{-1}\,\bm{E}_{r}\right)\right]\right\} \nonumber \\
\times\mathrm{vec}\left[\left(\bm{E}_{r}^{\dagger}\widehat{\bm{R}}_{0}^{-1}\,\bm{E}_{r}\right)^{-1}\,\bm{E}_{r}^{\dagger}\,\widehat{\bm{R}}_{0}^{-1}\,\bm{Z}_{d,0}\,\bm{C}^{\dagger}(\bm{C}\bm{C}^{\dagger})^{-1}\right]\\
=\mathrm{Tr}[\bm{Z}_{d,0}^{\dagger}\,\widehat{\bm{R}}_{0}^{-1}\,\bm{E}_{r}\,\widehat{\bm{\Gamma}}_{22}^{\circ}\,\bm{E}_{r}^{\dagger}\,\widehat{\bm{R}}_{0}^{-1}\,\bm{Z}_{d,0}\,\bm{P}_{\bm{C}^{\dagger}}]\label{eq: Durbin test (final form)}
\end{gather}
where we have exploited standard properties of $\mathrm{vec}(\cdot)$
in Eq.~(\ref{eq: Durbin test (final form)}).

Clearly, the last result\emph{ coincides with the Rao Test for the
 I-GMANOVA model}, as apparent from comparison with Rao statistic
reported in Eq.~(\ref{eq: Rao_final form (implicit)}) in the manuscript.

\section{Proofs of useful equalities\label{sec: Appendix _ Useful equalities}}

\subsection*{Proof of Eqs. (\ref{eq: MIS_E1}) and (\ref{eq: MIS_E2})}

Hereinafter we provide the proof of Eqs. (\ref{eq: MIS_E1}) and (\ref{eq: MIS_E2}),
which are fundamental for proving CFARness of GLR, Wald and LH statistics.
We start by observing that: 
\begin{gather}
(\bm{Z}_{{\scriptscriptstyle W1}}\bm{V}_{c,1})^{\dagger}\,\bm{P}_{\bm{A}_{0}}^{\perp}(\bm{Z}_{{\scriptscriptstyle W1}}\bm{V}_{c,1})\nonumber \\
=\bm{Z}_{c}^{\dagger}\left\{ \bm{S}_{c}^{-1}-(\bm{S}_{c}^{-1}\bm{E}_{t})(\bm{E}_{t}^{\dagger}\bm{S}_{c}^{-1}\bm{E}_{t})^{-1}\bm{E}_{t}^{\dagger}\,\bm{S}_{c}^{-1}\right\} \bm{Z}_{c}\label{eq: MIS_GLRT eq1 proof}
\end{gather}
Analogously, we write:
\begin{gather}
(\bm{Z}_{{\scriptscriptstyle W1}}\bm{V}_{c,1})^{\dagger}\,\bm{P}_{\bm{A}_{1}}^{\perp}(\bm{Z}_{{\scriptscriptstyle W1}}\bm{V}_{c,1})\nonumber \\
=\bm{Z}_{c}^{\dagger}\left\{ \bm{S}_{c}^{-1}-(\bm{S}_{c}^{-1}\bm{A})(\bm{A}^{\dagger}\bm{S}_{c}^{-1}\bm{A})^{-1}\bm{A}^{\dagger}\,\bm{S}_{c}^{-1}\right\} \bm{Z}_{c}\label{eq: MIS_GLRT eq2 proof}
\end{gather}
Before proceeding further, we define the following partitioning for
matrix $\bm{S}_{c}^{-1}$ as:
\begin{equation}
\bm{S}_{c}^{-1}=\begin{bmatrix}\bm{S}^{11} & \bm{S}^{12} & \bm{S}^{13}\\
\bm{S}^{21} & \bm{S}^{22} & \bm{S}^{23}\\
\bm{S}^{31} & \bm{S}^{32} & \bm{S}^{33}
\end{bmatrix}\,.
\end{equation}
Furthermore, $\bm{S}^{ij}$, $(i,j)\in\{1,2,3\}\times\{1,2,3\}$,
is a sub-matrix whose dimensions can be obtained replacing $1$, $2$
and $3$ with $t$, $r$ and $(N-J)$, respectively. First, exploiting
$\bm{E}_{t}$ structure leads to:
\begin{gather}
(\bm{E}_{t}^{\dagger}\bm{S}_{c}^{-1}\bm{E}_{t})=\bm{S}^{11},\quad\quad\;(\bm{S}_{c}^{-1}\bm{E}_{t})=\begin{bmatrix}\bm{S}^{11}\\
\bm{S}^{21}\\
\bm{S}^{31}
\end{bmatrix}\,.
\end{gather}
Accordingly, the  matrix within curly brackets in Eq. (\ref{eq: MIS_GLRT eq1 proof})
can be rewritten as:
\begin{gather}
\left\{ \bm{S}_{c}^{-1}-(\bm{S}_{c}^{-1}\bm{E}_{t})(\bm{E}_{t}^{\dagger}\bm{S}_{c}^{-1}\bm{E}_{t})^{-1}\bm{E}_{t}^{\dagger}\,\bm{S}_{c}^{-1}\right\} =\nonumber \\
\begin{bmatrix}\bm{0}_{t\times t} & \bm{0}_{t\times(N-t)}\\
\bm{0}_{(N-t)\times t} & \bm{S}_{2}^{-1}
\end{bmatrix}\,,
\end{gather}
where we have denoted:
\begin{equation}
\bm{S}_{2}\triangleq\begin{bmatrix}\bm{S}_{22} & \bm{S}_{23}\\
\bm{S}_{32} & \bm{S}_{33}
\end{bmatrix}\,.
\end{equation}
Secondly, after substitution in Eq. (\ref{eq: MIS_GLRT eq1 proof}),
we obtain:
\begin{gather}
(\bm{Z}_{{\scriptscriptstyle W1}}\bm{V}_{c,1})^{\dagger}\,\bm{P}_{\bm{A}_{0}}^{\perp}(\bm{Z}_{{\scriptscriptstyle W1}}\bm{V}_{c,1})=\bm{Z}_{23}^{\dagger}\,\bm{S}_{2}^{-1}\,\bm{Z}_{23}\,;\label{eq: prefinal}
\end{gather}
where $\bm{Z}_{23}\triangleq\begin{bmatrix}\bm{Z}_{2}^{T} & \bm{Z}_{3}^{T}\end{bmatrix}^{T}$.
Finally, by exploiting the block inverse expression of $\bm{S}_{2}$
in Eq. (\ref{eq: prefinal}), it follows that
\begin{gather}
(\bm{Z}_{{\scriptscriptstyle W1}}\bm{V}_{c,1})^{\dagger}\,\bm{P}_{\bm{A}_{0}}^{\perp}(\bm{Z}_{{\scriptscriptstyle W1}}\bm{V}_{c,1})=\bm{Z}_{2.3}^{\dagger}\,\bm{S}_{2.3}^{-1}\,\bm{Z}_{2.3}+\bm{Z}_{3}^{\dagger}\,\bm{S}_{33}^{-1}\,\bm{Z}_{3}\,,\label{eq: Appendix - Equality (1)}
\end{gather}
which proves Eq. (\ref{eq: MIS_E1}). Similarly, exploiting $\bm{A}$
structure, it can be shown that:
\begin{gather}
\bm{A}^{\dagger}\bm{S}_{c}^{-1}\bm{A}=\begin{bmatrix}\bm{S}^{11} & \bm{S}^{12}\\
\bm{S}^{21} & \bm{S}^{22}
\end{bmatrix};\quad\bm{S}_{c}^{-1}\bm{A}=\begin{bmatrix}\bm{S}^{11} & \bm{S}^{12}\\
\bm{S}^{21} & \bm{S}^{22}\\
\bm{S}^{31} & \bm{S}^{32}
\end{bmatrix}\,.
\end{gather}
Accordingly, we can rewrite the matrix within the curly brackets in
Eq. (\ref{eq: MIS_GLRT eq2 proof}) as: :
\begin{gather}
\left\{ \bm{S}_{c}^{-1}-(\bm{S}_{c}^{-1}\bm{A})(\bm{A}^{\dagger}\bm{S}_{c}^{-1}\bm{A})^{-1}\bm{A}^{\dagger}\,\bm{S}_{c}^{-1}\right\} \nonumber \\
=\begin{bmatrix}\bm{0}_{J\times J} & \bm{0}_{J\times(N-J)}\\
\bm{0}_{(N-J)\times J} & \bm{S}_{33}^{-1}
\end{bmatrix}\,.
\end{gather}
Finally, gathering the above results leads to:
\begin{gather}
(\bm{Z}_{{\scriptscriptstyle W1}}\bm{V}_{c,1})^{\dagger}\,\bm{P}_{\bm{A}_{1}}^{\perp}(\bm{Z}_{{\scriptscriptstyle W1}}\bm{V}_{c,1})=\bm{Z}_{3}^{\dagger}\,\bm{S}_{33}^{-1}\,\bm{Z}_{3}\label{eq: Appendix - Equality (2)}
\end{gather}
which proves Eq. (\ref{eq: MIS_E2}).\emph{ }

\subsection*{Proof of Eqs. (\ref{eq: MIS_E3}) and (\ref{eq: MIS_E4})}

Hereinafter we provide a proof of Eqs. (\ref{eq: MIS_E3}) and (\ref{eq: MIS_E4}),
which are fundamental for proving CFARness of Rao (Durbin) statistic.
Firstly, it can be shown that:
\begin{align}
\widehat{\bm{R}}_{0}^{-1/2}\,\bm{P}_{\bar{\bm{A}}_{0}}^{\perp}\widehat{\bm{R}}_{0}^{-1/2} & =\begin{bmatrix}\bm{0}_{t\times t} & \bm{0}_{t\times(N-t)}\\
\bm{0}_{(N-t)\times t} & \widehat{\bm{R}}_{0,2}^{-1}
\end{bmatrix}\,;\\
\widehat{\bm{R}}_{0}^{-1/2}\,\bm{P}_{\bar{\bm{A}}_{1}}^{\perp}\,\widehat{\bm{R}}_{0}^{-1/2} & =\begin{bmatrix}\bm{0}_{J\times J} & \bm{0}_{J\times(N-J)}\\
\bm{0}_{(N-J)\times J} & \widehat{\bm{R}}_{0,33}^{-1}
\end{bmatrix}\,;
\end{align}
where we have defined the following partitioning:
\begin{equation}
\widehat{\bm{R}}_{0}=\begin{bmatrix}\widehat{\bm{R}}_{0,11} & \widehat{\bm{R}}_{0,12} & \widehat{\bm{R}}_{0,13}\\
\widehat{\bm{R}}_{0,21} & \widehat{\bm{R}}_{0,22} & \widehat{\bm{R}}_{0,23}\\
\widehat{\bm{R}}_{0,31} & \widehat{\bm{R}}_{0,32} & \widehat{\bm{R}}_{0,33}
\end{bmatrix},\quad\widehat{\bm{R}}_{0,2}\triangleq\begin{bmatrix}\widehat{\bm{R}}_{0,22} & \widehat{\bm{R}}_{0,23}\\
\widehat{\bm{R}}_{0,32} & \widehat{\bm{R}}_{0,33}
\end{bmatrix},
\end{equation}
where $\widehat{\bm{R}}_{0,ij}$, $(i,j)\in\{1,2,3\}\times\{1,2,3\}$,
is a sub-matrix whose dimensions can be obtained replacing $1$, $2$
and $3$ with $t$, $r$ and $(N-J)$, respectively. Then, exploiting
the above results we obtain:
\begin{eqnarray}
(\bm{Z}_{{\scriptscriptstyle W0}}\bm{V}_{c,1})^{\dagger}\,\bm{P}_{\bar{\bm{A}}_{0}}^{\perp}(\bm{Z}_{{\scriptscriptstyle W0}}\bm{V}_{c,1}) & = & \bm{Z}_{23}^{\dagger}\,\widehat{\bm{R}}_{0,2}^{-1}\,\bm{Z}_{23}\label{eq: R1_intermediate}\\
(\bm{Z}_{{\scriptscriptstyle W0}}\bm{V}_{c,1})^{\dagger}\,\bm{P}_{\bar{\bm{A}}_{1}}^{\perp}\,(\bm{Z}_{{\scriptscriptstyle W0}}\bm{V}_{c,1}) & = & \bm{Z}_{3}^{\dagger}\,\widehat{\bm{R}}_{0,33}^{-1}\,\bm{Z}_{3}\label{eq: R_2 intermediate}
\end{eqnarray}
Furthermore, it is not difficult to show, starting from Eq. (\ref{eq: Closed form MLE covariance H_0}),
that
\begin{eqnarray}
\widehat{\bm{R}}_{0,2} & = & K^{-1}\,[\bm{S}_{2}+\bm{Z}_{23}\,\bm{Z}_{23}^{\dagger}]\,,\\
\widehat{\bm{R}}_{0,33} & = & K^{-1}\,[\bm{S}_{33}+\bm{Z}_{3}\,\bm{Z}_{3}^{\dagger}]\,.
\end{eqnarray}
Finally, after substitution into Eqs. (\ref{eq: R1_intermediate})
and (\ref{eq: R_2 intermediate}) and application of Woodbury identity,
the claimed result is obtained.

\subsection*{Proof of Eqs. (\ref{eq: MIS_E5}) and (\ref{eq: MIS_E6})}

In what follows, we provide a proof of Eqs. (\ref{eq: MIS_E5}) and
(\ref{eq: MIS_E6}), which are exploited in Sec. \ref{sub: CFARness Gradient statistic}
for proving CFARness of Gradient statistic. We first observe that
the following equalities hold:
\begin{gather}
(\bm{Z}_{{\scriptscriptstyle W1}}\bm{V}_{c,1})^{\dagger}\,\bm{P}_{\bm{A}_{0}}^{\perp}\,\bm{S}_{c}^{1/2}\,\widehat{\bm{R}}_{0}^{-1/2}(\bm{Z}_{{\scriptscriptstyle W0}}\bm{V}_{c,1})\nonumber \\
=(\bm{Z}_{{\scriptscriptstyle W0}}\bm{V}_{c,1})^{\dagger}\,\bm{P}_{\bar{\bm{A}}_{0}}^{\perp}(\bm{Z}_{{\scriptscriptstyle W0}}\,\bm{V}_{c,1})\label{eq: CFARness GT (1st eq)}\\
(\bm{Z}_{{\scriptscriptstyle W1}}\bm{V}_{c,1})^{\dagger}\,\bm{P}_{\bm{A}_{1}}^{\perp}\,\bm{S}_{c}^{1/2}\,\widehat{\bm{R}}_{0}^{-1/2}(\bm{Z}_{{\scriptscriptstyle W0}}\bm{V}_{c,1})\nonumber \\
=K\,\left[(\bm{Z}_{{\scriptscriptstyle W1}}\bm{V}_{c,1})^{\dagger}\bm{P}_{\bm{A}_{1}}^{\perp}\,(\bm{Z}_{{\scriptscriptstyle W1}}\bm{V}_{c,1})\right.\nonumber \\
-(\bm{Z}_{{\scriptscriptstyle W1}}\bm{V}_{c,1})^{\dagger}\bm{P}_{\bm{A}_{1}}^{\perp}\,\bm{Z}_{{\scriptscriptstyle W1}}\bm{V}_{c,1}\left(\bm{I}_{M}+(\bm{Z}_{{\scriptscriptstyle W1}}\bm{V}_{c,1})^{\dagger}\,\bm{P}_{\bm{A}_{0}}^{\perp}\bm{Z}_{{\scriptscriptstyle W1}}\bm{V}_{c,1}\right)^{-1}\nonumber \\
\left.\times(\bm{Z}_{{\scriptscriptstyle W1}}\bm{V}_{c,1})^{\dagger}\,\bm{P}_{\bm{A}_{0}}^{\perp}\,\bm{Z}_{{\scriptscriptstyle W1}}\bm{V}_{c,1}\right]\label{eq: CFARness GT (2nd eq)}
\end{gather}
where we have used  Eq.~(\ref{eq: Lem1 R0}) of Lem. \ref{lem: ML covariance estimates properties}
in deriving the right-hand side  of Eq.~(\ref{eq: CFARness GT (1st eq)}).
Differently, Eq. (\ref{eq: CFARness GT (2nd eq)}) is obtained exploiting
the following steps: ($i$) use of $\widehat{\bm{R}}_{0}$ explicit
expression given by  Eq. (\ref{eq: Closed form MLE covariance H_0}),
($ii$) application of   Woodbury Identity to $\widehat{\bm{R}}_{0}^{-1}$
and ($iii$) simplification through the use of  the equality $\bm{P}_{\bm{A}_{1}}^{\perp}\bm{P}_{\bm{A}_{0}}^{\perp}=\bm{P}_{\bm{A}_{1}}^{\perp}$.
Finally, by exploiting\emph{ }Eq.~(\ref{eq: MIS_E3}) into Eq.~(\ref{eq: CFARness GT (1st eq)})
and Eqs. (\ref{eq: MIS_E1}) and (\ref{eq: MIS_E2}) into Eq.~(\ref{eq: CFARness GT (2nd eq)}),
we demonstrate the considered equalities.

\section{Simulation results showing specific coincidence results\label{sec: Simresults}}

In this section we confirm, through numerical results, the statistical
equivalence results obtained among the considered detectors for specific
adaptive detection scenarios. We remark that the simulation parameters
(i.e., the the structure of the covariance $\bm{R}$ and the generation
process for unknown signal matrix $\bm{B}$) are the same as those
used in the manuscript (as well as the Monte Carlo setup) and thus
are not reported for the sake of brevity.

\begin{figure}
\centering{}\includegraphics[width=0.98\columnwidth]{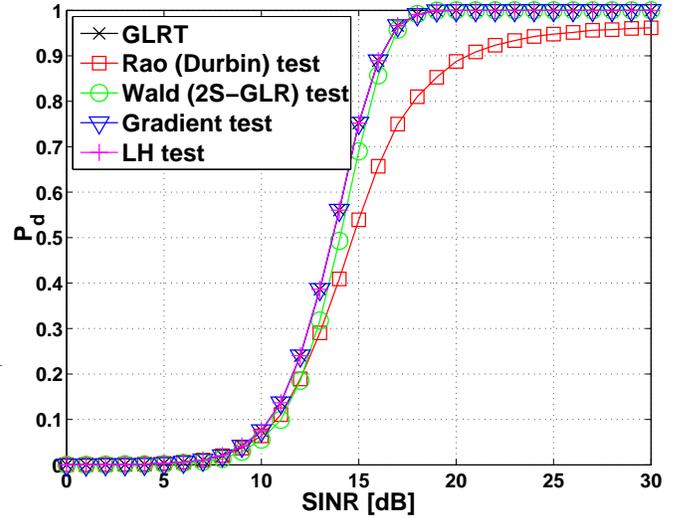}\protect\caption{$P_{d}$ vs. $\rho$ for all the considered detectors; vector subspace
detection with point-like ($M=1$) signal ($r=2$) and interference
($t=4$). Parameters $K=13$ and $N=8$. \label{fig: vector subspace point targets and interference} }
\end{figure}

First, in Fig. \ref{fig: vector subspace point targets and interference}
we show $P_{d}$ vs. $\rho$ (given $P_{fa}=10^{-4}$) for a setup
with point-like signal and interference ($M=1$) where the signal
belongs to a two-dimensional vector suspace ($r=2$) while the interference
to a four-dimensional vector subspace ($t=4$). We assume that each
column of $\bm{Z}$ is a vector of $N=8$ elements and $K=13$ samples
are assumed. It is apparent the statistical equivalence among GLR,
Gradient and LH tests.

\begin{figure}
\centering{}\includegraphics[width=1\columnwidth]{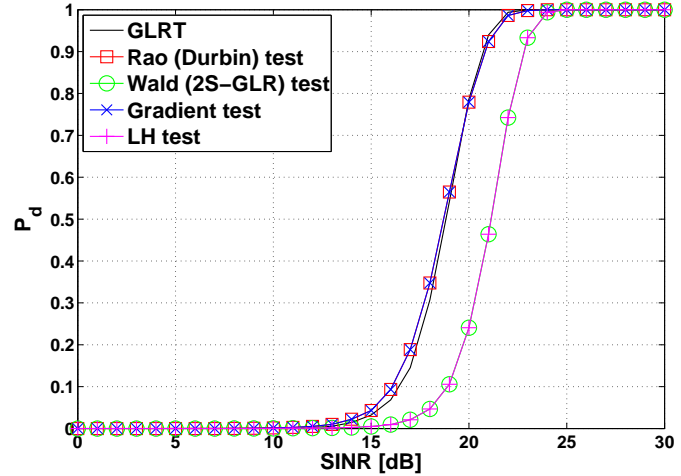}\protect\caption{$P_{d}$ vs. $\rho$ for all the considered detectors; Multidimensional
signals ($r=8$, $t=0$ and $N=8$). Parameters $K=24$ and $M=8$.
\label{fig: multidim signals} }
\end{figure}

Similarly, in Fig. \ref{fig: multidim signals} we show $P_{d}$ vs.
$\rho$ (given $P_{fa}=10^{-4}$) for a setup with multidimensional
signals ($N=r=8$) where $M=8$ and $K=24$. It is apparent the statistical
equivalence between Rao and Gradient tests and between Wald and LH
tests. 

\begin{figure}
\centering{}\includegraphics[width=0.98\columnwidth]{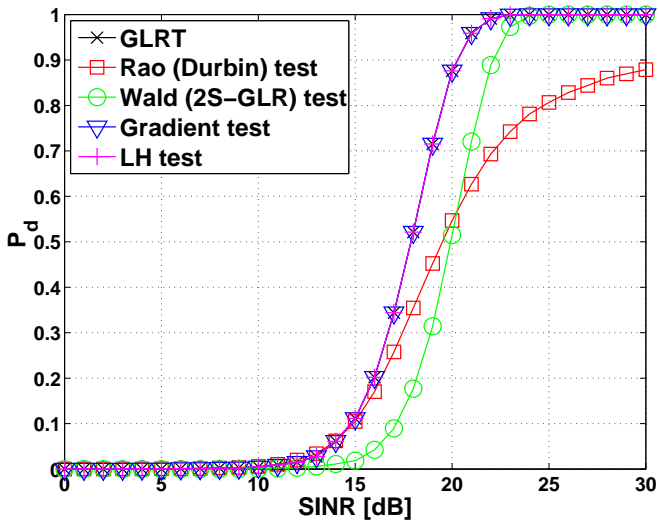}\protect\caption{$P_{d}$ vs. $\rho$ for all the considered detectors; Range-spread
targets ($M=8$) with rank-one subspace ($r=1$) and no-interference
($t=0$). Parameters $K=24$ and $N=8$. \label{fig: range-spread target} }
\end{figure}

Finally, in Fig. \ref{fig: range-spread target} we report $P_{d}$
vs. $\rho$ (given $P_{fa}=10^{-4}$) for a range-spread target ($M=8$)
with a rank-one signal subspace ($r=1$) and no interference ($t=0$),
where $K=24$ and $N=8$. It is apparent the statistical equivalence
among GLR, Gradient and LH tests.

\bibliographystyle{IEEEtran}
\bibliography{IEEEabrv,bib_adapt_det}

\end{document}